\documentclass[journal]{IEEEtran} \newcommand{\figsza}{0.99} \newcommand{\figszb}{0.99}  \newcommand{\figszc}{0.99} \newcommand{\figszd}{0.99} \newcommand{\figsze}{1.7} \newcommand{\figszf}{0.75}  \newcommand{\figszg}{0.95}  \newcommand{\figszh}{0.99} 

\usepackage[cmex10]{amsmath}
\usepackage{amsfonts}
\usepackage{amssymb}
\usepackage{amsopn}
\usepackage{graphicx}
\usepackage{subfigure}
\usepackage{epsfig}
\usepackage{color}
\usepackage{setspace}
\usepackage{bbm}
\usepackage{algorithm}
\usepackage{algorithmic}

\newtheorem{result:D0}{Proposition}
\newtheorem{result:cum}[result:D0]{Theorem}
\newtheorem{inequality}[result:D0]{Lemma}
\newtheorem{result:conn}[result:D0]{Proposition}
\newtheorem{result:conn:inequality}[result:D0]{Proposition}
\newtheorem{result:cyc}[result:D0]{Proposition}
\newtheorem{result:SDR}[result:D0]{Proposition}
\newtheorem{soln:special:indiv}[result:D0]{Proposition}
\newtheorem{result:indiv:conn}[result:D0]{Proposition}
\newtheorem{result:skewness}[result:D0]{Lemma}
\newtheorem{result:CG:indiv}[result:D0]{Proposition}

\newtheorem{app:indiv:dist}[result:D0]{Lemma}
\newtheorem{app:indiv:conn}[result:D0]{Lemma}

\graphicspath{{fig/}}

\newcommand{\ud}{\,\mathrm{d}}

\newcommand{\tr}{\text{Tr }}
\newcommand{\bo}[1]{\boldsymbol{#1} } 
\newcommand{\tbo}[1]{\widetilde{\boldsymbol{#1}}}
\newcommand{\sups}[1]{\ensuremath{^{\textrm{#1}}}}

\newcommand{\indicator}[1]{\mathbbm{1}_{\left[ {#1} \right] }}
\newcommand{\specialcell}[2][c]{\begin{tabular}[#1]{@{}c@{}}#2\end{tabular}}
\newcommand{\bs}{\begin{smallmatrix}}
\newcommand{\es}{\end{smallmatrix}}

\begin{document}

\markboth{UNDER REVIEW IN IEEE TRANSACTIONS ON INFORMATION THEORY}{Kar \MakeLowercase{\textit{et al.}}: Linear Coherent Estimation with Spatial Collaboration}

\title{Linear Coherent Estimation with Spatial Collaboration}

\author{Swarnendu~Kar,~\IEEEmembership{Student~Member,~IEEE,}
        and~Pramod~K.~Varshney,~\IEEEmembership{Fellow,~IEEE}
\thanks{S. Kar and P. K. Varshney are with the Department
of Electrical Engineering and Computer Science, Syracuse University, Syracuse,
NY, 13244 USA. E-mail: swkar@syr.edu.}
\thanks{Part of this work is accepted for presentation for ISIT-2012, IEEE International Symposium of Information Theory, July 1--6, 2012, Cambridge, MA, USA. }
\thanks{This research was partially supported by the National Science Foundation under Grant No. $0925854$ and the Air Force Office of Scientific Research under Grant No. FA-9550-10-C-0179. }
}


\maketitle

\begin{abstract}
A power constrained sensor network that consists of multiple sensor nodes and a fusion center (FC) is considered, where the goal is to estimate a random parameter of interest. In contrast to the distributed framework, the sensor nodes may be partially connected, where individual nodes can update their observations by (linearly) combining observations from other adjacent nodes. The updated observations are communicated to the FC by transmitting through a coherent multiple access channel. The optimal collaborative strategy is obtained by minimizing the expected mean-square-error subject to power constraints at the sensor nodes. Each sensor can utilize its available power for both collaboration with other nodes and transmission to the FC.  Two kinds of constraints, namely the cumulative and individual power constraints are considered. The effects due to imperfect information about observation and channel gains are also investigated. The resulting performance improvement is illustrated analytically through the example of a homogeneous network with equicorrelated parameters. Assuming random geometric graph topology for collaboration, numerical results demonstrate a significant reduction in distortion even for a moderately connected network, particularly in the low local-SNR regime.
\end{abstract}


\section{Introduction}
\begin{figure}[htb]
\centering
    \includegraphics[width=\figsza \columnwidth]{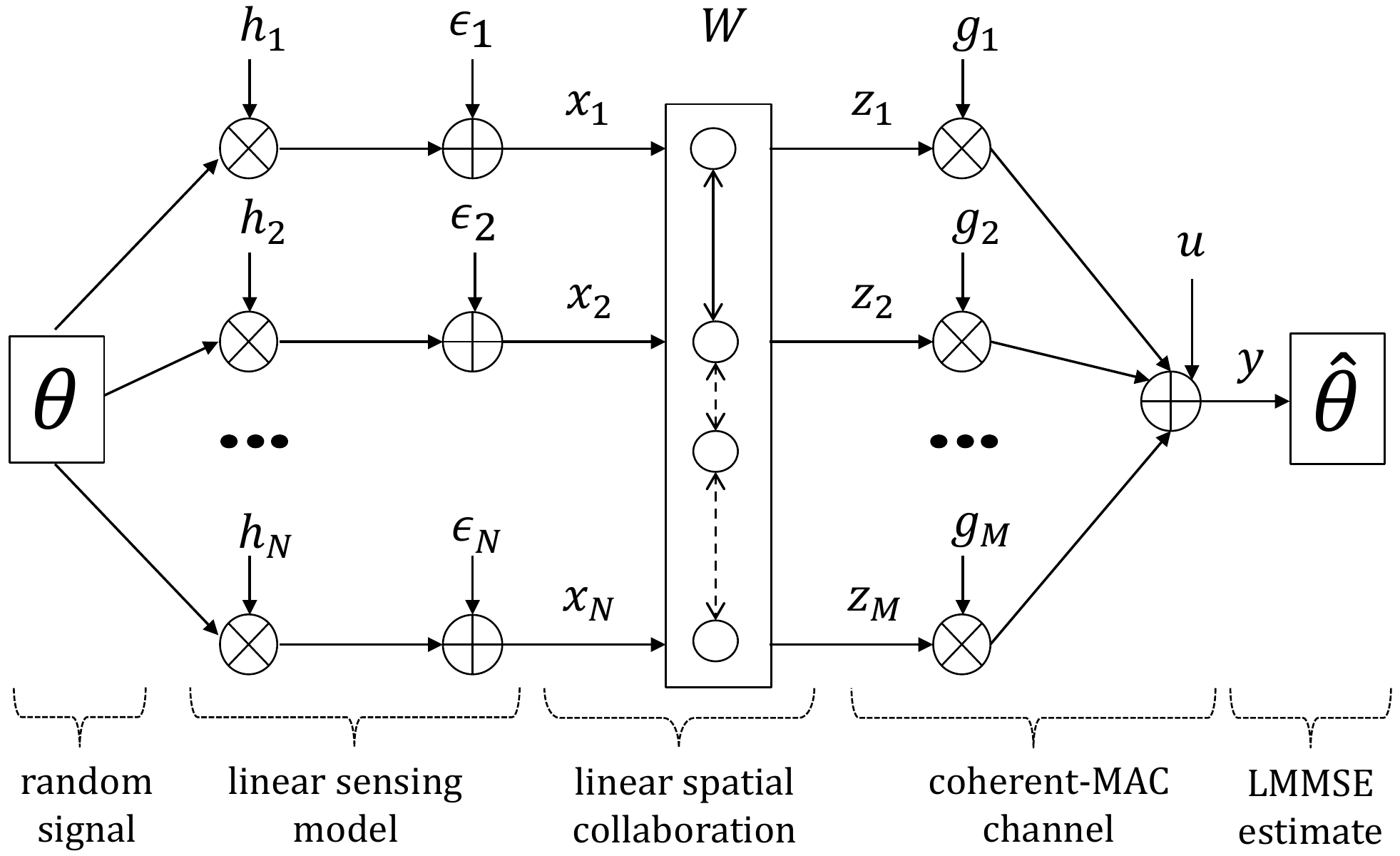}
  \caption{Sensor network performing collaborative estimation.}
  \label{fig:schematic}
\end{figure}

Wireless sensor networks consist of spatially distributed battery-powered sensors that monitor certain environmental conditions and often cooperate to perform specific signal processing tasks such as detection, estimation and classification \cite{Akyildiz02}. In this paper, we consider a network that is deployed for the purpose of estimating a common random parameter of interest. After observing noisy versions of the parameter, the sensors can share their observations among other neighboring nodes, an act referred to as \emph{collaboration} in this paper (following \cite{Fang09}). The observations from the neighbors are linearly combined using appropriate weights and then transmitted to the fusion center (FC) through a coherent multiple access channel (MAC). The FC receives the noise-corrupted signal and makes the final inference. The schematic diagram of such a system is shown in Figure \ref{fig:schematic} (we will introduce the notations and describe each block later in Section \ref{sec:probform}). 

The individual sensor nodes are battery powered and hence the network is power limited. The power constraints can be described by the following two situations 1) \emph{Cumulative}: Here the total power-usage in the network (summed across all the nodes) has to be below a pre-specified limit. 2) \emph{Individual}: Here each node has their own power constraint as dictated by the capacity of their batteries. The performance analysis of a \emph{cumulative-constraint} problem is usually simpler and more insightful since we have only one constraint to take care of. The \emph{individual-constraint} problem is more practical from an implementation perspective but also more difficult to analyze. In the absence of a power limit, the sensors could collaborate with all the other nodes, make the inference in the network, and transmit the estimated parameter to the FC without any further distortion (by using infinite transmission power). This is similar to the centralized inference situation, where the error in estimate is only due to the noisy observation process. However, with limited power availability, both collaboration and transmission have to be performed judiciously, so as to maximize the quality of inference at the FC. In this paper, we study the following problems. For a cumulative power constraint, we study the optimal allocation of power resources among various nodes and tasks (namely collaboration and transmission) so as to achieve the best estimation performance at the FC. Regarding individual power constraints, the goal is to allocate power between collaboration and transmission at each node.

In the absence of collaboration, this problem reduces to the class of distributed inference/beamforming algorithms. In \emph{distributed estimation}, the objective is to coordinate all the sensors so that without communicating with one-another, they collectively maximize the quality of inference at the FC. The quality of inference can be quantified by either the mean-square error (MSE) (in case of random signals) or Cram\'er-Rao lower bound (in case of deterministic unknown signals). Distributed estimation has been extensively researched both from analog \cite{Xiao08},\cite{Cui07},\cite{Sayed07} and digital \cite{Rib06},\cite{Li09},\cite{KarTSP12} encoding perspectives. Examples of analog encoding include the amplify-and-forward (AF) scheme, where the nodes amplify the raw observations and transmit to the FC by either forming a  coherent beam in a multiple access channel \cite{Xiao08} or using their own dedicated links with the FC for transmission (sometimes referred to as orthogonal MAC) \cite{Cui07}. The AF framework appears extensively in the literature \cite{Xiao08},\cite{Sayed07},\cite{Choi11} due to its simplicity of implementation in complex networks and provable information-theoretic-optimality properties for simple networks \cite{Gastpar08}. In another research direction, quantization of the observations may be performed prior to transmission  \cite{Luo05},\cite{Rib06}. The quantized observations are then communicated to the FC using digital communication, where further information may be lost due to channel errors \cite{Wu09}. Another closely related field of research in communication theory is \emph{distributed beamforming in relay networks} \cite{Havary08}, \cite{Jing09}, \cite{Li11}, where the objective is to maximize the signal-to-noise ratio (SNR), rather than minimize the estimation error, at the FC. These two problems are sometimes related, as one might imagine. When the observation and channel gains are perfectly known, the SNR and MSE functions are monotonically related and the two problems are equivalent. However, in the presence of observation and channel gain uncertainties, the SNR and MSE functions are different.

Though distributed inference/beamforming has been widely studied, research regarding collaborative estimation is relatively nascent. When the transmission channels are orthogonal and cost-free collaboration is possible within a fully connected sensor network, the optimal strategy is to perform the inference in the network and use the best available channel to transmit the estimated parameter \cite{Fang09}. In a rate-constrained framework considered in \cite{Kar11}, it was shown that spatial collaboration can be used to whiten the observation space, thereby enabling efficient resource allocation when the noise is correlated. In a preliminary version of this paper \cite{KarISIT12}, we have considered an extension of the AF framework, where sensors are able to linearly combine the observations from neighboring nodes before transmitting to the FC. We obtained the optimal cumulative power-distortion tradeoff when a fixed but otherwise cost-free collaborative topology is used to transmit over a coherent MAC channel. In this paper, we extend the problem formulation in three new directions, namely a) consideration of individual power constraints, b) consideration of imperfect information about observation and channel gains (the second order statistics are assumed to be known), and c) consideration of finite costs associated with collaboration. The primary contributions of this paper are as follows
\begin{itemize}
\item Extending the amplify-and-forward framework to formulate and analyze the problem of estimation with spatial collaboration
\item Defining a metric called \emph{collaboration gain}, that quantifies the worthiness of spatial collaboration as a tool to enhance the estimation performance
\item Demonstrating that for a fixed but otherwise cost-free (ideal) collaborative topology, the resulting optimization problem reduces to an eigen-decomposition problem for the cumulative-constraint case. For the individual-constraint case, accurate numerical solution can be obtained by solving several semi-definite feasibility problems. We investigate both the cases further by
deriving/analyzing the optimal achievable distortion and the corresponding weights for some special collaborative topologies like the distributed (no-connections), partially connected cycles and fully connected cases. We also derive the explicit expression of collaboration gain for a homogeneous network with identical channel and observation gains and equicorrelated observation noise. In particular, we demonstrate that collaboration is particularly effective in a certain power regime that depends on various factors like the \emph{skewness} (or variability) of power-availability in the network, uncertainty in observation and channel gains, and the correlation of the measurement noise.
\item Addressing the design of collaborative topologies where finite costs are involved in collaboration. We suggest an efficient algorithm that uses the results for the fixed-topology but cost-free case to find locally optimal solutions for the finite-cost case.
\end{itemize}

The rest of the paper is organized as follows. In Section \ref{sec:probform}, we formulate the problem after describing each block of the system in Figure \ref{fig:schematic}. We define  ``collaboration gain" ($\textsf{CG}$) that is normalized with respect to the operating region ($\textsf{CG}\in [0,1]$). It summarizes the efficacy of collaboration across various problem conditions. In Section \ref{sec:ideal}, we solve the optimal transmission-power allocation problem for a fixed but otherwise cost-free collaborative topology, (i.e., extend the results of \cite{KarISIT12} to address points (a) and (b) in the previous paragraph). We also derive explicit expressions for collaboration gain for a homogeneous network with equicorrelated noise. In Section \ref{sec:finite}, we address the problem with finite collaboration costs and suggest a greedy algorithm to obtain a locally optimal solution in polynomial time. Concluding remarks are presented in Section \ref{sec:conc}.

\section{Problem Formulation} \label{sec:probform}
\subsection{Linear Sensing Model}
We consider the scenario where the parameter of interest is a scalar random variable with known statistics, specifically, Gaussian distributed with zero mean and variance $\eta^2$. The observations at the sensor nodes $n=1,2,\ldots,N$ are governed by the linear model $x_n = \widetilde{h}_n \theta+\epsilon_n$, where $\widetilde{h}_n$ is the observation gain and $\epsilon_n$ is the measurement noise. The second order statistics of the observation gain $\widetilde{\bo h}=[\widetilde{h}_1,\widetilde{h}_2,\ldots,\widetilde{h}_N]^T$ is assumed to be 
\begin{align}
\mathbb E \, \widetilde{\bo h} =\bo{h}, \quad \textsf{var} \, \widetilde{\bo h}=\bo{\Sigma}_\textsf{h}.
\end{align} 
The measurement noise $\bo \epsilon=[\epsilon_1,\epsilon_2,\ldots,\epsilon_N]^T$ is assumed to be zero-mean, Gaussian with (spatial) covariance $\textsf{var} \, \bo \epsilon=\bo \Sigma$. \emph{Perfect knowledge of the observation model statistics $\bo{h}, \bo{\Sigma}_\textsf{h}$ and $\bo \Sigma$ is assumed.} In vector notation, we have
\begin{align}
\bo x = \widetilde{\bo h} \theta + \bo \epsilon, \label{def:x}
\end{align}
where $\bo x=[x_1,x_2,\ldots,x_N]^T$ denotes the observations. 


\subsection{Linear Spatial Collaboration}
We consider an extension of the analog amplify-and-forward scheme as our encoding and modulation framework for communication to the fusion center. In the basic amplify-and-forward scheme, each node transmits a weighted version of its own observation, say $W_n x_n$, with resulting power $W_n^2 \mathbb E[x_n^2]$. Such a scheme is appealing and often-used \cite{Cui07},\cite{Xiao08},\cite{Fang09} due to two reasons, 1) \emph{Uncoded nature:} Does not require block coding across time and hence efficient for low-latency systems, 2) \emph{Optimal in select cases:} For a memoryless Gaussian source transmitted through an additive white Gaussian noise (AWGN) channel (Figure \ref{fig:schematic} with $N=1$), an amplify-and-forward scheme helps achieve the optimal power-distortion tradeoff in an information-theoretic sense (see Example $2.2$ in \cite{Gastpar02}). The optimality of linear coding has also been established \cite{Gastpar03} for distributed estimation over a coherent MAC (Figure \ref{fig:schematic} without spatial collaboration).

In general, all the $N$ data-collecting sensor nodes in a network may not have the ability to communicate with the FC. In that case, they may still pass their information (through the act of collaboration) to another node which has a communication link with the FC (see the network in Figure \ref{fig:tx:example}, for example). We assume that the nodes are ordered in such a way that the first $M$ nodes (where $M\le N$) are able to communicate with the FC. Let the availability of collaborative links among the various nodes be represented by the $M\times N$ zero-one adjacency matrix (not necessarily symmetric) $\bo A$, where $A_{mn}\in \{0,1 \}$. An entry $A_{mn}=1$ signifies that node $n$ shares its observation with node $m$. \emph{Sharing of this observation is assumed to be realized through a reliable communication link that  consumes power $C_{mn}$, regardless of the actual value of observation}. The $M\times N$ matrix $\bo C$ describes all the costs of collaboration among various sensors and is assumed to be known. Since each node is trivially connected to itself, $A_{mm}=1$ and $C_{mm}=0$. We denote the set of all $\bo A$-sparse matrices as
\begin{align}
\mathcal S_A\triangleq \{\bo W\in\mathbb R^{M\times N}:W_{mn}=0 \text{ if } A_{mn}=0 \}.
\end{align}
Corresponding to an adjacency matrix $\bo A$ and an $\bo A$-sparse matrix $\bo W$, we define \emph{collaboration} in the network as individual nodes being able to linearly combine local observations from other collaborating nodes, 
\begin{align}
z_m=\sum_{\bs n=1,\ldots,N \\ A_{mn}=1 \es } W_{mn} x_n, \; m=1,\ldots,M.
\end{align} 
In effect, the network is able to achieve a one-shot spatial transformation $\bo W: \bo x\rightarrow \bo z$ of the form\footnote{It is worth emphasizing our assumption that though collaboration incurs a fixed cost (in terms of power consumed that could otherwise have been used for transmission), it is otherwise reliable, in the sense that the act of collaboration does not incur any errors. This can be implemented by, say communicating in a digital framework with sufficient precision and ensuring sufficient channel coding to counter the channel noise. We abstract this process by assigning a cost to the link when it is required to be used. An interesting problem which is worthy of research but beyond the scope of this paper, is to assume possibly erroneous collaboration, where $\bo z=\bo W \bo x+ \bo \zeta$ (say) and errors incurred during collaboration ($\bo \zeta$) decrease with collaboration power.}
\begin{align}
\bo z=\bo W \bo x, \quad \bo W\in\mathcal S_A. \label{def:z}
\end{align}
We refer to $\bo W$ as the matrix containing \emph{collaboration weights}. It may be noted that, 1) \emph{Particularization:} When $\bo W$ is a diagonal matrix (equivalently, $\bo A$ is the identity matrix $\bo I_M$), our collaborative scheme simplifies to the basic amplify-and-forward relay strategy as in \cite{Sayed07},\cite{Xiao08},  2) \emph{Collaboration cost:} Any collaboration involving $\bo W\in \mathcal S_A$ is achieved at the expense of power
\begin{align}
Q_{\bo A,m}\triangleq \sum_{n=1}^N C_{mn} A_{mn}, \label{def:QAn}
\end{align}
at node $m$, and cumulatively
\begin{align}
Q_{\bo A}\triangleq \sum_{m=1}^M Q_{\bo A,m}, \label{def:QA}
\end{align}
for the entire network, and 3) \emph{Transmission cost:} The power required for transmission of encoded message $z_m$ at node $m$ is,
\begin{align}
P_{\bo W,m} &\triangleq  \mathbb E_{\theta, \widetilde{\bo h}, \bo \epsilon}\left[z_m^2; \bo W\right]= \left[\bo W \bo E_\textsf{x} \bo W^T\right]_{m,m}, \mbox{ where} \label{def:PWm} \\
\bo E_\textsf{x} &\triangleq  \mathbb E_{\theta, \widetilde{\bo h}, \bo \epsilon}\left[\bo x \bo x^T\right]=\bo \Sigma+\eta^2(\bo h\bo h^T+\bo\Sigma_\textsf{h}).  \label{def:Ex}
\end{align}
Consequently, the cumulative transmission power in the network is 
\begin{align}
P_{\bo W}&\triangleq \sum_{m=1}^M P_{\bo W,m}=\tr\left[\bo W \bo E_\textsf{x} \bo W^T\right]. \label{def:PW}
\end{align}



\subsection{Coherent Multiple Access Channel}
The transformed observations $\bo z$ are assumed to be transmitted to the fusion center through a coherent MAC channel. In practice, a coherent MAC channel can be realized through \emph{transmit beamforming} \cite{Mudumbai09}, where sensor nodes simultaneously transmit a common message (in our case, all $z_m$-s are scaled versions of a common $\theta$) and the phases of their transmissions are controlled so that the signals constructively combine at the FC. Denote the channel gain at node $m$ by $\widetilde{g}_m$. The second order statistics of the channel $\widetilde{\bo g} \triangleq [\widetilde g_1, \widetilde g_2,\ldots, \widetilde g_M]$ is assumed to be,
\begin{align}
\mathbb E \,\widetilde{\bo g} =\bo{g}, \quad \textsf{var} \,\widetilde{\bo g}=\bo{\Sigma}_\textsf{g},
\end{align}
and the noise of the coherent MAC channel $u$ is assumed to be a zero-mean AWGN with variance $\xi^2$. \emph{Perfect knowledge of the channel statistics $\bo g$, $\bo \Sigma_\textsf{g}$ and $\xi^2$ is assumed.} The output of the coherent MAC channel (or the input to the fusion center) is 
\begin{subequations}
\begin{align}
y&=\widetilde{\bo g}^T\bo W \bo x+u, \quad u\sim\mathcal N(0,\xi^2)  \label{def:y} \\
 &=\underbrace{\widetilde{\bo g}^T \bo W \widetilde{\bo h}}_{\text{net gain}} \theta + \underbrace{\widetilde{\bo g}^T \bo W \bo \epsilon +u}_{\text{net zero-mean noise}}. \label{y:spread}
\end{align}
\end{subequations}

\subsection{Linear Minimum Mean Square Estimation}
Having received $y$, the goal of the fusion center is to obtain an accurate estimate $\widehat \theta$ of the original random parameter $\theta$. We restrict our attention to linear estimators of the form $\widehat \theta=a y$, where $a$ is a fixed constant subject to design. We consider the mean square error (MSE) as the distortion metric 
\begin{align}
\mathcal D_{\bo W}(a) \triangleq \mathbb E_{\theta,\widetilde{\bo h},\bo \epsilon,\widetilde{\bo g}, u}\left[(\theta-a y)^2; \bo W \right]. \label{def:DWa}
\end{align}  
From the theory of linear minimum mean square estimation (LMMSE, see \cite{Kay93}, Chapter 12), we readily obtain that
\begin{subequations}
\begin{align}
a_{\textsf{LMMSE}}&\triangleq \arg \min_{a} \mathcal D_{\bo W}(a)= \frac{\mathbb E[y\theta]}{\mathbb E[y^2]}, \mbox{ and} \label{lmmse:est} \\
D_{\bo W} &\triangleq \mathcal D_{\bo W}(a_{\textsf{LMMSE}})=\eta^2-\frac{(\mathbb E[y\theta])^2}{\mathbb E[y^2]}, \label{def:DW}
\end{align}
\end{subequations}
where the above expectations are w.r.t. all random variables $\{\theta,\widetilde{\bo h},\bo \epsilon,\widetilde{\bo g}, u\}$. From \eqref{def:y} and \eqref{y:spread}, we obtain 
\begin{align}
\begin{split}
\mathbb E\left[y^2\right]&=\tr\left[\bo E_\textsf{g} \bo W \bo E_\textsf{x} \bo W^T\right]+\xi^2, \mbox{ and}\\
\mathbb E\left[y\theta\right]&=\eta^2 \bo g^T \bo W \bo h, \mbox{ where}
\end{split} \\
\bo E_\textsf{g} &\triangleq \mathbb E\left[\widetilde{\bo g} \widetilde{\bo g}^T\right]=\bo g \bo g^T+\bo \Sigma_\textsf{g}, \label{def:Eg}
\end{align}
where $\bo E_\textsf{x}$ is defined in \eqref{def:Ex}.

\emph{Remark: Perfect observation gain and channel state information (OGI and CSI):}  When the observation and channel gains are precisely known, i.e., $\bo \Sigma_\textsf{h}=\bo \Sigma_\textsf{g}=0$, \eqref{y:spread} reduces to a linear Gaussian model conditioned on $\theta$, 
\begin{align}
y|\theta\sim \mathcal N(\bo g^T \bo W \bo h \theta,\, \bo g^T \bo W \bo \Sigma \bo W^T \bo g +\xi^2),  \label{y:CSI}
\end{align}
and hence the LMMSE estimator is also the minimum mean square estimator (MMSE) \cite{Kay93}, 
\begin{align}
\widehat \theta_{\textsf{LMMSE}}:=a_{\textsf{LMMSE}} y = \mathbb E_{\theta,\bo \epsilon, u} [\theta|y]=:\widehat \theta_{\textsf{MMSE}},
\end{align}
i.e., $\widehat \theta_{\textsf{LMMSE}}$ minimizes the distortion over all possible estimators (not just within the linear class). 

We know from the theory of MMSE estimation (see \cite{Kay93, VanTrees07}) that the optimal distortion $D_\textsf{MMSE}$ is related to the Fisher Information (FI),
\begin{align}
J\triangleq -\mathbb E_{\theta} \left\{ \mathbb E_{ y|\theta} \left\{ \frac{\ud^2 \theta}{\ud \theta^2} p(\bo y|\theta) \right\} \right\}, \label{def:FI}
\end{align}
by $D_\textsf{MMSE}\ge \left( \frac{1}{\eta^2}+J \right)^{-1}$ in general and that equality holds for linear Gaussian models of the form \eqref{y:CSI}. Though the FI (Equation \eqref{def:FI}) results in
\begin{align} 
J=\frac{\left( \bo g^T \bo W \bo h \right)^2}{ \bo g^T \bo W \bo \Sigma \bo W^T \bo g +\xi^2 } \label{def:J:both}
\end{align}
for the case of perfect OGI and perfect CSI (the linear Gaussian model in \eqref{y:CSI}), the FI is difficult to derive for cases when the observation and channel gains are uncertain. In fact, this is the main reason why we consider LMMSE estimation (which is suboptimal in general but easier to compute) rather than MMSE estimation (which is optimal but difficult to compute). 

For the purposes of notation in this paper, for all cases (whether we have perfect OGI/CSI or not), we would find it convenient to work with the quantity
\begin{align}
J_{\bo W}\triangleq \frac{1}{D_{\bo W}}-\frac{1}{\eta^2}. \label{def:JD}
\end{align}
as a surrogate for the LMMSE distortion $D_{\bo W}$ (as in \eqref{def:DW}). Note that $D_{\bo W}$ and $J_{\bo W}$ are monotonically related and that minimizing $D_{\bo W}$ is equivalent to maximizing $J_{\bo W}$. Motivated by the preceding discussions, we would refer to $J_{\bo W}$ as the \emph{equivalent Fisher Information}, or sometimes simply FI or even distortion, for the sake of brevity.

\subsection{Problem Statement}
The design of the collaboration weights $\bo W$ is critical since it affects both the power requirements and estimation performance of the entire application. Specifically, the following quantities depend on $\bo W$, 1) the resources required to collaborate, i.e., $Q_{\textsf{nz}(\bo W),m}$\footnote{Definition of operators $\textsf{nz}(\cdot)$, $\textsf{zero}(\cdot)$, and $\textsf{nnz}(\cdot)$: The operator $\textsf{nz}:\mathbb R^{N\times N}\rightarrow \{0,1\}^{N\times N}$ is used to specify the non-zero elements of a matrix. If $W_{ij}\neq 0$, then $\left[\textsf{nz}( \bo W)\right]_{ij}=1$, else $\left[\textsf{nz} (\bo W)\right]_{ij}=0$. Similarly, the operator $\textsf{zero}:\mathbb R^{N\times N}\rightarrow \{0,1\}^{N\times N}$ is used to specify the zero elements of a matrix, $\left[\textsf{zero}( \bo W)\right]_{ij}=1-\left[\textsf{nz}( \bo W)\right]_{ij}$. The operator $\textsf{nnz}:\mathbb R^{N\times N}\rightarrow \mathbb Z_{+}$ is used to specify the number of non-zero elements of a matrix.} for individual nodes or $Q_{\textsf{nz}(\bo W)}$ cumulatively for the network (see \eqref{def:QAn} and \eqref{def:QA}), 2) the resources required to transmit, i.e., $P_{\bo W,m}$ for individual nodes or $P_{\bo W}$ cumulatively for the network (see \eqref{def:PWm} and \eqref{def:PW}), and 3) the final distortion of the estimate at the FC, $D_{\bo W}$, provided by \eqref{def:DW}. In this paper, we address the problems of designing the collaboration matrix that minimizes the distortion subject to either 1) a system-wide cumulative power constraint,
\begin{equation}
\begin{aligned}
& \underset{\bo W}{\text{minimize}} & & D_{\bo W} \\
&                            \text{subject to} & & P_{\bo W}+Q_{\text{\textsf{nz}}(\bo W)}\le P^\textsf{C},
\end{aligned} \label{prob:cum}
\end{equation}
or 2) power constraints at individual sensor nodes,
\begin{equation}
\begin{aligned}
& \underset{\bo W}{\text{minimize}} & & D_{\bo W} \\
&                            \text{subject to} & & P_{\bo W,m}+Q_{\text{\textsf{nz}}(\bo W),m}\le P^\textsf{C}_m, \, m=1,\ldots,M.
\end{aligned} \label{prob:indiv}
\end{equation}
We note that problem  \eqref{prob:indiv} (with $M$ individual power constraints) is more realistic from a deployment point of view, since various nodes in a network can possess significantly different power sources, based on age of deployment or make/type of the batteries. However,  problem \eqref{prob:indiv} is significantly more difficult than problem \eqref{prob:cum}, which has only one cumulative power constraint. Problem \eqref{prob:cum} is more important from a system design and analysis point of view, since it is more tractable analytically and as a result, reveals significant insights on the various system level tradeoffs.

\subsection{Solution Methodology}
Both problems \eqref{prob:cum} and \eqref{prob:indiv}, in general, have no known procedure that efficiently computes (in polynomial-time) the globally optimal solution(s). However, \emph{for the special case when the entries of the collaboration cost matrix $\bo C$ are either zero or infinity, $C_{ij}\in\{0,\infty\}$, we will show that globally optimal solutions for both the problems can be obtained using efficient numerical techniques, and for problem \eqref{prob:cum}, even a closed-form solution can be derived}. 

Physically, this special case corresponds to the situation when the topology of a network is fixed (and hence not subject to design) and communication among neighbors is relatively inexpensive compared to communication with the FC. Let $\bo A=\text{\textsf{zero}}(\bo C)$ denote the permitted adjacency matrix for such a  situation. Hence, the collaboration costs vanish, and problems \eqref{prob:cum} and \eqref{prob:indiv} simplify to
\begin{equation}
\begin{aligned}
& \underset{\bo W\in \mathcal S_{\bo A}}{\text{minimize}} & & D_{\bo W} \\
&                                                             \text{subject to} & & P_{\bo W}\le P^\textsf{C},\; \mbox{ and}
\end{aligned} \label{prob:ideal:cum}
\end{equation}
\begin{equation}
\begin{aligned}
& \underset{\bo W\in \mathcal S_{\bo A}}{\text{minimize}} & & D_{\bo W} \\
&                            \text{subject to} & & P_{\bo W,m}\le P^\textsf{C}_m, \, m=1,\ldots,M.
\end{aligned} \label{prob:ideal:indiv}
\end{equation}
respectively, which are optimization problems in $\textsf{nnz}(\bo A)$ variables. Since problems \eqref{prob:ideal:cum} and \eqref{prob:ideal:indiv}, which will be solved in Section \ref{sec:ideal}, arise out of the assumption of zero-cost for collaboration, we would refer to them as \emph{ideal-collaborative} problems.

For the general case, when the topology is flexible and collaboration incurs a finite cost, a polynomial-time sub-optimal algorithm is proposed in Section \ref{sec:finite} where the bigger problem is broken into smaller sub-problems, where several ideal-collaborative problems of the form \eqref{prob:ideal:cum} or \eqref{prob:ideal:indiv} are solved at each iteration. Specifically, we start from the distributed topology $\bo A=\left[ \bo I_M | \bo 0 \right]$, and follow a greedy algorithm to augment the collaborative topology with the most power-efficient link at each iteration. 

\subsection{Performance Metric - Collaboration Gain} \label{sec:def:CG}
\begin{figure}[hbt]
\centering
    \includegraphics[width=\figszb \columnwidth]{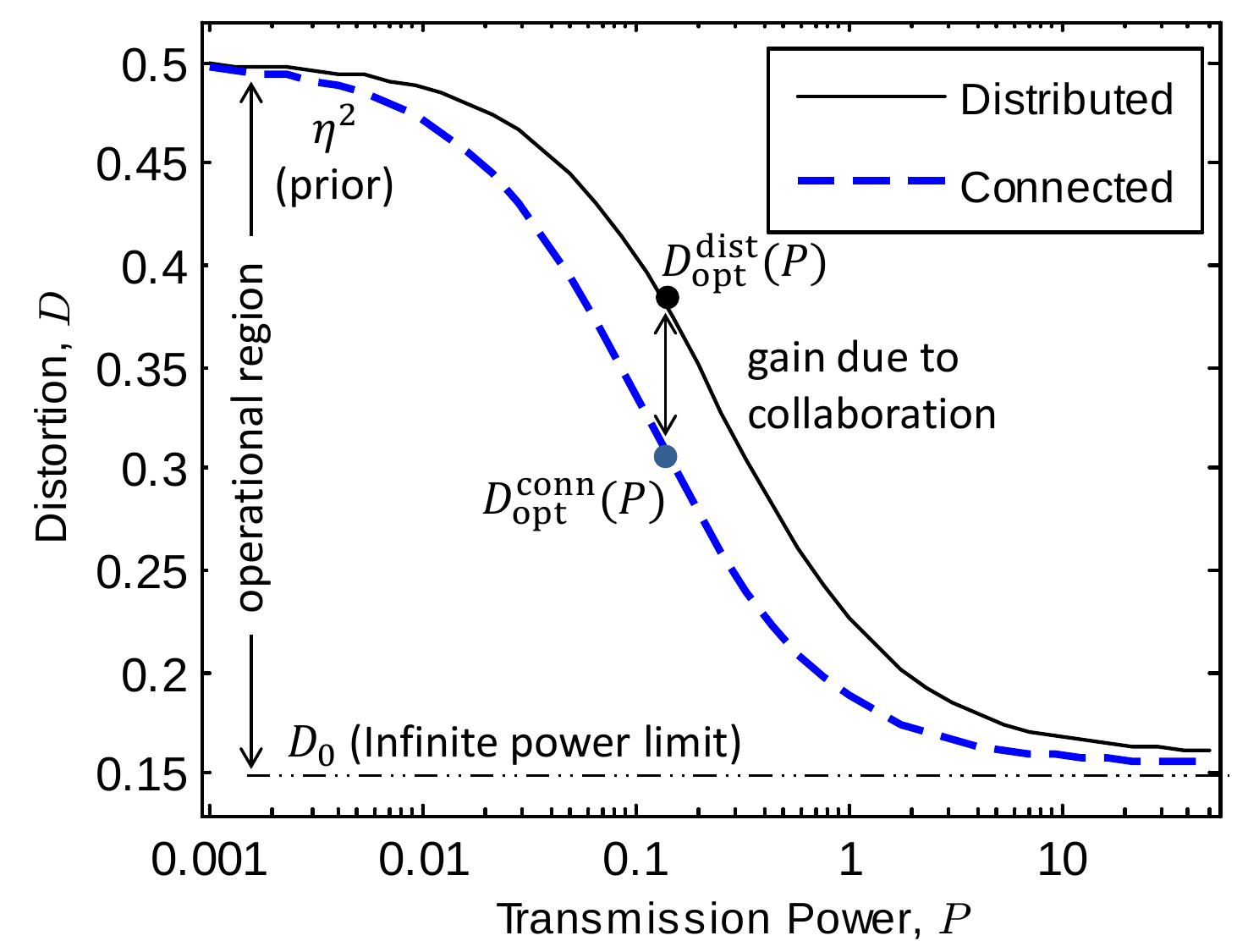}
  \caption{A typical power-distortion curve illustrating collaboration gain.}
  \label{fig:cooperative:snr}
\end{figure}

Let the optimal solutions to problems \eqref{prob:cum} and \eqref{prob:indiv} be denoted by $D_{\textsf{opt}}(\bo P^\textsf{C})$, where (note bold notation) $\bo P^\textsf{C}=P^\textsf{C}$ for a cumulative-constraint (problem \eqref{prob:cum}) and $\bo P^\textsf{C}=[P_1^\textsf{C},P_2^\textsf{C},\ldots,P_M^\textsf{C}]^T$ for individual constraints (problem \eqref{prob:indiv}). Generally, the distortion $D_{\textsf{opt}}(\bo P^\textsf{C})$ depends on a number of problem conditions other than $\bo P^\textsf{C}$, which includes 1) the source variance $\eta^2$, 2) the noise variance $\bo \Sigma$, 3) the observation gain statistics $\{\bo h, \bo \Sigma_\textsf{h}\}$, 4) the coherent channel gain statistics $\{\bo g, \bo \Sigma_\textsf{g}\}$, and 5) the power needed to collaborate (matrix $\bo C$). However, to assess and compare the benefits of spatial collaboration for a wide-range of problem conditions, we seek a metric that is normalized with respect to the operational region. Towards that goal, we define a few quantities. 1) Let 
\begin{align}
D_0 \triangleq D_{\textsf{opt}}\left(\bo P^\textsf{C}\rightarrow \bo \infty; \bo A=\bo 1 \bo 1^T \right), \label{def:D0}
\end{align}
denote the optimal distortion that can be obtained with arbitrary collaboration and without any power constraints (the actual value of $D_0$ will be derived later, see \eqref{J0} for an early preview). 
2) Let
\begin{align} 
D_\textsf{opt}^\textsf{dist}(\bo P^\textsf{C}) \triangleq D_{\textsf{opt}}\left(\bo P^\textsf{C}; \bo A=\left[ \bo I_M | \bo 0 \right] \right)
\end{align}
denote the optimal distortion for the distributed scenario, i.e., transmission power is optimally allocated among sensors and there is no collaboration among them. 3) Also let
\begin{align} 
D_\textsf{opt}^\textsf{conn}(\bo P^\textsf{C}) \triangleq D_{\textsf{opt}}\left(\bo P^\textsf{C}; \bo A=\bo 1 \bo 1^T \right)
\end{align}
denote the optimal distortion for the fully connected collaborative topology. Note that 
\begin{align}
\underbrace{D_0}_{\begin{smallmatrix} 
  \text{Infinite power} \\ 
  \text{Full collab.} 
\end{smallmatrix}} \le \; \underbrace{D_\textsf{opt}^\textsf{conn}(\bo P^\textsf{C})}_{\begin{smallmatrix} 
  \text{Finite power} \\ 
  \text{Full collab.} 
\end{smallmatrix}} \le \; \underbrace{D_\textsf{opt}^\textsf{dist}(\bo P^\textsf{C})}_{\begin{smallmatrix} 
  \text{Finite power} \\ 
  \text{No collab.} 
\end{smallmatrix}} \;\le \underbrace{\eta^2}_{\begin{smallmatrix} 
  \text{Zero power} \\ 
  \text{(prior only)}
\end{smallmatrix}}, \label{CG:inequality}
\end{align}
where $\eta^2$ is the worst-case distortion that corresponds to the prior information only. Equation \eqref{CG:inequality} is illustrated in Figure \ref{fig:cooperative:snr}, where a typical  operational region is depicted alongwith the power-distortion tradeoff for distributed and connected topologies for the cumulative-constraint problem\footnote{As we shall see later, for the cumulative-constraint problem, all the available power must be used at optimality, i.e., $P_\textsf{opt}=P^\textsf{C}$, and hence the subscript $(\textsf{C})$ is dropped from $P^\textsf{C}$ in Figure \ref{fig:cooperative:snr}.}.  The goal of any estimation application is to close as much of the performance gap ($\eta^2-D_0$) as possible using limited resources and spatial collaboration is a tool that enables efficient allocation of those resources. We are now in a position to define as \emph{Collaboration Gain} ($\textsf{CG}$), the following normalized (centered and scaled) metric,
\begin{align}
\textsf{CG}=\frac{D_\textsf{opt}^\textsf{dist}(\bo P^\textsf{C})-D_\textsf{opt}^\textsf{conn}(\bo P^\textsf{C})}{\eta^2-D_0}. \label{def:CG}
\end{align}

Note that $0\le \textsf{CG} \le 1$, which means that efficacy of collaboration can now be summarized for a wide range of problem conditions. For example, if for a problem $\textsf{CG}=0.01$ (say), we might conclude that collaboration is not sufficiently beneficial for that particular problem. On the other hand, if $\textsf{CG}=0.2$ (say), we would conclude that spatial collaboration closes the realizable performance gap by $20\%$ and hence, may be worth considering.

\section{Main Results - Ideal Collaborative power allocation:}  \label{sec:ideal}
In this section, we consider the situation when the entries of the collaboration cost matrix $\bo C$ are either zero or infinity, $C_{ij}\in\{0,\infty\}$, i.e., we will solve problems \eqref{prob:ideal:cum} and \eqref{prob:ideal:indiv}, where the topology $\bo A=\textsf{zero}(\bo C)$ is assumed to be fixed and not subject to design.

\subsection{Explicit formulation w.r.t. non-zero weights}

From \eqref{def:DW} and \eqref{def:JD}, we note that minimizing the distortion $D_{\bo W}$ is equivalent to maximizing the equivalent Fisher Information,
\begin{align}
J_{\bo W}=\frac{\left(\bo g^T \bo W \bo h\right)^2}{\tr\left[\bo E_\textsf{g} \bo W \bo E_\textsf{x} \bo W^T\right]-\eta^2 \left(\bo g^T \bo W \bo h\right)^2+\xi^2}, \label{def:JW}
\end{align}
where $D_{\bo W}$ and $J_{\bo W}$ are related by Equation \eqref{def:JD}. Part of the numerator and denominator of $J_{\bo W}$ and also the expressions for power (\eqref{def:PWm} for individual and \eqref{def:PW} for cumulative) are quadratic functions of the non-zero elements in $\bo W$. To see that explicitly, we concatenate the elements of $\bo W$ (column-wise, only those that are allowed to be non-zero), in $\bo w=[w_1,w_2,\ldots,w_L]^T$. For $l=1,2,\ldots,L$, define indices $m_l$ and $n_l$ such that $w_l=\bo W_{m_l,n_l}$. Further, we define $L\times L$ matrices $\bo \Omega_{\textsf{JN}}$, $\bo \Omega_{\textsf{JD}}$, $\bo \Omega_{\textsf P,m}$, $\bo \Omega_{\textsf P}\triangleq \sum_{m=1}^M \bo \Omega_{\textsf P,m}$ and $L\times N$ matrix $\bo G$ such that the following identities,
\begin{subequations}
\begin{align}
\bo g^T \bo W &=\bo w^T \bo G,\\
J_{\bo W}&=\frac{\overbrace{\left(\bo g^T \bo W \bo h\right)^2}^{=\;\bo w^T \bo \Omega_{\textsf{JN}} \bo w}}{\underbrace{\tr\left[\bo E_\textsf{g} \bo W \bo E_\textsf{x} \bo W^T\right]-\eta^2 \left(\bo g^T \bo W \bo h\right)^2}_{=\;\bo w^T \bo \Omega_{\textsf{JD}} \bo w}+\xi^2}, \\
P_{\bo W}&=\underbrace{\tr\left[\bo W\bo E_{\textsf x} \bo W^T\right]}_{\bo w^T \bo \Omega_{\textsf{P}} \bo w}=\sum_{m=1}^M \underbrace{\left[\bo W\bo E_{\textsf x} \bo W^T\right]_{m,m}}_{\bo w^T \bo \Omega_{\textsf{P},m} \bo w}
\end{align}
\end{subequations}
are satisfied. Precisely, the elementwise descriptions for all the matrices are as follows,
\begin{align}
\begin{split}
\left[\bo G\right]_{l,n}&=\left\{ \begin{array}{rl}
g_{m_l},              & n=n_l, \\
0,                        & \mbox{otherwise}
\end{array} \right. , \\
\left[\bo \Omega_{\textsf{JN}}\right]_{k,l}&=g_{m_k} g_{m_l} h_{n_k} h_{n_l} \; \Leftrightarrow \; \bo \Omega_\textsf{JN}=\bo G\bo h\bo h^T \bo G^T,\\
\left[\bo \Omega_{\textsf{JD}}\right]_{k,l}&=\left[\bo E_\textsf{g}\right]_{m_k,m_l} \left[\bo E_\textsf{x}\right]_{n_k,n_l}-\eta^2 \left[\bo \Omega_{\textsf{JN}}\right]_{k,l}, \; \mbox{and} \\
\left[\bo \Omega_{\textsf P,m}\right]_{k,l}&=\left\{ \begin{array}{rl}
\left[\bo E_{\textsf x}\right]_{n_k,n_l},              & m_k=m_l=m, \\
0,                        & \mbox{otherwise}
\end{array} \right. ,
\end{split} \label{def:OPOJ}
\end{align}
for $k,l=1,2,\ldots, L$, $n=1,2,\ldots, N$ and $m=1,2,\ldots,M$. 

\begin{figure}[htb]
\centering
    \includegraphics[width=\figszc \columnwidth]{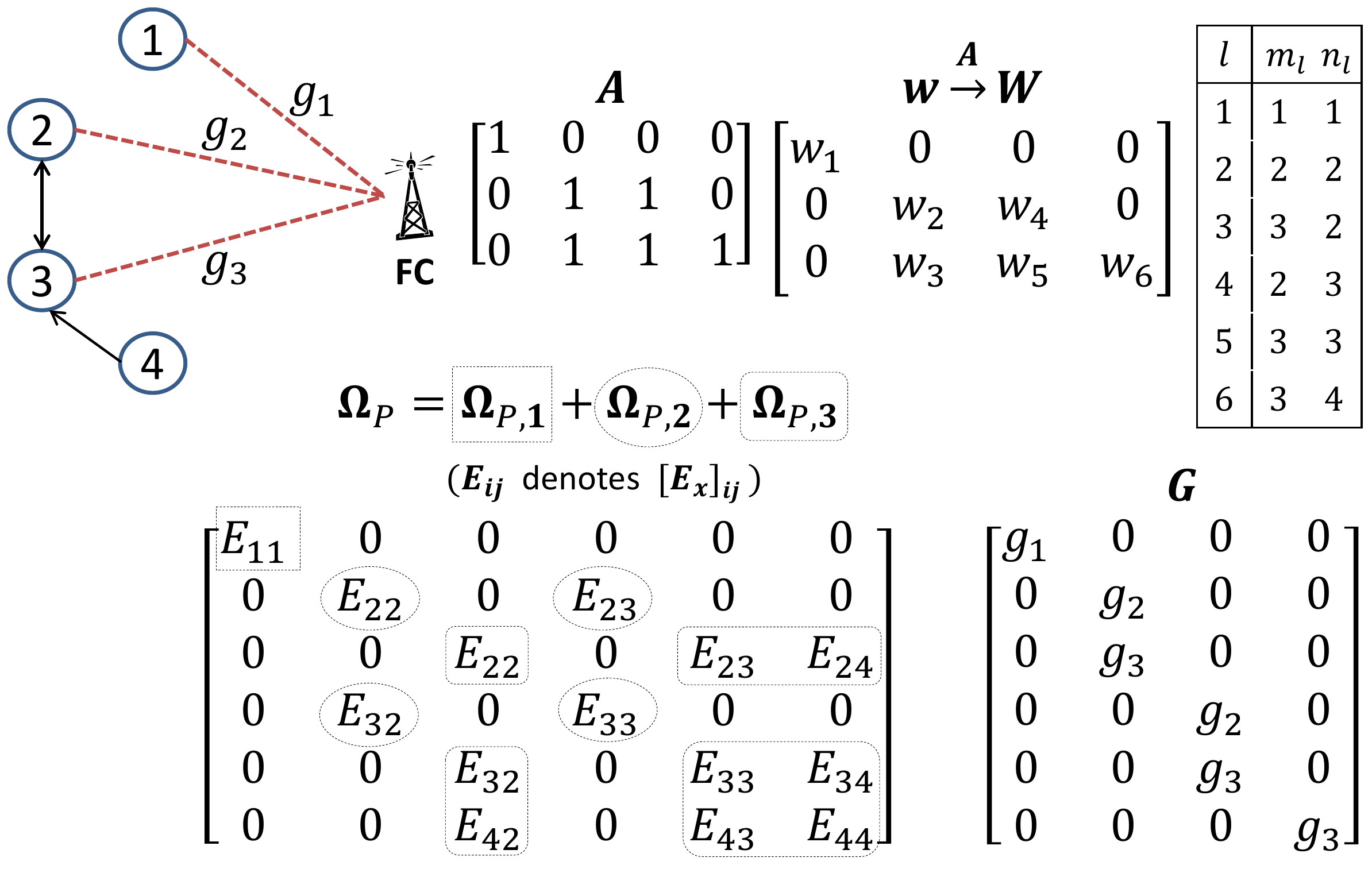}
  \caption{Matrices for problem formulation in explicit form - an example.  }
  \label{fig:tx:example}
\end{figure}

Though $\bo \Omega_{\textsf{JN}}$ is rank-1 (as described above), in general, there are no compact expressions for the matrices $\bo \Omega_{\textsf{JD}}$ and $\bo \Omega_{\textsf P,m}$. We illustrate some relevant matrix definitions ($\bo \Omega_{\textsf{P},m}$ and $\bo G$ in particular) through an example, in Figure \ref{fig:tx:example}, with $N=4$ data-collection nodes, $M=3$ communicating nodes and $3$ collaborating links, resulting in a total of $L=6$ non-zero coefficients in the collaboration matrix $\bo W$.

For some special cases and regular topologies, more compact expressions for $\bo \Omega_{\textsf{JD}}$ may be derived. For the special case when perfect channel state information is available ($\bo \Sigma_\textsf{g}=0$), it is easy to see that 
\begin{align}
\bo \Omega_{\textsf{JD}}&=\bo G\left(\bo E_\textsf{x} -\eta^2 \bo h \bo h^T \right) \bo G^T \nonumber  \\
&=\bo G \widetilde{\bo \Sigma} \bo G^T, \; \mbox{ where }
\widetilde{\bo \Sigma}\triangleq \bo \Sigma +\eta^2 \bo \Sigma_\textsf{h}. \label{Omega:JD:CSI}
\end{align} 
This simplification will also be useful later in our discussion.

With the help of these definitions, the objective function (Fisher Information) is simplified as
\begin{align}
J_{\bo w} \triangleq \frac{\bo w^T \bo \Omega_{\textsf{JN}} \bo w}{\bo w^T \bo \Omega_{\textsf{JD}} \bo w+\xi^2},  \label{def:Jw}
\end{align}
and problems \eqref{prob:ideal:cum} and \eqref{prob:ideal:indiv} are re-written as
\begin{equation}
\begin{aligned}
& \underset{\bo w}{\text{maximize}} & &  J_{\bo w}\\
&                           \text{subject to} & & \bo w^T \bo \Omega_{\textsf P} \bo w\le P^\textsf{C}, \mbox{ and}
\end{aligned} \label{prob:ideal:Q:cum}
\end{equation}
\begin{equation}
\begin{aligned}
& \underset{\bo w}{\text{maximize}} & & J_{\bo w} \\
&                            \text{subject to} & & \bo w^T \bo \Omega_{\textsf P,m} \bo w\le P^\textsf{C}_m, \, m=1,\ldots,M.
\end{aligned} \label{prob:ideal:Q:indiv}
\end{equation}
respectively, which are both optimization problems with the same fractional-quadratic objective \eqref{def:Jw} and single (or multiple) quadratic constraint(s). Solution of problems \eqref{prob:ideal:Q:cum} and \eqref{prob:ideal:Q:indiv} will be provided in Sections \ref{sec:cum} and  \ref{sec:indiv} respectively.

\subsection{Cumulative power constraint} \label{sec:cum}
 Since multiplying $\bo w$ by a scalar $\alpha>1$ (strictly) increases both $J_{\bo w}$ and power $\bo w^T \bo \Omega_\textsf{P} \bo w$ (and for $\alpha<1$, strictly decreases them), problem \eqref{prob:ideal:Q:cum} is equivalent to its converse formulation, where power is minimized subject to a maximum distortion constraint (represented by $J^\textsf{C}$),
\begin{equation}
\begin{aligned}
& \underset{\bo w}{\text{minimize}} & & \bo w^T \bo \Omega_\textsf{P} \bo w \\
&                                                             \text{subject to} & & J_{\bo w}\ge J^\textsf{C},
\end{aligned} \label{prob:ideal:Q:cum:conv}
\end{equation}
in the sense that the optimal solutions $J_{\text{opt}}(P^\textsf{C})$ (of \eqref{prob:ideal:Q:cum}) and $P_{\text{opt}}(J^\textsf{C})$ (of \eqref{prob:ideal:Q:cum:conv}) are inverses of one another. Moreover, the optimal solutions hold with active constraints (satisfying equalities $P=P^\textsf{C}$ for\eqref{prob:ideal:Q:cum} and $J=J^\textsf{C}$ for \eqref{prob:ideal:Q:cum:conv}). From \eqref{def:Jw}, problem \eqref{prob:ideal:Q:cum:conv} is further equivalent to,
\begin{align}
\begin{aligned}
& \underset{\bo w}{\text{minimize}} & & \bo w^T \bo \Omega_\textsf{P} \bo w \\
&                            \text{subject to} & & \bo w^T \left( J \bo \Omega_\textsf{JD}-\bo \Omega_\textsf{JN} \right) \bo w  +J \xi^2 \le 0,
\end{aligned} \label{prob:ideal:Q:cum:conv:2}
\end{align}
which is a quadratically constrained quadratic program (QCQP) in  $L \triangleq \text{\textsf{nnz}}(\bo A)$ variables. Note in \eqref{prob:ideal:Q:cum:conv:2} that, though $\bo \Omega_\textsf{P}$ is positive definite (it is composed of blocks of another positive definite matrix $\bo E_\textsf{x}$, see Figure \ref{fig:tx:example}, for example), the matrix $J \bo \Omega_\textsf{JD}-\bo \Omega_\textsf{JN}$ is not, and hence problem \eqref{prob:ideal:Q:cum:conv:2} is not convex. However, a QCQP with exactly one constraint (as in problem \eqref{prob:ideal:Q:cum:conv:2}) still satisfies strong duality 
 (for a background, see Appendix B, \cite{Boyd04}) and hence the optimal solution to \eqref{prob:ideal:Q:cum:conv:2} satisfies the Karush-Kuhn-Tucker (KKT) conditions
\begin{align}
\left( \bo \Omega_\textsf{P} +\mu \left( J \bo \Omega_\textsf{JD}-\bo \Omega_\textsf{JN} \right) \right) \bo w=0.
\end{align}
Together with the following active constraint conditions at optimality
\begin{align}
P=\bo w^T \bo \Omega_\textsf{P} \bo w,\; \mbox{ and }\; \bo w^T \left( J \bo \Omega_\textsf{JD}-\bo \Omega_\textsf{JN} \right) \bo w+J \xi^2=0,
\end{align}
which implies $\mu=\frac{P}{J\xi^2}$, the solution to problem \eqref{prob:ideal:Q:cum:conv:2} (equivalently, problems \eqref{prob:ideal:Q:cum:conv}, \eqref{prob:ideal:Q:cum} and \eqref{prob:ideal:cum}) is summarized below.

\begin{result:cum} \emph{(Power-Distortion tradeoff for Linear Coherent Ideal-Collaborative Estimation)} \label{result:cum:lbl}
For a given topology $\bo A$, let $J\in\left(0, \lambda_{\mathcal G,\textsf{max}}\left(\bo \Omega_\textsf{JN},\bo \Omega_\textsf{JD} \right) \right)$\footnote{Definitions of eigenvalue related operators: The operators $\lambda(\bo P)$ and $\bo v(\bo P)$ denote the solution(s) to the ordinary eigenvalue problem $\bo P\bo v=\lambda \bo v$. Operator $\lambda_{\textsf{max}}(\cdot )$ denote the maximum among all real eigenvalues and $\lambda^{\textsf{pos}}_{\textsf{min}}(\cdot )$ denote the minimum among all positive eigenvalues (i.e., the positive eigenvalue that is closest to, but different from, zero). The operators $\lambda_{\mathcal G }(\bo P,\bo Q)$ and $\bo v_{\mathcal G }(\bo P,\bo Q)$  denote the solution(s) to the generalized eigenvalue problem $\bo P\bo v=\lambda \bo Q \bo v$. Operators $\lambda_{\mathcal G,\textsf{max}}(\cdot,\cdot )$ and $\lambda^{\textsf{pos}}_{\mathcal G,\textsf{min}}(\cdot,\cdot )$ are similarly defined as $\lambda_{\textsf{max}}(\cdot )$ and $\lambda^{\textsf{pos}}_{\textsf{min}}(\cdot )$ respectively. Note that, when $\bo Q$ is full-rank, then $\lambda_{\mathcal G }(\bo P,\bo Q)=\lambda (\bo Q^{-1}\bo P)$.}. The optimal tradeoff between distortion (represented by $J$) and cumulative transmission power $P$ and also the optimal weights $\bo w$ that achieve that tradeoff, are obtained through the solution of the generalized eigenvalue problem,
\begin{align}
\left(\frac{\bo \Omega_\textsf{P}}{P_\xi}-\frac{\bo \Omega_\textsf{JN}}{J}+\bo \Omega_\textsf{JD}\right) \bo w=0, \mbox{ where } P_\xi\triangleq \frac{P}{\xi^2}.
\end{align}
In particular, the function $J_\textsf{opt}(P):\,\left(0,\infty\right) \rightarrow \left(0,J_0^{\bo A}\right)$ and its inverse $P_\textsf{opt}(J)$ are
\begin{subequations}
\begin{align}
J_\textsf{opt}(P)&=\lambda_{\mathcal G,\textsf{max}}\left(\bo \Omega_\textsf{JN},\bo \Omega_\textsf{JD}+\frac{\bo \Omega_\textsf{P}}{P_\xi}\right), \; \mbox{and } \label{JP:cum}\\
P_\textsf{opt}(J)&=\lambda_{\mathcal G,\textsf{min}}^\textsf{pos}\left(\bo \Omega_\textsf{P},-\bo \Omega_\textsf{JD}+\frac{\bo \Omega_\textsf{JN}}{J}\right) \xi^2, \label{PJ:cum}
\end{align}
\end{subequations}
respectively. This optimal tradeoff is achieved when the weights of collaboration matrix is an appropriately scaled version of the (generalized) eigenvector corresponding to \eqref{JP:cum} (or \eqref{PJ:cum}, since they are equivalent), say $\bo v_\textsf{opt}$. That is, $\bo w_\textsf{opt}=c \bo v_\textsf{opt}$, where the scalar $c$ is such that $\bo w_\textsf{opt}^T\bo \Omega_\textsf{P} \bo w_\textsf{opt}=P$.
\end{result:cum}

Theorem \ref{result:cum:lbl} is important since it helps to (numerically) compute the power-distortion tradeoff for arbitrary problem conditions (like topology, noise covariance, second-order statistics of the observation and channel gains). Since the numerical complexity for eigenvalue problems is roughly cubic in the size of the problem (see for example \cite{Pan99}), the complexity of computing the (cumulative) power-distortion tradeoff is $\mathcal O(L^3)$, where (recall that) $L=\textsf{nnz}(\bo A)$ is the number of non-zero collaboration weights.

Corresponding to the example topology in Figure \ref{fig:tx:example} and randomly chosen system parameters $\bo h, \bo \Sigma$ and $\bo g$, a typical power-distortion tradeoff curve is shown in Figure \ref{fig:cooperative:snr} (bold line). Theorem \ref{result:cum:lbl} can be simplified further for several specific scenarios, allowing deeper insight into the power-distortion tradeoff as it relates to the problem parameters. The first obvious simplification is because of the rank-1 property of $\bo \Omega_\textsf{JN}$. The only non-zero generalized eigenvalue is (provided the inverse exists)
\begin{subequations}
\begin{align}
J_\textsf{opt}(P)=\bo h^T\bo G^T\left(\bo \Omega_\textsf{JD}+\frac{\bo \Omega_\textsf{P}}{P_\xi} \right)^{-1} \bo G \bo h&,   \label{Jopt} \\
\mbox{with eigenvector }\bo w_\textsf{opt} \propto \left(\bo \Omega_\textsf{JD}+\frac{\bo \Omega_\textsf{P}}{P_\xi} \right)^{-1} \bo G \bo h&. \label{Wopt}
\end{align} 
\end{subequations}
Equation \eqref{Jopt} explicitly shows the effect of finite-power constraint $P_\xi$ on the distortion. Some other insightful examples are discussed next. 

\emph{Example 1: } For the case of \emph{perfect CSI} ($\bo \Sigma_{\textsf g}=0$), we note from \eqref{Omega:JD:CSI} that $\bo \Omega_\textsf{JD}=\bo G \widetilde{\bo \Sigma} \bo G^T$. It follows that
\begin{align}
&J_\textsf{opt}(P)=\lambda_{\mathcal G, \textsf{max}}\left(\bo G \bo h \bo h^T \bo G^T, \bo G  \widetilde{\bo \Sigma} \bo G^T +\frac{\bo \Omega_\textsf{P}}{P_\xi}\right) \label{Jopt:CSI:gev:orig} \\
&\; \stackrel{\text(a)}{=}\lambda_{\mathcal G, \textsf{max}}\left( \bo h \bo h^T, \widetilde{\bo \Sigma} +\frac{\bo \Gamma_\textsf{P}}{P_\xi}\right),\; \bo \Gamma_\textsf{P}\triangleq \left(\bo G^T \bo \Omega_\textsf{P}^{-1} \bo G\right)^{-1}  \label{Jopt:CSI:gev} \\
&\; \stackrel{\text(b)}{=}\bo h^T  \left(\widetilde{\bo \Sigma}+\frac{\bo \Gamma_\textsf{P}}{P_\xi} \right)^{-1} \bo h, \label{Jopt:CSI} 
\end{align}
assuming all the inverses exist. Step (a) reduces the size of the eigenvalue problem from $L$ to $N$ yet preserving the non-zero eigenvalues. Note that the corresponding generalized eigenvectors of problems \eqref{Jopt:CSI:gev:orig} (say $\bo v_L$) and \eqref{Jopt:CSI:gev} (say $\bo v_N$) are related by $\bo v_L=\bo \Omega_\textsf{P}^{-1}\bo G \bo \Gamma_\textsf{P} \bo v_N$. Step (b) describes the only non-zero generalized value of problem \eqref{Jopt:CSI:gev}, since $\bo h \bo h^T$ is rank-1. Note that optimal collaboration weights are provided by $\bo w_\textsf{opt} \propto\bo \Omega_\textsf{P}^{-1}\bo G \tbo \Sigma \left( \tbo \Sigma+\frac{\bo \Gamma_\textsf{P}}{P_\xi} \right)^{-1} \bo h$.

When, in addition to \emph{perfect CSI}, we also have \emph{perfect OGI} ($\bo \Sigma_{\textsf h}=0$), Equation \eqref{Jopt:CSI} further simplifies to 
\begin{align}
J_\textsf{opt} (P)&=\bo h^T\left(\bo \Sigma+\frac{\bo \Gamma_\textsf{P}}{P_\xi} \right)^{-1} \bo h, \label{Jopt:both}
\end{align}
which was obtained in a preliminary version of this paper \cite{KarISIT12}. Equation \eqref{Jopt:both} can be compared to the centralized case, where measurements $x_n$ are directly observed through gains $\bo h$ and measurement noise with variance $\bo \Sigma$, for which the Fisher Information is $J_\textsf{cent}\triangleq \bo h^T \bo \Sigma^{-1} \bo h$, which is also the infinite power limit of \eqref{Jopt:both}. One can think of the additional quantity  $\frac{\bo \Gamma_\textsf{P}}{P_\xi}$ in \eqref{Jopt:both}, which factors in the effect of channels, the collaboration topology and finite transmission power, as equivalent to the variance of an additional noise that is added to the measurement noise.

When, in addition to \emph{perfect OGI and CSI}, we also have a \emph{distributed} topology ( $\bo A=\bo I_M$) and the measurement noise is {uncorrelated} ($\bo \Sigma$ is diagonal), we can proceed as follows. We have $\bo w=\textsf{diag}(\bo W)$\footnote{Definition of operators $\textsf{diag}(\cdot)$ and  $\textsf{vec}(\cdot)$: While operating on a matrix, $\textsf{diag}:\mathbb R^{M\times N}\rightarrow \mathbb R^{\textsf{min}(M,N)}$ is used to extract the diagonal elements. While operating on a vector, $\textsf{diag}:\mathbb  R^M\rightarrow \mathbb R^{M\times M}$ is used to construct a matrix by specifying only the diagonal elements, the other elements being zero. The vectorization operator $\textsf{vec}:\mathbb R^{M\times N}\rightarrow \mathbb R^{MN}$ stacks up all the elements of a matrix column-by-column. }. 
This means that $\bo \Omega_\textsf{P}=\textsf{diag}(\textsf{diag}(\bo \Sigma+\eta^2\bo h\bo h^T))$ is a diagonal matrix with $m$\sups{th} element as $\sigma_m^2+\eta^2 h_m^2$, and $\bo G=\textsf{diag}(\bo g)$. Consequently, $\bo \Gamma_\textsf{P}$ is a diagonal matrix  with $m$\sups{th} element as $\frac{\sigma_m^2+\eta^2 h_m^2}{g_m^2}$. Hence, the optimal power-distortion tradeoff in Equation \eqref{Jopt:both} further simplifies to (define $\sigma_m^2\triangleq \Sigma_{m,m}$ and $\gamma_m\triangleq \frac{\eta^2 h_m^2}{\sigma_m^2}$)
\begin{align}
&J_\textsf{opt}(P)=\sum_{m=1}^M \frac{h_m^2}{\sigma_m^2}\left[1+\frac{1+\gamma_m}{P_\xi g_m^2}\right]^{-1},
 \label{Jopt:both:dist}
\end{align} 
which was also obtained in \cite{Xiao08}. Since for the centralized case, we have $J_\textsf{cent}=\sum_{m=1}^M \frac{h_m^2}{\sigma_m^2}$, Equation \eqref{Jopt:both:dist} indicates the exact fractions of individual Fisher Information that ``reaches" the receiver. When subjected to a network-wide power constraint, information from the more informative (higher $\gamma_m$) and less reliable (lower $g_m$) sensor undergoes a higher degree of ``attenuation". While a higher observation gain $h_m$ clearly carries more information, it also requires quadratically higher power to transmit in an amplify-and-forward framework such as ours (note that $P_m=w_m^2 (\sigma_m^2+\eta^2 h_m^2)$). Similarly, a lower magnitude of channel gain $g_m$ implies that quadratically higher transmission power is needed to compensate for the channel. Hence, in the optimal tradeoff \eqref{Jopt:both:dist}, it turns out that information from higher-$h_m$ and lower-$g_m$ sensors are attenuated by a larger factor.

\emph{Example 2:} When the network is fully \emph{connected} ($\bo A=\bo 1 \bo 1^T$), we proceed from \eqref{Jopt} and obtain the following result. 

\begin{result:conn} \label{result:conn:lbl}
The optimal solution for Example 2 is
\begin{align}
\begin{split}
&J_\textsf{opt}(P)= \widetilde{J} \left[1+\frac{1+\eta^2  \widetilde{J}}{\mathcal G}\right]^{-1} \mbox{ and } W_\textsf{opt} \propto   \bo u \bo v^T,\\
&\mbox{where } \widetilde{J} \triangleq \bo h^T \tbo \Sigma^{-1} \bo h, \; \mathcal G \triangleq \bo g^T \tbo \Sigma_\textsf{g}^{-1} \bo g, \; \bo u=\tbo \Sigma_\textsf{g}^{-1} \bo g,  \\
&\bo v=\tbo \Sigma^{-1} \bo h,\; \tbo \Sigma \triangleq \bo \Sigma + \eta^2 \bo \Sigma_\textsf{h} \mbox{ and } \tbo \Sigma_\textsf{g} \triangleq \bo \Sigma_\textsf{g} +\frac{\bo I}{P_\xi}.
\end{split} \label{Jopt:conn}
\end{align}
Also, for the special case with \emph{perfect OGI} and \emph{perfect CSI} (when $\bo \Sigma_\textsf{h}=\bo \Sigma_\textsf{g}=0$), the resulting distortion is information theoretically optimal.
\end{result:conn}
\begin{IEEEproof} See Appendix \ref{app:conn}. 
\end{IEEEproof}

This example is important since a connected topology makes use of all possible collaboration links and hence Equation \eqref{Jopt:conn} gives the LMMSE performance limit for a network with cumulative transmission power constraint. From \eqref{Jopt:conn}, we are now in a position to compute the optimal achievable Fisher Information $J_0$ (and equivalently the distortion $D_0$), by letting the power go to infinity,
\begin{align}
J_0= \widetilde{J} \left[1+\frac{1+\eta^2  \widetilde{J}}{\mathcal G_0}\right]^{-1}, \mbox{ where } \mathcal G_0 \triangleq \bo g^T \bo \Sigma_\textsf{g}^{-1} \bo g. \label{J0}
\end{align}
As discussed earlier, $D_0$ (definition in \eqref{def:D0}) forms a lower bound for distortion in a network  and is useful to characterize the operational region and subsequently define the collaboration gain (Equation \eqref{def:CG}).


From Equation \eqref{Jopt:conn}, it can be explicitly seen that for a fully connected topology, more observation or channel uncertainty always deteriorate the estimation performance, a notion that is intuitive but still was not completely evident in the discussion so far. The following result formalizes this notion.

\begin{result:conn:inequality} \label{result:conn:inequality:lbl}
If $\bo \Sigma_\textsf{h,1} \succ \bo \Sigma_\textsf{h,2}$ and $\bo \Sigma_\textsf{g,1} \succ \bo \Sigma_\textsf{g,2}$ (here $\bo A \succ \bo B$ implies that $\bo A-\bo B$ is a symmetric positive definite matrix), then for a connected topology,
\begin{align}
J_\textsf{opt,1}(P) < J_\textsf{opt,2}(P).
\end{align}
\end{result:conn:inequality}
\begin{IEEEproof}  This is established by re-writing Equation \eqref{Jopt:conn} as  
\begin{align}
J_\textsf{opt}(P)&= \left[\left( 1+\frac{1}{\mathcal G} \right)\frac{1}{\widetilde{J}}+\frac{\eta^2}{\mathcal G}\right]^{-1},
\end{align}
which shows explicitly that $J_\textsf{opt}(P)$ is a monotonically increasing function of $\widetilde{J}$  and $\mathcal G$. Recall that $\widetilde J=\bo h^T \left( \bo \Sigma + \eta^2 \bo \Sigma_\textsf{h} \right)^{-1} \bo h$ and $\mathcal G=\bo g^T \bo \Sigma_\textsf{g}^{-1} \bo g$.  Assuming $\bo \Sigma_\textsf{h,1} \succ \bo \Sigma_\textsf{h,2}$ and $\bo \Sigma_\textsf{g,1} \succ \bo \Sigma_\textsf{g,2}$, it is sufficient to show that a) $\widetilde{J}_\textsf{1} < \widetilde{J}_\textsf{2}$ and b) $\mathcal G_\textsf{1} < \mathcal G_\textsf{2}$, both of which are evident from the inequality involving positive definite matrices in Lemma \ref{inequality:lbl} (see Appendix \ref{app:inequality}).
 \end{IEEEproof} 

The last two examples provided insight on the distributed and fully connected topologies respectively. The following example addresses a partially connected topology and shows how the distortion decreases with an increase in links available for collaboration.

\begin{figure}[htb]
\centering
    \includegraphics[width=\figszd \columnwidth]{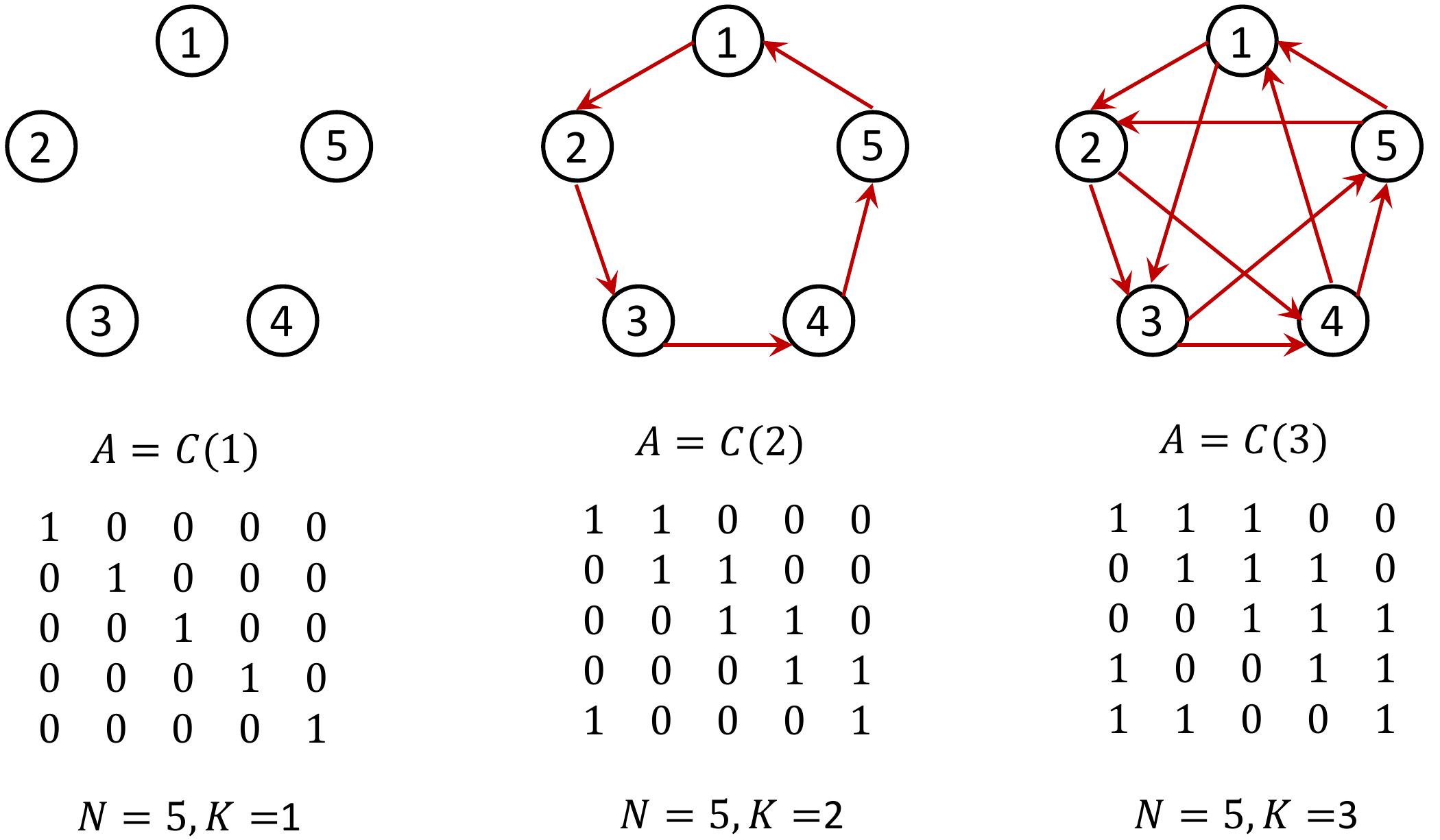}
  \caption{Directed cycle graphs, $(K-1)$ connected.}
  \label{fig:cooperative:cycle}
\end{figure}

\emph{Partially connected cycle graphs: } In Figure \ref{fig:cooperative:cycle}, we display a class of graphs, namely the $(K-1)$ connected directed cycle, for $K=1,2,\ldots,M$, in which each node shares its observations with the next $K-1$ nodes. The adjacency topology of such a graph will be denoted as $\bo A=\mathcal C(K)$. Note that $K=1$ denotes the distributed scenario while $K=M$ denotes the fully connected scenario.

\emph{Example 3:} We assume that the collaborative topology is a $(K-1)$ connected directed cycle ($\bo A=\mathcal C(K)$) and the channel gain and uncertainties are such that the network is \emph{homogeneous and equicorrelated}. In particular, we denote a) the \emph{expected} observation and channel gains by $h_0^2$ and $g_0^2$, b) the observation and channel gain uncertainties by $\alpha_\textsf{h}$ and $\alpha_\textsf{g}$, and c) measurement noise variance and correlation by $\sigma^2$ and $\rho$, and thereby assume
\begin{align}
\begin{split}
\bo h&=h_0\sqrt{\alpha_\textsf{h}} \bo 1, \bo \Sigma_\textsf{h}=h_0^2 (1-\alpha_\textsf{h}) \bo I,\\
\bo g&=g_0\sqrt{\alpha_\textsf{g}} \bo 1, \bo \Sigma_\textsf{g}=g_0^2 (1-\alpha_\textsf{g}) \bo I, \mbox{ and}\\
\bo \Sigma&=\sigma^2 \left((1-\rho)\bo I+\rho\bo 1\bo 1^T\right). \label{prob:homo}
\end{split}
\end{align}
These assumptions provide an analytically tractable example that is representative of a broad range of problem conditions and will also be used in the subsequent discussions. In addition to covering the partial collaboration $(1\le K \le N)$ regime, note that $\alpha_\textsf{h}=1$ implies perfect OGI,  $\alpha_\textsf{g}=1$ implies perfect CSI and $\rho=0$ implies uncorrelated measurement noise. Equation \eqref{Jopt} can be simplified further for this example, to obtain the following result.
\begin{result:cyc} \label{result:cyc:lbl}
The optimal solution for Example 3 is
\begin{align}
\begin{split}
&J_\textsf{opt}(P)=\frac{h_0^2}{\sigma^2} \left[\rho_N+\widetilde{\alpha}_\textsf{h} \left(\rho_N+\frac{\gamma}{N}\right) \right.\\
&\left.+\frac{1}{N}\left(\widetilde{\alpha}_\textsf{g}+\frac{1}{P_\xi g_0^2 \alpha_\textsf{g}}\right)\left\{\gamma+\rho_K +\widetilde{\alpha}_\textsf{h} \left(\rho_K+\frac{\gamma}{K}\right)\right\}\right]^{-1}, \\
&\mbox{where } \rho_t\triangleq \rho+\frac{1-\rho}{t},\; \gamma\triangleq\frac{\eta^2h_0^2}{\sigma^2},\mbox{ and } \widetilde{\alpha}\triangleq\frac{1}{\alpha}-1.
\end{split} \label{Jopt:cyc}
\end{align}
\end{result:cyc}
\begin{IEEEproof} See Appendix \ref{app:cycle}.   \end{IEEEproof}
Equation \eqref{Jopt:cyc} helps us quantify the efficacy of collaboration in a partially connected network. For example, when the measurement noise is uncorrelated ($\rho=0$) and we have perfect OGI ($\alpha_\textsf{h}=1$, equivalently $\widetilde \alpha_\textsf{h}=0$) and perfect CSI ($\alpha_\textsf{g}=1$, equivalently $\widetilde \alpha_\textsf{g}=0$), Equation \eqref{Jopt:cyc} reduces to 
\begin{align}
J=\frac{h_0^2 g_0^2}{\sigma^2 \xi^2} \frac{N P }{\gamma + \frac{1}{K}}.
\end{align}
For the high-($N P$) regime, (i.e., when $J$ is large and $D\approx J^{-1}$), we can now compare the power requirements of a distributed topology (say $P^\textsf{dist}$, for $K=1$) with that of a $(K-1)$ connected topology (say $P^{\mathcal C(K)}$), provided an identical distortion performance is desired. Since
$\frac{P^\textsf{dist} }{\gamma + 1}=\frac{P^{\mathcal C(K)} }{\gamma + \frac{1}{K}}$, this implies that the relative savings in power is
\begin{align}
\frac{P^\textsf{dist}  - P^{\mathcal C(K)} }{P^\textsf{dist} }=\frac{1-\frac{1}{K}}{\gamma+1},
\end{align}
which depends on the local-SNR $\gamma$. For example, when the local-SNR is large, say $\gamma=100$, then even a fully connected network (large $K$) can provide only $1\%$ power savings. On the other hand, if the local-SNR is small, say $\gamma=1$, then even a $1$-connected network (for which $K=2$) can provide $25\%$ efficiency in power savings. The conclusion is that one needs to be judicious in the design of collaborative topologies, especially when there are overhead costs associated with it. The design of collaborative topologies with finite collaboration cost will be discussed later in Section \ref{sec:finite}.

\subsection{Individual power constraints} \label{sec:indiv}
In the previous subsection, we have discussed the solution to the ideal-collaborative power allocation problem with a cumulative transmission power constraint. In this subsection, we consider the case when the individual nodes have separate power constraints. We recall problem \eqref{prob:ideal:Q:indiv},
\begin{equation}
\begin{aligned}
& \underset{\bo w}{\text{maximize}} & & J_{\bo w}=\frac{\bo w^T \bo \Omega_{\textsf{JN}} \bo w}{\bo w^T \bo \Omega_{\textsf{JD}} \bo w+\xi^2} \\
&                            \text{subject to} & & \bo w^T \bo \Omega_{\textsf P,m} \bo w\le P^\textsf{C}_m, \, m=1,\ldots,M.
\end{aligned} \label{prob:ideal:Q:indiv:2}
\end{equation}
Let $J_\textsf{opt}$ be the optimal solution of problem \eqref{prob:ideal:Q:indiv:2}, although it is not clear yet how this solution can be obtained. Unlike problem \eqref{prob:ideal:Q:cum} for which a closed from solution was derived in Section \ref{sec:cum}, in general, there are no known closed form expressions of $J_\text{opt}$. However, as we shall show below, $J_\text{opt}$ can still be precisely obtained using an efficient (polynomial time) numerical procedure. For some special cases though, (somewhat) closed form and insightful expressions for $J_\text{opt}$ can be obtained, which will be discussed later.

\subsubsection{Numerical Solution} \label{sec:indiv:numerical}
We would use the semi-definite relaxation (SDR) technique for quadratically constrained problems \cite{Havary08}, \cite{LuoSDP09} to solve problem \eqref{prob:ideal:Q:indiv:2}. The SDR technique is widely used in the literature since it can reduce an otherwise intractable problem to one with polynomial time complexity. However, the main drawback of SDR technique is that, in general, it can guarantee only a sub-optimal solution. \emph{But, in some special problems, which includes our problem at hand (as we shall establish later), the relaxation involved in SDR is exact and hence, the SDR technique becomes an efficient numerical tool to obtain the (precisely) optimal solution}. For more details on the SDR technique, including sub-optimality analysis for special classes of problems, the reader is referred to the book-chapter \cite{LuoSDP09}. 

We proceed using arguments similar to \cite{Havary08}, \cite{Zheng09}. Define
\begin{align}
\bo X \triangleq \bo w \bo w^T \in \mathbb R^{L\times L}, \label{def:WW}
\end{align}
so that problem \eqref{prob:ideal:Q:indiv:2} is equivalent to,
\begin{equation}
\begin{aligned}
& \underset{\bo X}{\text{maximize}} & & J_{\bo X}=\frac{\tr \left[\bo \Omega_\textsf{JN} \bo X\right]}{\tr \left[\bo \Omega_\textsf{JD} \bo X\right]+\xi^2} \\
&                            \text{subject to} & & \tr \left[\bo \Omega_{\textsf P,m} \bo X\right] \le P^\textsf{C}_m, \, m=1,\ldots,M, \\
&                     & & \textsf{rank}\; \bo X=1,\; \bo X \succeq 0, \\
\end{aligned} \label{prob:ideal:SDP:indiv}
\end{equation}
Problem \eqref{prob:ideal:SDP:indiv} is further equivalent to,
\begin{equation}
\begin{aligned}
& \text{find maximum} & & J \\
& \text{such that}         & & \mathcal X(J) \; \mbox{is nonempty}, 
\end{aligned} \label{prob:ideal:SDP:indiv:2}
\end{equation}
where $\mathcal X(J)$ is defined as the following (feasible) set,
\begin{align}
\mathcal X(J) \triangleq \left\{\bo X\left| \begin{array}{l}\tr \left[\left( J\bo \Omega_\textsf{JD}-\bo \Omega_\textsf{JN}\right) \bo X\right]+J \xi^2 \le 0, \\
\tr \left[\bo \Omega_{\textsf P,m} \bo X\right] \le P^\textsf{C}_m, \, m=1,\ldots,M, \\
\textsf{rank}\; \bo X=1,\; \bo X \succeq 0.
\end{array} \right\} \right.\label{def:XWJ}
\end{align}
Note that by definition, any $\bo X \in \mathcal X(J_\textsf{opt})$ will correspond to the optimal weights that maximize the Fisher Information. Also note that $\mathcal X(J_1) \supset \mathcal X(J_2)$ when $0\le J_1<J_2$. Therefore, assuming we can test the feasibility of $\mathcal X(J)$ for some $J$, a simple bisection search over $\left[ 0,J_0 \right]$ can potentially yield $J_\textsf{opt}$ with arbitrary accuracy\footnote{Arbitrary accuracy is only of theoretical interest, since the solution to the feasibility problem \eqref{prob:ideal:SDP:indiv:2} will have numerical errors. Hence, a more realistic stopping criterion is a fixed number of iterations, say $15$.} (since $J_\textsf{opt}<J_0$ for finite power). However, testing the feasibility of $\mathcal X(J)$ is a difficult problem.

Though the set of symmetric positive-semidefinite matrices is convex and the other $M+1$ inequalities in \eqref{def:XWJ} are also convex, $\mathcal X(J)$ is still not convex due to the rank constraint $\textsf{rank}\; \bo X=1$. Relaxing this constraint, we define,
\begin{align}
\mathcal X^\mathsf{R}(J) \triangleq \left\{\bo X\left| \begin{array}{l}\tr \left[\left( J\bo \Omega_\textsf{JD}-\bo \Omega_\textsf{JN}\right) \bo X\right]+J \xi^2 \le 0, \\
\tr \left[\bo \Omega_{\textsf P,m} \bo X\right] \le P^\textsf{C}_m, \, m=1,\ldots,M, \\
\bo X \succeq 0,
\end{array} \right\} \right.\label{def:XWJ:R}
\end{align}
which is now a convex set (superscript $\mathsf{R}$ stands for \emph{relaxation}). Because of this relaxation, we have $\mathcal X(J) \subset \mathcal X^\mathsf{R}(J)$ for all $J$. We denote the solution to the new problem,
\begin{equation}
\begin{aligned}
& \text{find maximum} & & J \\
& \text{such that}         & & \mathcal X^\mathsf{R}(J) \; \mbox{is nonempty}, 
\end{aligned} \label{prob:ideal:SDP:indiv:3}
\end{equation}
as $J^\mathsf{R}_\textsf{opt}$, so that $J_\textsf{opt} \le J^\mathsf{R}_\textsf{opt}$. Hence, in general, the solution of the relaxed problem \eqref{prob:ideal:SDP:indiv:3} only provides an upper bound to the solution of the original problem \eqref{prob:ideal:SDP:indiv:2}. However, the following result establishes the fact that the relaxation is tight, i.e., $J_\textsf{opt} = J^\mathsf{R}_\textsf{opt}$ for our specific problem.

\begin{result:SDR} \emph{(Semidefinite relaxation is tight): } \label{result:SDR:lbl}
Assume $\bo \Omega_\textsf{P}$ be positive definite. Then, for any feasible $J< J_\textsf{opt}$, $\mathcal X^\mathsf{R}(J)$ contains a rank-1 matrix.
\end{result:SDR}
\begin{IEEEproof} See Appendix \ref{app:SDR}. \end{IEEEproof}

The matrix $\bo \Omega_\textsf{P}$ is usually positive definite since it is composed of blocks of the matrix $\bo E_\textsf{x}=\mathbb E[\bo x \bo x^T]$. Since $\mathcal X^\mathsf{R}(J)$ contains a rank-$1$ matrix, $\mathcal X^\mathsf{R}(J) \cap \mathcal X(J)$ is non-empty, and feasibility of $\mathcal X^\mathsf{R}(J)$ also implies the feasibility of $\mathcal X(J)$, thereby making the relaxation tight (in terms $J_\text{opt}^\textsf{R}$ being equal to $J_\text{opt}$). 
Note that the solution to \eqref{prob:ideal:SDP:indiv:3} can also be obtained using a bisection search. However, feasibility test of $\mathcal X^\mathsf{R}(J)$ is a convex semi-definite programming problem and hence can be performed efficiently (in polynomial-time). The computational complexity\footnote{Generally, numerical techniques for semidefinite programming are iterative in nature, see \cite{Vanden96} for a detailed discussion. In the dual formulation of the problem, each iteration solves a linear problem in $M$ variables and $\frac{L(L+1)}{2}$ equations, with the resulting complexity being $\mathcal O(M^2 L^2)$. The number of such iterations is generally between $5$ to $20$ for many practical purposes (although theoretically, it is also a polynomial function).} of such a feasibility problem is roughly $\mathcal O(M^2 L^2)$, where $M$ is the number of sensors and $L$ is the number of non-zero collaboration weights. We have used the publicly available software \textsf{SeDuMi} as the optimization tool \cite{Sedumi99} for our numerical simulations.

\subsubsection{Closed Form solutions}
Though numerical solution of the general problem \eqref{prob:ideal:Q:indiv:2} can be obtained using the procedure outlined in Section \ref{sec:indiv:numerical}, (somewhat) closed form solutions can be obtained for some special cases. All the special cases discussed in this subsection will make use of the following optimization problem in its core,
\begin{equation}
\begin{aligned}
& \underset{\bo t}{\text{maximize}} & & F_{\bo t}=\frac{\left(\sum_{m=1}^M a_m t_m \right)^2 }{\sum_{m=1}^M b_m t_m^2 +\xi^2}, \\
&                            \text{subject to} & & 0\le t_m \le c_m, \, m=1,\ldots,M, \\
&                            \text{given that} & & a_m \mbox{ and } b_m \mbox{ are positive for all } m,
\end{aligned} \label{prob:special:indiv}
\end{equation}
which is known to have the following solution.
\begin{soln:special:indiv} (Solution of problem \eqref{prob:special:indiv}, see \cite{Jing09}):  \label{soln:special:indiv:lbl}
Order the sensors based on the parameter $d_m\triangleq \frac{a_m}{b_m c_m}$ such that, without loss of generality,
\begin{align}
d_{1} \ge d_{2} \ge \cdots \ge d_M. \label{d:order}
\end{align}
Define
\begin{align}
\Phi_k&\triangleq\frac{\sum_{m=1}^k b_m c_m^2 +\xi^2}{\sum_{m=1}^k a_m c_m}. \label{def:Phi}
\end{align}
Also, define $\widetilde m$ algorithmically as follows - keep checking in the decreasing order $\widetilde m=\{M,M-1,\ldots,2\}$ whether $\Phi_{\widetilde m-1} \frac{a_{\widetilde m}}{ b_{\widetilde m}} \ge c_{\widetilde m}$, and stop at the first instance this condition is satisfied. If $\Phi_{1} \frac{a_2}{ b_2} < c_2$, then $\widetilde m=1$. Then the solution to problem \eqref{prob:special:indiv} is $F_\textsf{opt}$ which is achieved when $\bo t=\bo t_\textsf{opt}$, where
\begin{align}
\begin{split}
F_\textsf{opt}&=\frac{\left(\sum_{m=1}^{\widetilde{m}} a_m c_m \right)^2 }{\sum_{m=1}^{\widetilde{m}} b_m c_m^2 +\xi^2}+\sum_{m=\widetilde{m}+1}^M \frac{a_m^2}{b_m}, \\
t_{\textsf{opt},m}&=\left\{ 
\begin{array}{rl} c_m, & m=1,\ldots,\widetilde m \\ 
\Phi_{\widetilde m} \frac{a_m}{b_m} < c_m, & m = \widetilde m+1,\ldots,M.
 \end{array} \right.
\end{split} \label{soln:indiv}
\end{align}
\end{soln:special:indiv}

Note that all the constraints are active (i.e., $\widetilde{m}=M$ or $t_{\textsf{opt},m}=c_m$ for all $m$) if and only if $\Phi_{M-1} d_M \ge 1$.

\emph{Notation:} Corresponding to constants $\bo a, \bo b$ and $\bo c$, we would denote the optimal solution \eqref{soln:indiv} of problem \eqref{prob:special:indiv} as $F_\textsf{opt}(\bo a,\bo b,\bo c)$. When mentioned in conjunction with $F_\textsf{opt}(\bo a,\bo b,\bo c)$, the corresponding value of $\bo t$ will be simply denoted as $\bo t_\textsf{opt}$, i.e., without the arguments, to avoid repetition.

Next, we provide some insightful examples. Some of these results will be used to derive the collaboration gain for homogeneous networks in Section \ref{sec:CG}. 

\emph{Example 4:}
For the problem with \emph{a) distributed topology, b) perfect information about observation and channel gains, and c) uncorrelated measurement noise}, problem \eqref{prob:ideal:Q:indiv:2} can be simplified as (note that $\bo w=\textsf{diag}(\bo W)$),
\begin{equation}
\begin{aligned}
& \underset{\bo w}{\text{maximize}} & & J_{\bo w}=\frac{\bo w^T \bo a \bo a^T \bo w}{\bo w^T \textsf{diag}(\bo b) \bo w+\xi^2} \\
&                            \text{subject to} & & w_m^2 \sigma_{\textsf{x},m}^2 \le P^\textsf{C}_m, \, m=1,\ldots,M,
\end{aligned} \label{prob:special:indiv:dist}
\end{equation}
where 
\begin{align}
\begin{split}
a_m&= g_m h_m,\; b_m= g_m^2\sigma_m^2, \; \sigma_m^2=\left[\bo \Sigma\right]_{m,m}, \\ 
\sigma_{\textsf{x},m}^2&= \sigma_m^2+\eta^2 h_m^2.
\end{split} \label{ab:dist}
\end{align}
Note that since $a_m$ and $b_m$ are positive, the value of $w_m$ at optimality has to be positive, hence the quadratic constraints in \eqref{prob:special:indiv:dist} reduce to the linear constraints as in \eqref{prob:special:indiv}. The rest of the problem is solved by defining  $c_m \triangleq \sqrt{\frac{P_m^\textsf{C}}{\sigma_m^2+\eta^2 h_m^2}}$ and applying Proposition \ref{soln:special:indiv:lbl}. We obtain
\begin{align}
J_\textsf{opt}\left( \bo P^\textsf{C} \right)=&\frac{\left(\sum_{m=1}^{\widetilde{m}} h_m g_m c_m \right)^2 }{\sum_{m=1}^{\widetilde{m}} \sigma_m^2 g_m^2 c_m^2 +\xi^2}+\sum_{m=\widetilde{m}+1}^M \frac{h_m^2}{\sigma_m^2}. \label{Jopt:both:indiv}
\end{align}

Equation \eqref{Jopt:both:indiv} is mathematically equivalent to the solution in \cite{Jing09}, which was obtained in the context of maximum-SNR beamforming. This is because, for perfect OGI and perfect CSI, maximum-SNR also implies minimum-MSE. However, in the presence of observation and channel gain uncertainties, the two problems are different. 

The optimal Fisher Information in \eqref{Jopt:both:indiv} can also be compared to that in the cumulative-constraint case. With individual constraints, some sensors (those with lower ``reliability-to-power ratio" $d_m= \frac{h_m \sqrt{\sigma_m^2+\eta^2 h_m^2}}{g_m \sigma_m^2 \sqrt{P_m^\textsf{C}}}$, precisely sensors $m=\widetilde m+1,\ldots,M$) effectively ``convey" the entirety of their individual Fisher Information to the FC while the other sensors ($m=1,\ldots,\widetilde m$) can convey only a fraction of their combined sum. In contrast, for the cumulative-constraint problem, different fractions of individual Fisher Information reach the FC. However, in the infinite-power limit, both the cases converge to the centralized Fisher Information $J_\textsf{cent}=\sum_{m=1}^M \frac{h_m^2}{\sigma_m^2}$. 

The previous example assumed perfect CSI and perfect OGI. The following extension of Example 3 (homogeneous network with equicorrelated parameters) illustrates the deterioration of performance with observation and channel gain uncertainties.

\emph{Example 5:} Consider a distributed topology with problem conditions similar to Example 3 (see \eqref{prob:homo}). To find the optimal distortion for this example, we proceed directly from \eqref{def:JW}, using the full-matrix notation
\begin{equation}
\begin{aligned}
& \underset{\begin{smallmatrix}\bo w \\ \bo W=\textsf{diag}(\bo w) \end{smallmatrix}}{\text{maximize}} & & J_{\bo w}=\frac{\left( \bo g^T \bo W \bo h \right)^2}{\tr\left[\bo E_\textsf{g} \bo W \bo E_\textsf{x} \bo W^T \right]-\eta^2 \left( \bo g^T \bo W \bo h \right)^2+\xi^2}, \\
&                            \text{subject to} & & \left[ \bo W \bo E_\textsf{x} \bo W^T \right]_{m,m} \le P^\textsf{C}_m, \, m=1,\ldots,M.
\end{aligned} \label{prob:ideal:indiv:toy}
\end{equation}
Noting the following identities,
\begin{align}
\begin{split}
\left( \bo g^T \bo W \bo h \right)^2 &= \alpha_\textsf{g} \alpha_\textsf{h} g_0^2 h_0^2 \left( \bo w^T \bo 1 \right)^2, \\
\tr\left[\bo E_\textsf{g} \bo W \bo E_\textsf{x} \bo W^T \right] &= g_0^2 \sigma_\textsf{x}^2 \left\{ (1-\alpha_\textsf{g} \alpha_\textsf{x}) \bo w^T \bo w \right. \\ 
&\qquad \qquad \qquad \left. +\alpha_\textsf{g} \alpha_\textsf{x} \left( \bo w^T \bo 1 \right)^2 \right\}, \\
\left[ \bo W \bo E_\textsf{x} \bo W^T \right]_{m,m} &=\sigma_\textsf{x}^2 w_m^2, \mbox{ where } \\
\sigma_\textsf{x}^2=\sigma^2+\eta^2 h_0^2, &\; \alpha_\textsf{x}=\frac{\rho \sigma^2+\alpha_\textsf{h} \eta^2 h_0^2 }{\sigma_\textsf{x}^2},
\end{split}
\end{align}
problem \eqref{prob:ideal:indiv:toy} is equivalent to,
\begin{equation}
\begin{aligned}
& \underset{\bo w}{\text{maximize}} & & F_{\bo w}=\frac{\alpha_\textsf{g} \alpha_\textsf{h} g_0^2 h_0^2 \left( \bo w^T \bo 1 \right)^2}{g_0^2 \sigma_\textsf{x}^2 (1-\alpha_\textsf{g} \alpha_\textsf{x}) \bo w^T \bo w+\xi^2}, \\
&                            \text{subject to} & & \sigma_\textsf{x}^2 w_m^2 \le P^\textsf{C}_m, \, m=1,\ldots,M,
\end{aligned} \label{prob:ideal:indiv:toy2}
\end{equation}
in the sense that $F_{\bo w}$ is monotonically related to $J_{\bo w}$ through $
J_{\bo w}=\left[ \frac{1}{F_{\bo w}} + \frac{\rho \sigma^2}{\alpha_\textsf{h} h_0^2} \right]^{-1}$. 
Problem \eqref{prob:ideal:indiv:toy2} is similar in form to problem \eqref{prob:special:indiv} and the solution can be obtained by applying Proposition \ref{soln:special:indiv:lbl}. We have
\begin{align}
&J_\textsf{opt}\left( \bo P \right)=\left[\frac{1}{F_\textsf{opt}\left(\bo a,\bo b,\bo c \right)}+\frac{\rho \sigma^2}{\alpha_\textsf{h} h_0^2}\right]^{-1}, \mbox{ where} \label{Jopt:homo:indiv} \\
&a_m= \sqrt{\alpha_\textsf{g} \alpha_\textsf{h}} g_0 h_0,\; b_m= \sigma_\textsf{x}^2 g_0^2 (1-\alpha_\textsf{g} \alpha_\textsf{x}),\;  c_m=\sqrt{\frac{P_m^\textsf{C}}{\sigma_\textsf{x}^2}}. \nonumber
\end{align}
When all the constraints are active, the solution is 
\begin{align}
\begin{split}
&J_\textsf{opt}\left( \bo P \right)=\left[\frac{g_0^2 (1-\alpha_\textsf{g} \alpha_\textsf{x}) \sum P_m+\xi^2}{\alpha_\textsf{g} \alpha_\textsf{h} g_0^2 h_0^2 \sigma_\textsf{x}^{-2}  \left( \sum \sqrt{P_m} \right)^2} + \frac{\rho \sigma^2}{\alpha_\textsf{h} h_0^2}\right]^{-1},
\end{split} \label{Jopt:homo:indiv:active}
\end{align}
from which it is evident that the Fisher Information decreases with more uncertainty in observation and channel gains (lower values of $\alpha_\textsf{h}$ and $\alpha_\textsf{g}$). This result will be useful later to derive the collaboration gain for this example.

The following example considers a fully connected network.

\emph{Example 6:} For the problem with \emph{a) fully connected topology ($\bo A=\bo 1\bo1^T$) and b) uncorrelated channel gain uncertainty ($\bo \Sigma_\textsf{g}$ is diagonal)}, we can start from \eqref{prob:ideal:indiv:toy}, where the variable of optimization is the entire matrix $\bo W$,
\begin{equation}
\begin{aligned}
& \underset{\bo W}{\text{maximize}} & & J_{\bo W}=\frac{\left( \bo g^T \bo W \bo h \right)^2}{\tr\left[\bo E_\textsf{g} \bo W \bo E_\textsf{x} \bo W^T \right]-\eta^2 \left( \bo g^T \bo W \bo h \right)^2+\xi^2}, \\
&                            \text{subject to} & & \left[ \bo W \bo E_\textsf{x} \bo W^T \right]_{m,m} \le P^\textsf{C}_m, \, m=1,\ldots,M.
\end{aligned} \label{prob:ideal:indiv:full}
\end{equation}
Problem \eqref{prob:ideal:indiv:full} can be simplified further based on the assumption of diagonal $\bo \Sigma_\textsf{g}$. However, the analysis is relegated to Appendix \ref{app:indiv:conn} and we state just the result here. 
\begin{result:indiv:conn} \label{result:indiv:conn:lbl} The optimal solution for Example 6 is
\begin{align} 
\begin{split}
&J_\textsf{opt}\left( \bo P^\textsf{C} \right)=\widetilde{J} \left[1+\frac{1+\eta^2  \widetilde{J}}{ F_\textsf{opt}\left(\bo a,\bo b,\bo c \right) }\right]^{-1}, \mbox{ where}\\
&\;\bo a= \bo g,\; \bo b=\textsf{diag}\left(\bo \Sigma_\textsf{g} \right), \mbox{ and } c_m=\sqrt{P_m^\textsf{C}},
\end{split} \label{Jopt:indiv:conn}
\end{align}
and $\widetilde{J} = \bo h^T \tbo \Sigma^{-1} \bo h$, $\tbo \Sigma=\bo \Sigma + \eta^2 \bo \Sigma_\textsf{h}$, as defined in \eqref{Jopt:conn}. Optimality is achieved when the corresponding weights are
\begin{align}
\bo W_\textsf{opt}=\kappa \bo t_\textsf{opt} \bo v^T,\;  \bo v=\widetilde{\bo \Sigma}^{-1} \bo h,\; \kappa=\frac{1}{\sqrt{\widetilde{J}(1+\eta^2 \widetilde{J})}}. \label{Wopt:indiv:conn}
\end{align} 
\end{result:indiv:conn}

\begin{IEEEproof} See Appendix \ref{app:indiv:conn}. \end{IEEEproof}

From the formula for optimal weights $\bo W_\textsf{opt}$ in \eqref{Wopt:indiv:conn}, we note that all the sensors have identical fusion rules (precisely, the vector $\bo v^T$) but they transmit using different transmission power (according to $\kappa \bo t_\textsf{opt}$). It may be surprising to note that even though all the sensors are transmitting the same information coherently, they may still refrain from using the maximum power available ($\bo t_{\textsf{opt},m} \le c_m$, in general). This is because of the uncertainty in the channel (which is captured by the diagonal entries of $\bo \Sigma_\textsf{g}$). If a channel is too uncertain, allocating a large amount of power to the sensor (even if the power is locally available) may not be helpful for inference. Indeed, all other parameters being constant, a higher magnitude of $\left[ \bo \Sigma_\textsf{g} \right]_{m,m}$ results in a lower value of $d_m$ (see problem \eqref{prob:special:indiv}) and hence the power constraint is more likely to be inactive at optimality. However, for perfect CSI ($\bo \Sigma_\textsf{g}=0$), all the sensors must transmit with maximum power available.

We are now in a position to explicitly derive the formula of collaboration gain for homogeneous networks. It must be mentioned here that explicit formulas of $\textsf{CG}$ for arbitrary problems are difficult to derive and we only provide the example of homogeneous networks in this paper. As mentioned in Section \ref{sec:def:CG}, collaboration gain is a useful metric that indicates the efficacy of spatial collaboration for estimation. 

\subsection{Collaboration Gain for a Homogeneous Network} \label{sec:CG}
This analysis of homogeneous networks is particularly illustrative since it is actually possible to derive closed-form expressions of $\textsf{CG}$ for a wide range of individual (and also cumulative) power constraints. From the closed form Equation of $\textsf{CG}$, we can actually predict the conditions for which collaboration will be particularly effective. Specifically, we will derive the formula for collaboration gain and analyze how it depends on the problem parameters like size of the network, noise correlation, observation/channel gain uncertainties and power constraints.

\subsubsection{Theoretical analysis for a homogeneous network} \label{sec:homo}
Let $P$ be the total power used in the network. For the cumulative-constraint problem, $P$ is a sufficient descriptor of the power constraints. However, for the individual-constraint problem, we have $M$ different power constraints $\{ P_1, P_2, \ldots, P_M\}$, which makes it difficult to analyze the problem theoretically. Fortunately, for a large subset of those problems, only $2$ summary-descriptors of $\{P_1, P_2, \ldots, P_M\}$ suffice to characterize the collaboration gain -- namely, the cumulative power $P$, and a \emph{skewness} parameter $\kappa$, defined as,
\begin{align}
\kappa \triangleq \frac{ P_\textsf{sq} }{P},\; P_\textsf{sq} \triangleq \left(\sum_{m=1}^M \sqrt{P_m} \right)^2,\; P = \sum_{m=1}^M P_m. \label{def:kappa}
\end{align}
As a example, we refer to Equation \eqref{Jopt:homo:indiv:active}, where the distortion is seen to depend on only $\sum P_m$ (which is $P$) and $\left(\sum \sqrt{P_m} \right)^2$ (which is $\kappa P$). It is easy to see (applying Cauchy-Schwartz inequality) that 
$1\le \kappa \le M$, where the limiting values have distinct significance. The value $\kappa=1$ implies that only one of the sensors has all the power ($P_1=P$, say) while other sensors have no power at all  $P_2=\cdots=P_M=0$. On the other hand, the value $\kappa=M$ implies that all sensors have equal power allocated to them $P_1=\cdots=P_M=\frac{P}{M}$. Define $\kappa_M$ (a normalized\footnote{The parameter $\kappa_M$ is similar in structure to the Chiu-Jain fairness metric \cite{Chiu89} used for congestion control in computer networks.} version of $\kappa$) as 
\begin{align}
\kappa_M=\frac{\kappa -1}{M-1} \in [0,1]. \label{def:kappa:M}
\end{align}
A stronger rationale behind naming the parameter $\kappa$ as \emph{skewness} is provided by the following result, which shows that $\kappa$ is monotonically related to another quantity that explicitly enables power allocation in a skewed manner.

\begin{result:skewness}[Parameter $\kappa$ is an indicator of skewness] \label{result:skewness:lbl}
Assume that power available at the $M$ nodes are
\begin{align}
P_1=P_0, P_2=\varrho P_0,\ldots, P_M=\varrho^{M-1} P_0,\; \varrho\in [0,1], \label{P:alloc}
\end{align}
so that $\varrho$ is an explicit measure of dissimilarity (or skewness) among the power available at various nodes. Then (recall definition of $\kappa$ in \eqref{def:kappa}),
\begin{align}
\kappa(\varrho) = \frac{\left( 1+\sqrt{\varrho}+\cdots+\left( \sqrt{\varrho} \right)^{M-1} \right)^2}{1+\varrho+\cdots+ \varrho^{M-1}}
\end{align}
is a strictly monotonically increasing function of $\varrho$.  Note that $\kappa(\varrho=0)=1$ and $\kappa(\varrho=1)=M$. 
\end{result:skewness}
\begin{IEEEproof} See Appendix \ref{app:skewness}. \end{IEEEproof}

The following proposition provides an explicit formula for the collaboration gain in a homogeneous network with equicorrelated parameters for both the individual-constraint and cumulative-constraint problems. We would use Examples 3, 5 and 6 to establish this result, the derivation of which is relegated to Appendix \ref{app:CG:indiv}.

\begin{result:CG:indiv} \emph{(Collaboration gain for a homogeneous network)}: \label{result:CG:indiv:lbl} 
For the individual-constraint problem, let the node indices be arranged (without loss of generality) in such a way so that $P_1^\textsf{C}\le \cdots \le P_M^\textsf{C}$. Let $P$ and $\kappa$ denote the summary-descriptors of $\{ P_1^\textsf{C}, \ldots, P_M^\textsf{C} \}$ as per Equation \eqref{def:kappa}. Assume,
\begin{align}
P_M^\textsf{C} \le \left( \frac{\sum_{m=1}^{M-1} P_m^\textsf{C}+\frac{\xi^2}{g_0^2(1-\alpha_\textsf{g} \alpha_\textsf{x}) } }{\sum_{m=1}^{M-1} \sqrt{P_m^\textsf{C}} } \right)^2. \label{CG:indiv:cond}
\end{align}
Then, all the power constraints are active at optimality, $P_m=P_m^\textsf{C},\, \forall m$,  for both distributed and connected topologies, and the collaboration gain is
\begin{align}
\begin{split}
&\textsf{CG}=\frac{\left(1-\frac{1}{M}\right) \left( \frac{\kappa}{M}\frac{1+(M-1)\alpha_\textsf{g}}{1+(\kappa-1)\alpha_\textsf{g}} \right) \left(1-\frac{\kappa-1}{M-1}\frac{\alpha_\textsf{g}}{1+P_\textsf{g}^{-1}}\right)\left(1-\alpha_\textsf{x}\right)}{\left(1+\frac{1}{P_\textsf{g}\left(1+(\kappa-1)\alpha_\textsf{g}\right)}\right)\left(1+\frac{(\kappa-1)\alpha_\textsf{g}}{1+P_\textsf{g}^{-1}}\alpha_\textsf{x} \right)}, \\
&\mbox{where}\; P_\textsf{g}= \frac{P g_0^2}{\xi^2},\; \alpha_\textsf{x}= \frac{\rho+\gamma \alpha_\textsf{h}}{1+\gamma},\; \gamma=\frac{\eta^2 h_0^2}{\sigma^2}.
\end{split} \label{CG:indiv}
\end{align}
For the cumulative-constraint problem, the collaboration gain is given by Equation \eqref{CG:indiv} by setting $\kappa=M$.
\end{result:CG:indiv}

\begin{IEEEproof} See Appendix \ref{app:CG:indiv}.  \end{IEEEproof}

Several remarks about Proposition \eqref{result:CG:indiv:lbl} are in order. Condition \eqref{CG:indiv:cond} is basically an assumption that the power constraints are \emph{not too skewed}. In Equation \eqref{CG:indiv} note that all the quantities in parenthesis in the numerator denote quantities less than $1$, while those in the denominator are greater than one, so this reaffirms the notion that collaboration gain is always less than $1$. Furthermore,

a) \emph{Dependence on local-SNR $\gamma$, noise correlation $\rho$, and observation gain uncertainty $\alpha_\textsf{h}$:} Collaboration gain increases with a decrease in $\alpha_\textsf{x}$, which means that $\textsf{CG}$ increases as a) Noise correlation decreases (smaller $\rho$), b) observation gain uncertainty increases (smaller $\alpha_\textsf{h}$), and c) local SNR decreases (smaller $\gamma$), provided $\rho<\alpha_\textsf{h}$  (which is typically true for a problem involving moderately correlated noise and sufficiently certain observation gain). Hence, collaboration is more effective when the local-SNR is small, measurement noise is uncorrelated and observation gains are uncertain.

b) \emph{Dependence on (normalized) total power $P_\textsf{g}$, power skewness $\kappa$ and  channel gain uncertainty $\alpha_\textsf{g}$:} To understand the effect of $P_\textsf{g}, \kappa$ and  $M$ on collaboration gain, we simplify Equation \eqref{CG:indiv} by considering the large-$M$ asymptotic regime. Note that in general, an infinite number of sensors implies that the Fisher Information is infinite and distortion is zero for both distributed and connected topologies, which is a trivial regime to consider. Towards the goal of analyzing regimes that incur only finite distortion in the asymptotic limit, we consider two cases, as listed below.

First, we would consider the \emph{fixed-$\kappa$-large-$P$ regime}, which signifies that the effective number of powered nodes are not increasing with $M$ (recall that $\kappa=1$ implies that only one node has all the power, regardless of $M$), i.e., most of the nodes are auxilliary nodes that provide their information to the powered nodes which then communicate with the fusion center. Hence, even with a large transmission power the distortion at the FC is finite, which makes this regime non-trivial. From \eqref{CG:indiv}, we note that
\begin{align}
\lim_{\begin{smallmatrix}  
M\rightarrow \infty \\ P_\textsf{g}\rightarrow \infty \end{smallmatrix}} \textsf{CG}
=\frac{(1- \alpha_\textsf{x})\alpha_\textsf{g} \kappa}{\left( 1+\alpha_\textsf{g}(\kappa-1) \right) \left( 1+\alpha_\textsf{x} \alpha_\textsf{g}(\kappa-1) \right)}. \label{CG:indiv:regime1}
\end{align}
With the additional technical assumption that $\alpha_\textsf{g}>\frac{1}{1+\alpha_\textsf{x}}$ (which basically means that the channel gains are sufficiently certain), the collaboration gain (Equation \eqref{CG:indiv:regime1}) decreases as $\kappa$ increases, and hence is maximum when $\kappa=1$, at which point 
\begin{align}
\textsf{CG}=(1- \alpha_\textsf{x})\alpha_\textsf{g}.
\end{align}
That $\textsf{CG}$ decreases with $\alpha_\textsf{x}$ has already been discussed in the previous remark, and also applies for this regime. We conclude that in the fixed-$\kappa$-large-$P$ regime, collaboration is highly effective if there are only a few powered sensors (small $\kappa$), and the channel gains (higher $\alpha_\textsf{g}$) are fairly certain.

Next, we consider the \emph{fixed-$(\kappa/M)$-finite-$(P\times M)$ regime}. Here, the normalized skewness parameter $\kappa_M$ is kept constant, which means that effectively, the number of powered sensors $\kappa$ increase linearly with $M$ (note the contrast with previous regime). To keep the effective Fisher Information (and resulting distortion) finite, the total available power $P$ must scale inversely proportional\footnote{Another asymptotic domain that is conceptually different but otherwise would yield identical results is to keep the power $P$ constant and let the channel gains $\bo g$ scale at the rate of $\frac{1}{\sqrt{M}}$. The argument here is that, though the network size increases with $M$, the channel capacity of the effective multiple-input-single output (MISO) channel induced by the $M$ sensors, precisely $\frac{1}{2}\log\left(1+\frac{P\sum_{m=1}^M g_m^2}{\xi^2}\right)$, is held constant. } to $M$. Let $P_{M}\triangleq P_\textsf{g} M$ denote the normalized total power, which is a finite constant.  From \eqref{CG:indiv}, we can derive that
\begin{align}
\max_{P_{M}} \lim_{\begin{smallmatrix}  
M\rightarrow \infty \\  P_\textsf{g} M=P_{M}                                           \end{smallmatrix}} \textsf{CG} = \frac{1-\sqrt{\alpha_\textsf{x}}}{1+\sqrt{\alpha_\textsf{x}} }, \; \mbox{when} \; P_{M}^{*}=\frac{1}{\kappa_M \alpha_\textsf{g}\sqrt{\alpha_\textsf{x}}}. \label{CGmax}
\end{align} 
This implies that collaboration is more effective for smaller $\alpha_\textsf{x}$, a fact that was established earlier as well. We conclude that in the fixed-$(\kappa/M)$-finite-$(P\times M)$ regime, collaboration is most effective when the normalized operating power is a particular finite quantity, precisely $P_M^{*}=\frac{1}{\kappa_M \alpha_\textsf{g} \sqrt{\alpha_\textsf{x}}}$. Moreover, $P_M^{*}$ increases as the power constraints get more skewed (smaller $\kappa_M$) or the channel gains get more uncertain (smaller $\alpha_\textsf{g}$).

\begin{figure*}[htb]
\centering
    \includegraphics[width=\figsze \columnwidth]{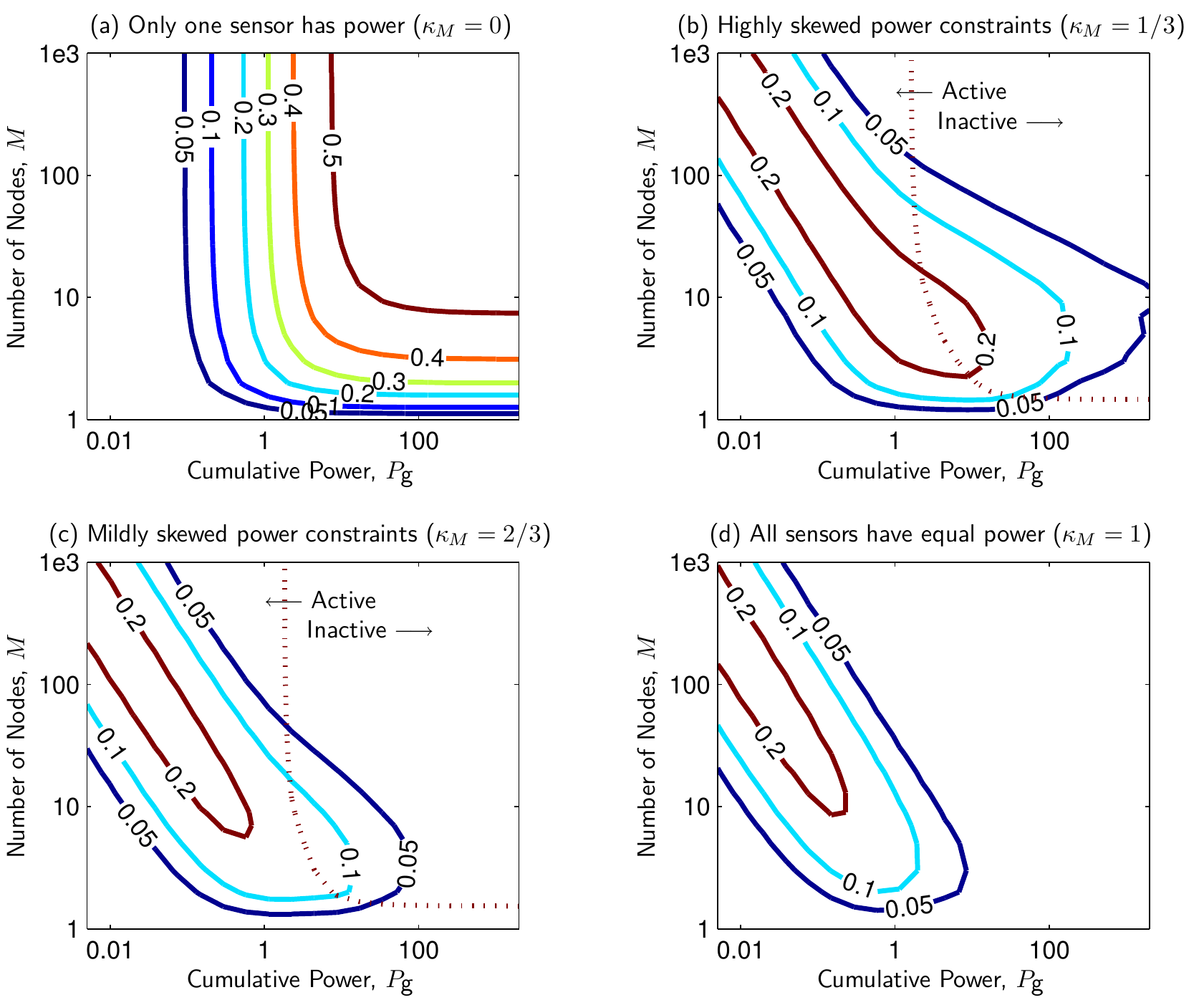}
  \caption{Collaboration gain in a homogeneous sensor network with skewed power constraints. }
  \label{fig:CG:indiv}
\end{figure*}

The assertions in Proposition \ref{result:CG:indiv:lbl} and the subsequent discussion are illustrated in Figure \ref{fig:CG:indiv} for an example with the following problem parameters $\eta^2=\frac{1}{2}$, $\sigma^2=1$, $\rho=0.1$, $ g_0=h_0=1$ and $\alpha_\textsf{g}=\alpha_\textsf{h}=0.9$. Contours of the actual collaboration gain are computed numerically using procedure outlined in Section \ref{sec:indiv} and displayed for a wide range of cumulative power $P_\textsf{g}$, number of nodes $M$ and skewness of power constraints. To illustrate the fixed-$\kappa$-large-$P$ regime, we have used $\kappa=1$ (only one sensor has all the power) in Figure \ref{fig:CG:indiv}(a). To illustrate the fixed-$(\kappa/M)$-finite-$(P\times M)$ regime, we have used $\kappa_M=\{1/3,2/3,1\}$ in Figures \ref{fig:CG:indiv}(b), \ref{fig:CG:indiv}(c), and \ref{fig:CG:indiv}(d) respectively. To simulate a particular skewness $\kappa$, the local power constraints are generated in accordance with \eqref{P:alloc} by finding the corresponding value $\varrho$ through bisection search (recall that $\varrho$ and $\kappa$ are monotonically related, as per Lemma \ref{result:skewness:lbl}). The active-constraint condition of Equation \eqref{CG:indiv:cond} is depicted through the dotted line in Figures \ref{fig:CG:indiv}(b) and \ref{fig:CG:indiv}(c). The portion of the figure to the right of the dotted line suggests that one or more of the constraints are inactive, while the left half denotes the region where all the constraints are active and consequently Equation \eqref{CG:indiv} is the accurate measure of collaboration gain. No dotted lines appear in Figure \ref{fig:CG:indiv}(a) and \ref{fig:CG:indiv}(d) because all the constraints are trivially active in both the cases, although for different reasons. For $\kappa=1$, there is only one sensor with all the power and hence it must transmit with full power, which explains Figure \ref{fig:CG:indiv}(a). For $\kappa=M$, we have equal power allocation ($P_m^\textsf{C}=P^\textsf{C}$, say) and condition \eqref{CG:indiv:cond} reduces to $P^\textsf{C} \le \left(\sqrt{P^\textsf{C}} +\frac{\xi^2}{g_0^2(1-\alpha_\textsf{g} \alpha_\textsf{x}) (M-1) \sqrt{P^\textsf{C}} }  \right)^2$, which is trivially satisfied for all $P^\textsf{C}$.

For the problem parameters mentioned above, we calculate that $\alpha_\textsf{x}\approx 0.37$. For the fixed-$\kappa$-large-$P$ regime, theoretical justifications predict that the maximum gain possible (across various problem conditions) is $(1-\alpha_\textsf{x})\alpha_\textsf{g}\approx 0.57$, which can be confirmed from the contours toward the top-right corner of Figure \ref{fig:CG:indiv}(a). For the fixed-$(\kappa/M)$-finite-$(P\times M)$ regime, theoretical predictions yield that the maximum gain is $\frac{1-\sqrt{\alpha_\textsf{x}}}{1+\sqrt{\alpha_\textsf{x}} } \approx 0.24$, which can be validated from the innermost contours of Figures \ref{fig:CG:indiv}(b), \ref{fig:CG:indiv}(c) and \ref{fig:CG:indiv}(d). Note that the contours shifts to the left as $\kappa_M$ increases. This is due to the fact that the normalized power required to achieve maximum collaboration gain decreases as $\kappa_M$ become larger, a fact also discussed above. In conclusion, for this particular problem instance, upto $57\%$ (and $24\%$) of the distortion performance can be recovered using collaboration in the fixed-$\kappa$-large-$P$ regime (and fixed-$(\kappa/M)$-finite-$(P\times M)$) regimes. We have established this fact using both numerical results and theoretical insights.

\subsubsection{Random geometric graphs} \label{sec:ideal:rgg}
\begin{figure}[htb]
\begin{center}
    \includegraphics[width=\figszf \columnwidth,height=\figszf \columnwidth]{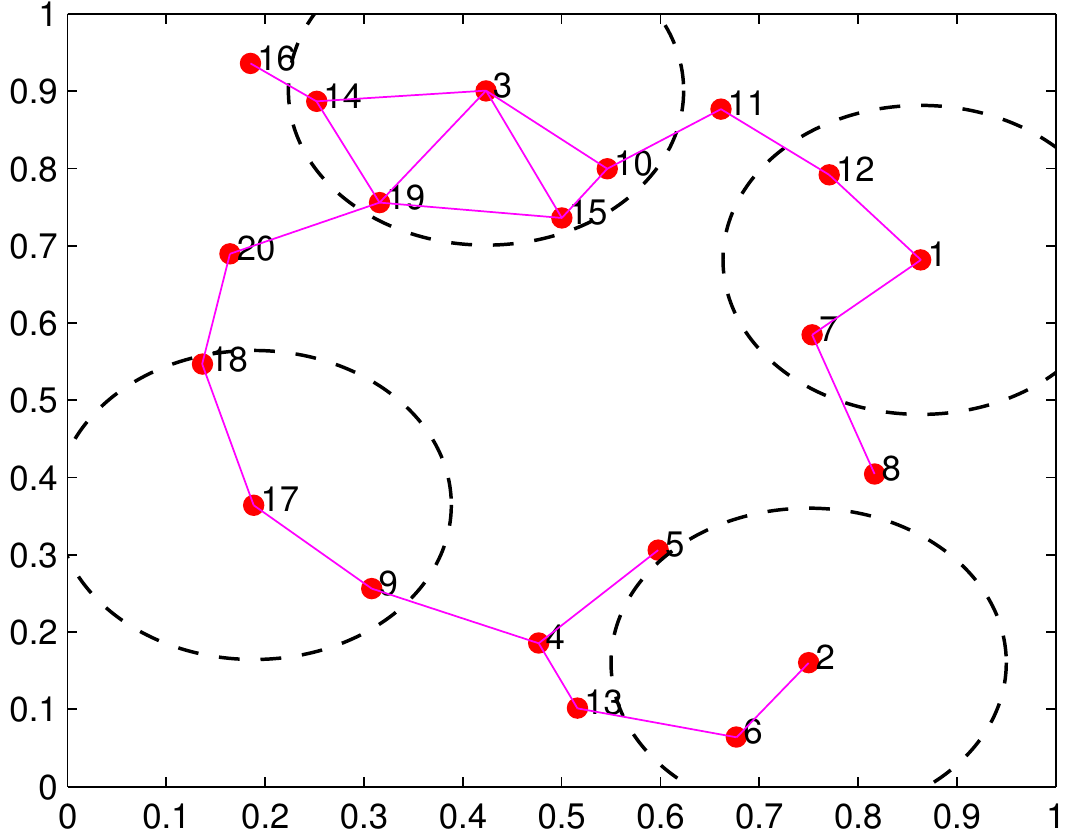}
  \caption{Random Geometric Graph with 20 nodes, used for example in Section \ref{sec:ideal:rgg}. Edges are shown for pairwise distance less than 0.2. The radius of collaboration is depicted for sensors 1,2,3 and 17.}
  \label{fig:rgg}
  \end{center}
\end{figure}

To demonstrate how the distortion decreases with increasing collaboration, we consider the following simulation setup. The spatial placement and neighborhood structure of the sensor network is modeled as a Random Geometric Graph, $\text{RGG}(N,r)$ \cite{Freris10}, where sensors are uniformly distributed over a unit square and bidirectional communication links are possible only for pairwise distances at most $r$, i.e., the adjacency matrix is
$\bo A$ such that $A_{i,j}=\indicator{d_{i,j}\le r}$. 
 Correspondingly, the cost matrix is a $\{0,\infty\}$ matrix with the $(i,j)$\sups{th} element being zero only if $d_{i,j}\le r$, otherwise being infinity. We assume $N=20$ sensor nodes and gradually increase the radius of collaboration from $r=0$ (signifying distributed topology) to $r=\sqrt{2}$ (signifying connected topology, since the sensors are placed in a unit square). The simulated sensor network is depicted in Figure \ref{fig:rgg}, with collaboration radius $r=0.2$. In general, for $0<r<\sqrt{2}$, the network is only partially connected as in Figure \ref{fig:rgg}. 

We simulate a homogeneous network with the following parameters $\sigma^2=1$, $\rho=0$ (independent noise), $g_0=h_0=1$ and $\alpha_\textsf{g}=\alpha_\textsf{h}=0.9$. To contrast the effect of prior uncertainty on collaboration gain, we simulate two different variance of the prior, $\eta^2=0.1$ and $0.5$. To illustrate the effect of power constraints, we simulate a wide range of both cumulative-power and skewness. We simulate three skewness conditions -- $\kappa_M=0.5, 0.75$ and $1$, the value of $1$ implying equal power allocation. The three values of cumulative-power that were simulated were a) $P_\textsf{g}=P^*=\frac{1}{0.75 \times 20}\frac{1}{\alpha_\textsf{g} \sqrt{\alpha_\textsf{x}}}$ (recall from the discussion in Section \ref{sec:homo} that, for $\kappa=0.75$, this is a high-$\textsf{CG}$ operating region), b) $P_\textsf{g}=P^*/4$ (depicting the low power regime), and c) $P_\textsf{g}=4 P^*$ (depicting the high power regime).

\begin{figure*}
\centering
\subfigure[$\eta^2=0.1$]{
\includegraphics[width=\figszg \columnwidth]{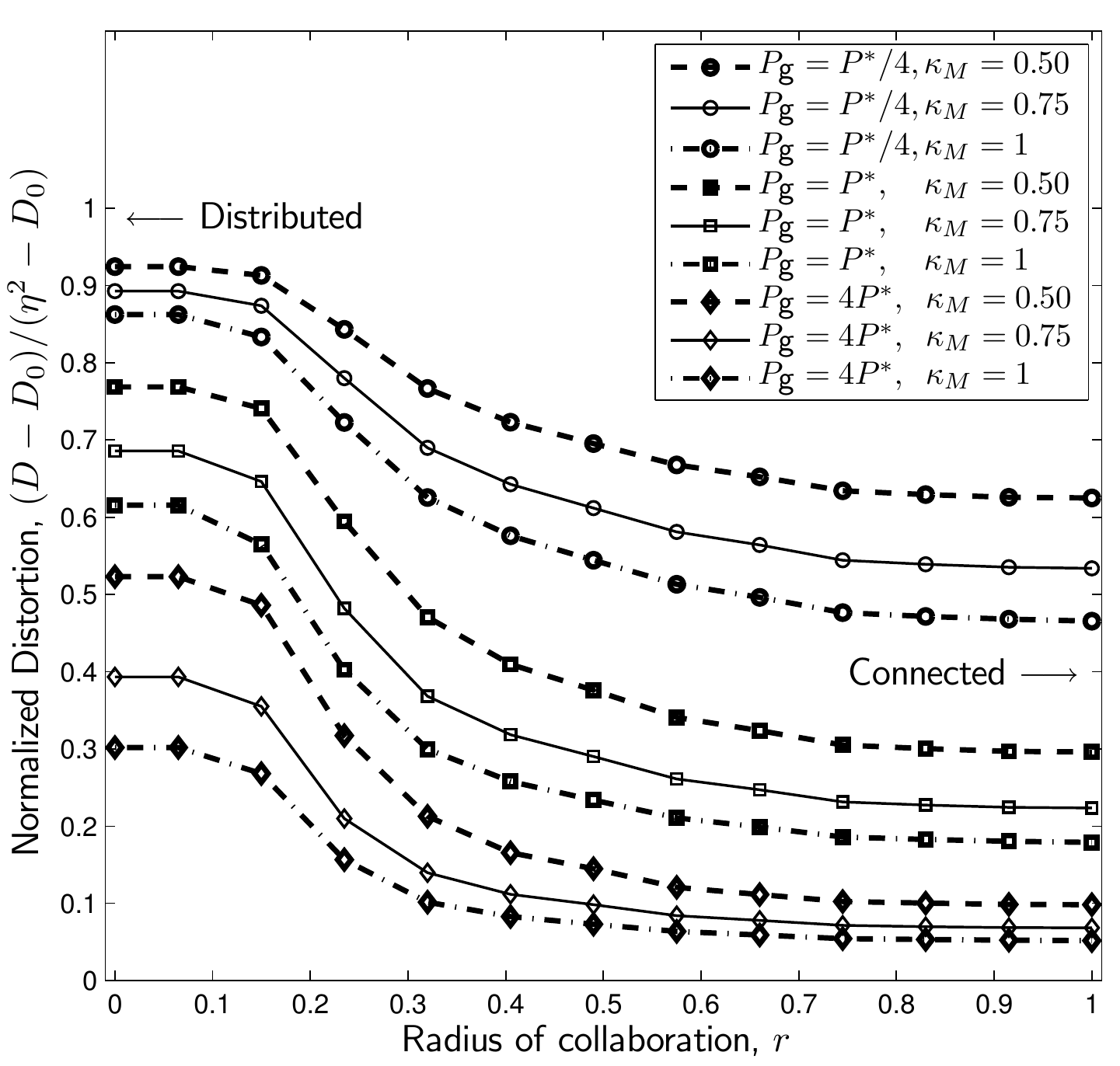}
\label{fig:rgg:e2:1}
}
\subfigure[$\eta^2=0.5$]{
\includegraphics[width=\figszg \columnwidth]{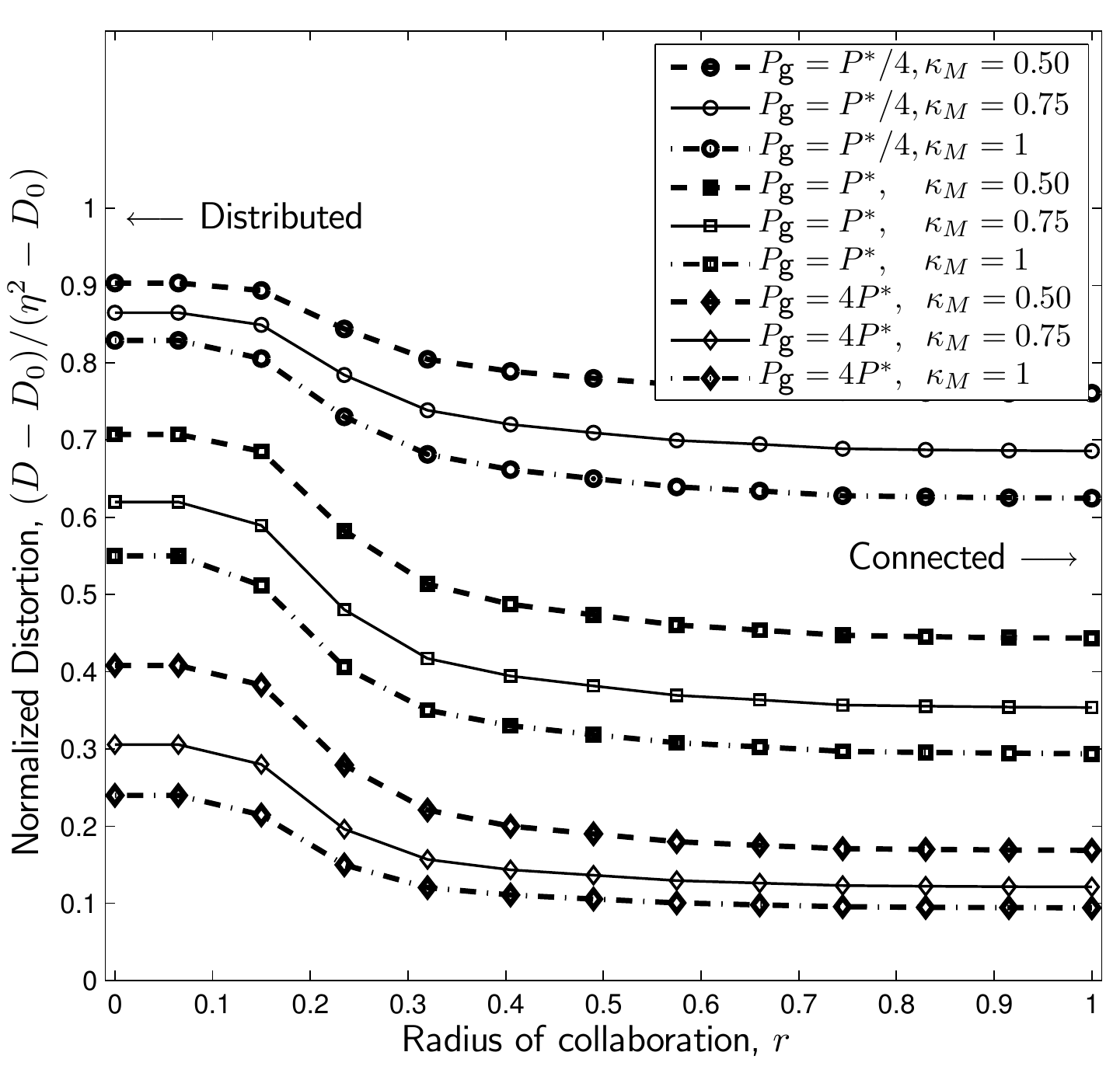}
\label{fig:rgg:e2:5}
}
\caption{Improved in distortion due to increasing collaboration among 20 nodes in a random geometric graph. }
\label{fig:rgg:e2}
\end{figure*}

The simulation results are depicted in Figures \ref{fig:rgg:e2}-(a) and (b) for the two values of prior uncertainty $\eta^2=0.1$ and $\eta^2=0.5$ respectively. Corresponding to these values of $\eta^2$, the infinite-power distortion $D_0$, the maximum possible collaboration gain $\textsf{CG*}$ and the corresponding operating power $P^*$ for $\kappa_M=0.75$, are calculated by using \eqref{Jopt:cyc} and \eqref{CGmax},
\begin{align}
\begin{matrix} 
\eta^2\downarrow   & D_0  & \textsf{CG*} & P^*(\kappa_M=0.75) \\
0.1                         & 0.07  & 0.29            & 0.13                          \\
0.5                         & 0.04  & 0.56            & 0.25.
\end{matrix}
\end{align}

With varying power availability and varying extent of collaboration, the resulting distortion varies between $D\in(D_0,\eta^2)$, where the ranges are $(0.07,0.1)$ and $(0.04,0.5)$, for the prior variance $\eta^2=0.1$ and $\eta^2=0.5$ respectively. To compare the effect of collaboration across the two different problem conditions, we depict the normalized distortion $\frac{D-D_0}{\eta^2-D_0}$ in Figures \ref{fig:rgg:e2}-(a) and (b). Note that in this normalized scale, the collaboration gain is simply the difference between the right-most (distributed) and the left-most (connected) ends of a curve. The efficacy of collaboration, as indicated by the downward slope of the curves, is clearly demonstrated in Figures \ref{fig:rgg:e2}-(a) and (b), where it is noted that a large part of the overall gain is achieved only with partial collaboration, i.e., the distortion tends to saturate beyond a collaboration radius of $r\approx0.4$. Hence, though the collaboration gain, as defined in \eqref{def:CG}, requires a fully connected topology, a large part of that gain can be realized with only a partially connected network. Also, we note that the efficacy of collaboration diminishes when the operating power is too small ($P_\textsf{g}=P^*/4$) or too large ($P_\textsf{g}=4 P^*$), indicating the fact that collaboration in a network should be used judiciously, specially when there are costs associated with it. Finally, we note that the efficacy of collaboration is higher when the prior has a lower variance (the curves in Figure \ref{fig:rgg:e2}-(a) has more downward slope than those in Figure \ref{fig:rgg:e2}-(b)). This observation is explained directly by the comments following Proposition \ref{result:CG:indiv:lbl}, where we argued that collaboration gain increases with decreasing local-SNR.

\section{Collaboration with finite costs} \label{sec:finite}
In Section \ref{sec:ideal}, we have solved the optimal collaboration problem for the situation when the cost of communication for each link is either zero or infinity, i.e., $C_{i,j}\in \{0,\infty \}$ (also termed as the ideal case). In this section, we address the general problem where communication may incur a non-zero but finite cost, i.e., $0 < C_{i,j} < \infty$ (also termed as the finite-cost case). Unlike the ideal case, finding the globally optimal solution for the finite-cost case is a difficult problem and there are no known numerical techniques that efficiently solve this problem. In this section, we outline an efficient (polynomial-time) numerical procedure that obtains a locally optimal solution to the finite-cost problem. We first describe our iterative solution for the individual-constraint problem. The cumulative-constraint problem follows from similar arguments and is described next. Lastly, solutions to both the problems are demonstrated using numerical simulations. 

\subsection{Individual power constraint}
We propose an iterative solution as follows. 
Let the collaboration topology and transmission power availability (for all $N$ nodes, in vector form) at iteration $i$ be denoted by $\bo A_i$ and $\bo P^\textsf{trans}_i$ respectively. Note that the transmission power availability is the difference between the original power constraint and the collaboration cost due to the topology $\bo A_i$. For the $n$\sups{th} node, it means
\begin{align}
\left[ \bo P^\textsf{trans}_i \right]_n=\left[ \bo P^\textsf{C}\right]_n-\sum_{m=1}^M \left[\bo A_i  \right]_{mn} \bo C_{mn}.
\end{align}
Note that for the auxiliary nodes $n=M+1,\ldots,N$, the transmission power availability $\left[ \bo P^\textsf{trans}_i \right]_n$ does not mean much, since they cannot transmit to the FC anyway. For those sensors, it is best if they use their entire resources for collaboration. 
Recall that the optimal distortion corresponding to any topology $\bo A$ and transmission power constraint $\bo P^\textsf{trans}$ can be obtained from the discussion in Section \ref{sec:ideal}. Denote such a distortion by $D_\textsf{opt}^{\bo A}\left( \bo P^\textsf{trans} \right)$. We start with a distributed topology, i.e., $\bo A_1=\left[\bo I_M | \bo 0 \right]$ and follow a greedy algorithm. At iteration $i$, we evaluate the distortion performance corresponding to all incremental topologies of the form $\bo A_i+ \bo E(m,n)$, where $\bo E(m,n)$ is an all-zero matrix except for the $(m,n)$\sups{th} element, which is $1$ (signifying an incremental $n\rightarrow m$ link). There are $M N-\textsf{nnz}(\bo A_i)$ such possibilities for selecting $\bo E(m,n)$, each corresponding to a link that is current not being used (equivalently $\left[ \bo A_i \right]_{mn}$ is zero). The number of such possibilities may even be lesser if there is not enough power to make a link possible. For example, if $C_{mn}>\left[ \bo P^\textsf{trans}_i \right]_{n}$, then the $n$\sups{th} node does not have sufficient power to use the $n\rightarrow m$ link for collaboration. Among all such possible links, let $n^*\rightarrow m^*$ denote the link that provides the best distortion performance. Then the iteration is concluded by augmenting the topology with the edge $n^*\rightarrow m^*$. Thus, in compact notations, each iteration is represented as,
\begin{align}
\begin{split}
(m^*,n^* )&=\arg \min_{
\bs m,n  \\ 
\left[A_i\right]_{mn}=0 \\ 
C_{mn} \le \left[ \bo P^\textsf{trans}_i \right]_{n} \es} 
D_\textsf{opt}^{\bo A_i+\bo E(m,n)}\left( \bo P^\textsf{trans}_i-\bo e_n C_{mn} \right), \\
\bo A_{i+1}&=\bo A_i+ \bo E_{m^*n^*},
\end{split} \label{iter:indiv}
\end{align}
where $\bo e_n$ is an all-zero vector with the exception of $n$\sups{th} element, which is $1$.
The iterations are terminated when one of the following conditions is satisfied, a) there is no feasible link, b) maximum number of iterations has exceeded a pre-specified limit or c) there is not enough increment in (relative) performance after a particular iteration, say, for a pre-specified $\delta$, if
\begin{align}
\frac{D_\textsf{opt}^{\bo A_{i+1}}\left( \bo P^\textsf{trans}_{i+1} \right)-
 D_\textsf{opt}^{\bo A_i}\left( \bo P^\textsf{trans}_i \right)}{\eta^2-D_0} \le \delta. \label{stop:indiv}
\end{align}
A rough estimate of computational complexity can be established as follows. Recall from Section \ref{sec:indiv} that computing $D_\textsf{opt}^{\bo A}(\bo P )$ is roughly $\mathcal O\left(M^2 L^2 \right)$, where $L=\textsf{nnz}(\bo A)$. Assume that the algorithm gets terminated after adding $\mathcal O(N)$ links, which is to say that each node communicates with a fraction of its neighbors at optimality. Since we start with the distributed topology we have $L=\mathcal O(N)$ for all the iterations. Since each of the iterations (as in \eqref{iter:indiv}) require approximately $M N$ function evaluations, the total complexity of the finite-cost collaborative power allocation problem is 
\begin{align}
\underbrace{\mathcal O(M^2 N^2)}_{\bs \text{evaluation} \\ \text{of } D_\textsf{opt} \es} \times \underbrace{\mathcal O(M N)}_{\bs \text{evaluations} \\ \text{per iteration} \es} \times \underbrace{\mathcal O(N)}_{\bs \text{number} \\ \text{of iterations} \es}=\mathcal O(M^3 N^4). \label{order:indiv}
\end{align}
It must be emphasized here that \eqref{order:indiv} is only a rough (but practical) measure of complexity and is not a rigorous bound. It assumes a fixed number of iterations to solve a semidefinite optimization problem (see discussion in Section \ref{sec:indiv}) and also sufficiently high collaboration costs so that only $\mathcal O(N)$ links are added starting from a distributed topology.

\subsection{Cumulative power constraint}
To solve the cumulative power-constraint problem, we proceed on similar lines. Let the cumulative power constraint be $P^\textsf{C}$ and the cumulative transmission power availability at iteration $i$ be denoted by $P^\textsf{trans}_i$, so that 
\begin{align}
P^\textsf{trans}_i = P^\textsf{C} -\sum_{m=1}^M \sum_{n=1}^N \left[\bo A_i  \right]_{mn} \bo C_{mn}.
\end{align}
Denote the optimal distortion corresponding to any topology $\bo A$ and transmission power constraint $P^\textsf{trans}$ (see Section \ref{sec:cum}) as $D_\textsf{opt}^{\bo A}\left(P^\textsf{trans} \right)$. Starting from a distributed topology, the best collaboration link $n^*\rightarrow m^*$ is selected at each iteration according to
\begin{align}
\begin{split}
(m^*,n^* )&=\arg \min_{
\bs m,n  \\ 
\left[A_i\right]_{mn}=0 \\ 
C_{mn} \le   P^\textsf{trans}_i \es} 
D_\textsf{opt}^{\bo A_i+\bo E(m,n)}\left( P^\textsf{trans}_i - C_{mn} \right), \\
\bo A_{i+1}&=\bo A_i+ \bo E_{m^*n^*},
\end{split} \label{iter:cum}
\end{align}
with the stopping criteria being similar to that described in the previous subsection. As regards to computational complexity, recall from \eqref{sec:cum} that the complexity of computing
$D_\textsf{opt}^{\bo A}(\bo P )$ is $\mathcal O\left(L^3 \right)$, where $L=\textsf{nnz}(\bo A)$. Assuming $\mathcal O(N)$ iterations (as in the previous case) the overall complexity for the finite-cost-collaborative cumulative-constraint problem is roughly
\begin{align}
\underbrace{\mathcal O(N^3)}_{\bs \text{evaluation} \\ \text{of } D_\textsf{opt} \es} \times \underbrace{\mathcal O(M N)}_{\bs \text{evaluations} \\ \text{per iteration} \es} \times \underbrace{\mathcal O(N)}_{\bs \text{number} \\ \text{of iterations} \es}=\mathcal O(M N^5).
\end{align}

\subsection{Numerical Simulations}
\begin{figure}[htb]
\centering
    \includegraphics[width=\figszh \columnwidth]{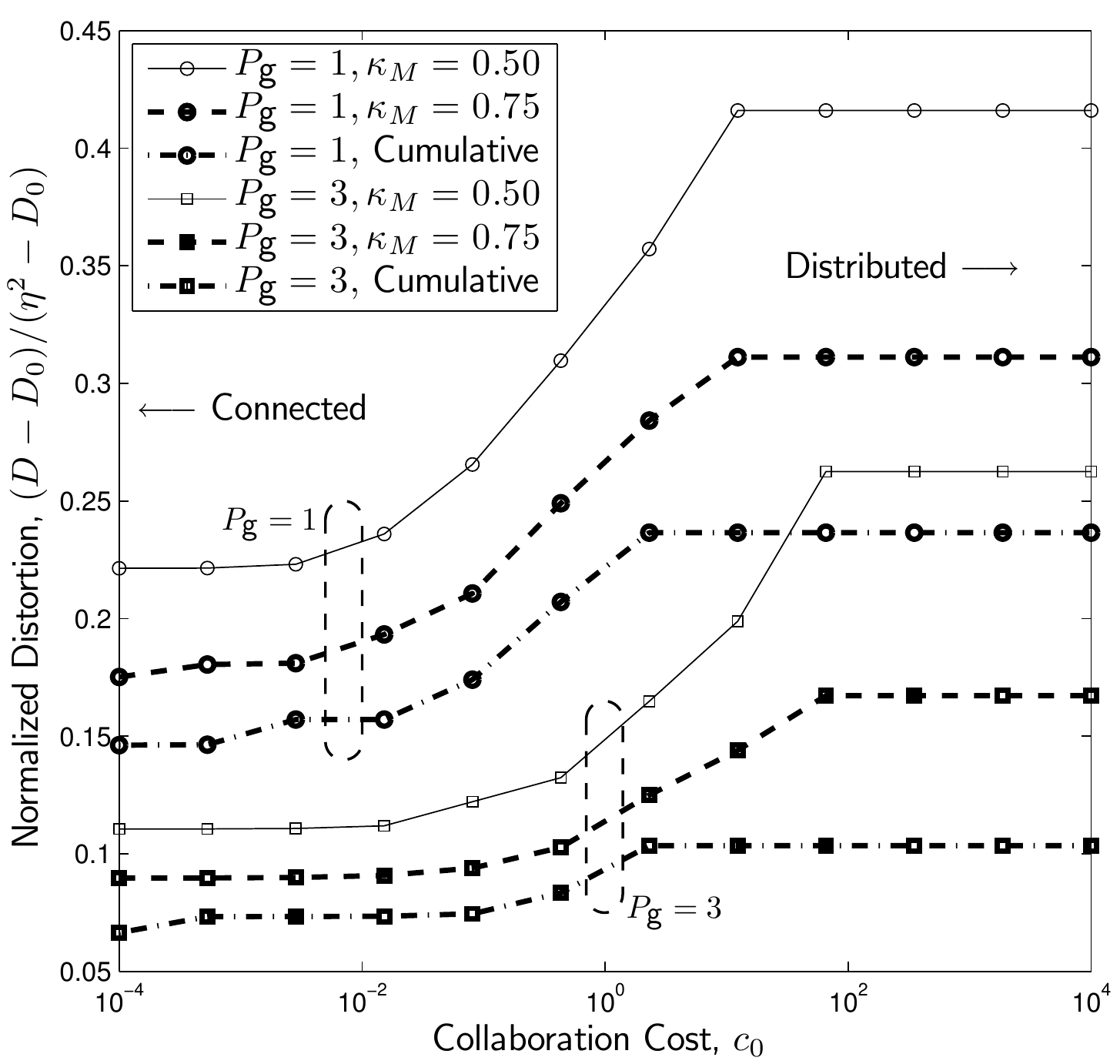}
  \caption{Estimation with finite collaboration cost - an example with 10 sensor nodes. }
  \label{fig:cost}
\end{figure}

To demonstrate the efficacy of collaboration in finite-cost scenarios, we consider a random geometric graph of $M=N=10$ nodes. As in the previous examples, we consider a homogeneous network with the following parameters $\eta^2=0.5$, $\sigma^2=1$, $\rho=0$ (independent noise), $g_0=h_0=1$ and $\alpha_\textsf{g}=\alpha_\textsf{h}=0.9$. The collaboration cost of link $m\rightarrow n$ is assumed to increase quadratically with the distance between nodes $m$ and $n$. This is because the gain of a wireless channel is often inversely proportional (upto a constant exponent) to the distance between a source and a receiver \cite{Gold05}. Consequently, to maintain a reliable communication link, the transmission power has to be scaled up accordingly. In particular, we assume
\begin{align}
C_{m,n}=c_0 d_{m,n}^2,
\end{align}
where $d_{m,n}$ denotes the distance between nodes $m$ and $n$ and $c_0$ is a constant of proportionality. For our numerical simulations, we consider a wide range of $c_0$, specifically $c_0\in [10^{-4},10^4]$, to depict the effect of collaboration cost on the distortion performance. A lower collaboration cost in effect allows the network to collaborate more and thereby reduces the distortion. We consider two magnitudes of total operating power (namely, $P_\textsf{g}=1$ and $3$) for both the individual-constraint and cumulative-constraint cases. For the individual-constraint case, we consider two skewness conditions for the power-constraint, namely $\kappa_M=0.5$ and $\kappa_M=0.75$. The corresponding distortion curves are shown in Figure \ref{fig:cost}. For very low values of the $c_0$, the distortion converges to that in a fully connected  network. Similarly, for very high values of the $c_0$, no links are selected for collaboration, and the network operates in a distributed manner. Since in our example, the network is homogeneous, equal power allocation among nodes is also the optimum power allocation for a cumulative-power-constrained problem. Consequently, the performance of the cumulative-constraint problem (dash-dotted lines) is always better than that with individual power constraints (bold and dashed lines). In conclusion, Figure \ref{fig:cost} shows that in a homogeneous network, the estimation performance improves with higher operating power, less skewed power constraints and lower collaboration cost among sensors, as expected.

\section{Conclusion} \label{sec:conc}
In this paper, we addressed the problem of collaborative estimation in a sensor network where sensors communicate with the FC using a coherent MAC channel. For the scenario when the collaborative topology is fixed and collaboration is cost-free, we obtained the optimal (cumulative) power-distortion tradeoff in closed-form by solving a QCQP problem. With individual power constraints, we have shown that the semidefinite relaxation technique can be used to obtain precisely optimal numerical results. Several special cases are presented as examples to highlight the problem conditions for which collaboration is particularly effective.  Through the use of both theoretical and numerical results, we established that collaboration helps to substantially reduce the distortion of the estimated parameter at the fusion center, specially in low local-SNR scenario. As future work, we wish to explore the collaborative estimation problem when the parameter to be estimated is a vector with correlated elements. The scenario when collaboration is erroneous, as mentioned earlier, is also important.


\appendices

\section{Proof of Proposition \ref{result:conn:lbl}} \label{app:conn}
Since $\bo w=\textsf{vec}(\bo W)$ for a connected topology, we obtain $\bo \Omega_\textsf{P}=\bo E_\textsf{x} \otimes \bo I$, $\bo G=\bo I \otimes \bo g$ and 
\begin{align}
\bo \Omega_\textsf{JD}&=\left( \bo E_\textsf{x} \otimes \bo E_\textsf{g} \right) -\eta^2 \bo G \bo h \bo h^T \bo G^T \nonumber \\
&=\left( \bo E_\textsf{x} \otimes \bo \Sigma_\textsf{g} \right) +\bo G \widetilde{\bo \Sigma} \bo G^T.
\end{align}
Substituting the appropriate values in \eqref{Jopt}, we have (assuming all the inverses exist)
\begin{align}
J_\textsf{opt}(P)&\stackrel{\text{(a)}}{=} \bo h^T \bo G^T \left( \bo G \tbo \Sigma \bo G^T + \tbo \Omega_\textsf{P} \right)^{-1} \bo G \bo h \nonumber \\
&\stackrel{\text{(b)}}{=} \bo h^T \left( \tbo \Sigma + \tbo \Gamma_\textsf{P} \right)^{-1} \bo h \nonumber \\
&\stackrel{\text{(c)}}{=} \bo h^T \left( \left(1+\frac{1}{\mathcal G} \right) \tbo \Sigma + \frac{1}{\mathcal G} \eta^2 \bo h\bo h^T \right)^{-1} \bo h \nonumber \\
&\stackrel{\text{(d)}}{=} \widetilde{J} \left[1+\frac{1+\eta^2  \widetilde{J}}{\mathcal G}\right]^{-1}, \label{Jopt:conn:app}
\end{align}
where step (a) follows by defining $\tbo \Omega_\textsf{P} \triangleq \bo E_\textsf{x} \otimes \tbo \Sigma_\textsf{g}$ where $\tbo \Sigma_\textsf{g}$ is already defined in \eqref{Jopt:conn}. Step (b) follows from arguments similar to those used in \eqref{Jopt:CSI:gev} and by defining  (and subsequently simplifying) $\tbo \Gamma_\textsf{P}$ as
\begin{align} 
\tbo \Gamma_\textsf{P} &\triangleq \left( \bo G^T \tbo \Omega_\textsf{P}^{-1} \bo G  \right)^{-1} \nonumber \\
&=\left( \left( \bo I \otimes \bo g^T \right) \left( \bo E_\textsf{x}^{-1} \otimes \tbo \Sigma_\textsf{g}^{-1} \right) \left( \bo I \otimes \bo g \right)  \right)^{-1} \nonumber \\
&= \frac{\bo E_\textsf{x}}{\mathcal G}, \; \mbox{ where } \mathcal G \triangleq \bo g^T \tbo \Sigma_\textsf{g}^{-1} \bo g,
\end{align}
step (c) follows from the fact that $\bo E_\textsf{x}=\tbo \Sigma+\eta^2 \bo h \bo h^T$, and step (d) follows from the definition $\widetilde J \triangleq \bo h^T \tbo \Sigma^{-1} \bo h$ and following identities involving rank-$1$ updated matrix inverses. For any scalars $\alpha \neq 0$ and $\beta$, vector $\bo p$ and invertible matrix $\bo Q$,
\begin{align}
&\left( \alpha \bo Q+\beta \bo p\bo p^T \right)^{-1}=\frac{\bo Q^{-1}}{\alpha}-\frac{\beta \bo Q^{-1}\bo p \bo p^T \bo Q^{-1}}{\alpha(\alpha+\beta Q_p)},\; \mbox{and} \label{identity:woodbury:mat} \\
&\bo p^T \left( \alpha \bo Q+\beta \bo p \bo p^T \right)^{-1}\bo p=\frac{Q_p}{\alpha+\beta Q_p},\; Q_p \triangleq \bo p^T \bo Q^{-1} \bo p.
\end{align}

From equations \eqref{Wopt} and discussion in Example 1, the optimal weights are,
\begin{align}
\bo w_\textsf{opt}&\propto \tbo \Omega_\textsf{P}^{-1} \bo G \tbo \Sigma \left( \tbo \Sigma+\tbo \Gamma_\textsf{P} \right)^{-1}\bo h, \quad \left(\mbox{from  \eqref{Wopt}} \right) \nonumber \\
&\propto \tbo \Omega_\textsf{P}^{-1} \bo G \bo h \nonumber \\
&=\left( \bo E_\textsf{x}^{-1} \otimes \tbo \Sigma_\textsf{g}^{-1} \right) \left( \bo h \otimes \bo g \right) \nonumber \\
&=\left( \bo E_\textsf{x}^{-1} \bo h \right) \otimes \left( \tbo \Sigma_\textsf{g}^{-1} \bo g \right)
\end{align}
which implies that $\bo W_{\text{opt}} \propto \tbo \Sigma_\textsf{g}^{-1} \bo g  \bo h^T \bo E_\textsf{x}^{-1}$. Step (a) follows the fact that $\left( \tbo \Sigma+\tbo \Gamma_\textsf{P} \right)^{-1}\bo h \propto \tbo \Sigma^{-1} \bo h$ (see \eqref{identity:woodbury:mat}). 

From Corollary $2.3.5$ of \cite{Gastpar07chapter}, the sum-rate required to encode a single-dimensional real-valued Gaussian source with variance $\eta^2$, observed through the vector $\bo h$ and Gaussian observation noise with covariance $\bo \Sigma$, in such a way that reconstruction incurs an average distortion of at most $D$, satisfies
\begin{align} 
R_{\text{tot}}\ge \frac{1}{2}\log \frac{\lambda}{D-D_0}, \mbox{ where } \lambda=\frac{\eta^4 J_0}{1+\eta^2 J_0}.
\end{align}
Since, for a fixed sum-power $P$, the sum-rate has to be lesser that the (centralized) capacity of the coherent MAC channel, i.e., $R_{\text{tot}}\le C$, where $C=\frac{1}{2}\log (1+\|\bo g \|^2 P_\xi)$, we obtain
\begin{align} 
1+\|\bo g \|^2 P_\xi\ge  \frac{\eta^4 J_0}{(D-D_0)(1+\eta^2 J_0)}.
\end{align}
Replacing $D$ by $J$ (recall, $J=\frac{1}{D}-\frac{1}{\eta^2}$) and after some algebra, we obtain,
\begin{align} 
J\le J_0 \left[1+\frac{1+\eta^2  J_0}{\| \bo g \|^2 P_\xi}\right]^{-1}.
\end{align}
But the right hand side is precisely the distortion achieved by a connected network (see   \eqref{Jopt:conn} and note that $\mathcal G=\| \bo g \|^2 P_\xi$ for $\bo \Sigma_\textsf{g}=0$). This establishes the information theoretic optimality.

\section{} \label{app:inequality}
\begin{inequality}[An inequality] \label{inequality:lbl}
For any $N$-dimensional vector $\bo p$ and $N\times N$ symmetric positive definite matrices $\bo A$ and $\bo B$,
\begin{align}
\frac{1}{\bo p^T \left(\bo A+\bo B\right)^{-1} \bo p} \ge \frac{1}{\bo p^T \bo A^{-1} \bo p}+\frac{1}{\bo p^T \bo B^{-1} \bo p}. \label{posdef:inequality}
\end{align}
\end{inequality}
\begin{IEEEproof} Since $\bo A,\bo B \in \mathcal S^{++}$, $\bo A^{-\frac{1}{2}}\bo B\bo A^{-\frac{1}{2}}\in \mathcal S^{++}$. Define by $\bo U$ and $\bo \Lambda$ the following eigendecomposition $\bo A^{-\frac{1}{2}}\bo B\bo A^{-\frac{1}{2}}=\bo U\bo \Lambda \bo U^T$. Hence $\lambda_n>0,\forall n$. Define $\bo q=\bo U^T\bo A^{-\frac{1}{2}} \bo h$. Note that 
\begin{align}\begin{split}
\bo q^T \bo q&=\bo p^T \bo A^{-1} \bo p,\\ 
\bo q^T \bo \Lambda^{-1}\bo q&=\bo p^T \bo B^{-1} \bo p, \mbox{ and }\\
\bo q^T (\bo I+\bo \Lambda)^{-1}\bo q&=\bo p^T \left(\bo A+\bo B\right)^{-1} \bo p.
\end{split}\end{align}
Hence, to prove \eqref{posdef:inequality}, it suffices to show that
\begin{align*}
\frac{1}{\sum_{n=1}^N\frac{q_n^2}{1+\lambda_n}} \ge \frac{1}{\sum_{n=1}^N q_n^2}+\frac{1}{\sum_{n=1}^N\frac{q_n^2}{\lambda_n}},
\end{align*}
or equivalently, with $a_n\triangleq \frac{1}{1+\lambda_n}$ and $b_n\triangleq\frac{1+\lambda_n}{\lambda_n}$,
\begin{align}
\sum_{n=1}^N q_n^2 \sum_{n=1}^N q_n^2 a_n b_n &\ge  \sum_{n=1}^N q_n^2 a_n \sum_{n=1}^N q_n^2 b_n. \label{chebyshev}
\end{align}
Since $\lambda_n>0,\forall n$, both $a_n$ and $b_n$ are decreasing functions of $\lambda_n$. Hence inequality \eqref{chebyshev} follows from the Chebyshev's (sum) inequality (page 240, Equation 1.4, \cite{Mitrinovic93}). Equality holds if and only if, for all indices $k$ for which $q_k\neq 0$ (denote such a set by $\text{\textsf{ixnz}}(\bo q)$), the eigenvalues are similar. That is, iff $\lambda_k=\lambda$, $\forall k\in\text{\textsf{ixnz}}(\bo q)$.
\end{IEEEproof}

\section{Proof of Proposition \ref{result:cyc:lbl}} \label{app:cycle}
We would use Equation \eqref{Jopt}. To start with, we note that $\bo G \bo h$ is a multiple of $\bo 1$ and both matrices $\bo \Omega_\textsf{JD}$ and $\bo \Omega_\textsf{P}$ has an eigenvector as $\bo 1$, where $\bo 1$ has dimension $L=M K$. In particular, careful inspection of \eqref{def:OPOJ} (the elements of matrices $\bo E_\textsf{g}$ and $\bo E_x$ take two distinct values, diagonal and otherwise) yields
\begin{align}
\begin{split}
&\bo G \bo h=g_0 h_0 \sqrt{\alpha_\textsf{g} \alpha_\textsf{h}} \bo 1, \\
&\bo \Omega_\textsf{P}  \bo 1=\sigma_\textsf{x}^2 \left( 1+(K-1) \alpha_\textsf{x} \right) \bo 1, \mbox{ and} \\
&\bo \Omega_\textsf{JD} \bo 1=\left[ \sigma_\textsf{x}^2 g_0^2 \left\{ 1+ (K-1)(\alpha_\textsf{x}+\alpha_\textsf{g}) \right. \right.\\
&\;\;\left.\left.+(M K-2K+1)\alpha_\textsf{x}\alpha_\textsf{g}\right\}- \eta^2 g_0^2 h_0^2 \alpha_\textsf{g} \alpha_\textsf{h} M K \right] \bo 1,
\end{split}
\end{align} 
where $\sigma_\textsf{x}^2=\sigma^2(1+\gamma)$, $\alpha_\textsf{x}=\frac{\rho +\gamma \alpha_\textsf{h}}{1+\gamma}$ and $\gamma=\frac{\eta^2 h_0^2}{\sigma^2}$. Based on the above equations, define scalars $\phi$, $\mu$, $\nu$ be such that $\bo G \bo h \bo h^T \bo G^T \bo 1=\phi \bo 1$, $\bo \Omega_\textsf{P} \bo 1=\mu\bo 1$ and $\bo \Omega_\textsf{JD} \bo 1=(\nu-\eta^2 \phi) \bo 1$, in particular,
\begin{align}
\begin{split}
&\phi=g_0^2 h_0^2 \alpha_\textsf{g} \alpha_\textsf{h} M K, \\
&\mu=\sigma_\textsf{x}^2 \left( 1+(K-1) \alpha_\textsf{x} \right), \mbox{ and} \\
&\nu=\sigma_\textsf{x}^2 g_0^2 \left\{ 1+ (K-1)(\alpha_\textsf{x}+\alpha_\textsf{g})+(M K-2K+1)\alpha_\textsf{x}\alpha_\textsf{g}\right\},
\end{split} \label{def:phi:mu:nu}
\end{align} 
From \eqref{Jopt}, we therefore obtain 
\begin{align}
J_\textsf{opt}(P)=\frac{\phi}{\nu+\frac{\mu}{P_\xi}-\eta^2 \phi }, \label{Jopt:cyc:app}
\end{align} 
which when simplified further leads to \eqref{Jopt:cyc}. Since $\bo 1$ is the corresponding eigenvector, we also have $\bo w_\textsf{opt}\propto \bo 1$, i.e., the sensors just average all the observations.

\section{Proof of Proposition \ref{result:SDR:lbl}} \label{app:SDR}
Our goal is to show that for any feasible $J<J_\textsf{opt}$, $\mathcal X^\mathsf{R}(J)$ contains a rank-1 matrix. Specifically, we will show that 
\begin{align}
\widetilde{\bo X}\triangleq \arg \max_{\bo X \in \mathcal X^\mathsf{R}(J)} \tr\left[\bo \Omega_\textsf{P} \bo X \right],
\end{align}
which is the global optimizer to the (convex) semi-definite optimization problem
\begin{equation}
\begin{aligned}
& \underset{\bo X}{\text{minimize}} & & \tr\left[\bo \Omega_\textsf{P} \bo X \right] \\
& \text{subject to} & & \tr \left[\left( J\bo \Omega_\textsf{JD}-\bo \Omega_\textsf{JN}\right) \bo X\right]+J \xi^2 \le 0, \\
&                         & & \tr \left[\bo \Omega_{\textsf P,m} \bo X\right] \le P^\textsf{C}_m, \, m=1,\ldots,M, \\
&                         & & -\bo X \preceq 0,
\end{aligned} \label{prob:ideal:SDP:OP:indiv}
\end{equation}
is rank-1. The Lagrangian of \eqref{prob:ideal:SDP:OP:indiv} is given by,
\begin{align}
\begin{split}
\mathcal L(\bo X,\alpha,\bo \beta,\bo Z)&=\tr\left[ \bo \Omega_\textsf{P} \bo X \right]\\
&+\alpha\left( \tr \left[\left( J\bo \Omega_\textsf{JD}-\bo \Omega_\textsf{JN}\right) \bo X\right]+J \xi^2 \right)  \\
&+ \sum_{m=1}^M \beta_m \left(\tr \left[\bo \Omega_{\textsf P,m} \bo X\right] - P^\textsf{C}_m\right) \\
&-\tr\left[\bo X \bo Z \right],
\end{split}
\end{align}
with the dual problem being,
\begin{equation}
\begin{aligned}
& \underset{\alpha,\, \bo \beta}{\text{maximize}} & & \alpha J \xi^2 -\sum_{m=1}^M \beta_m P_m^\textsf{C} \\
& \text{subject to} & & \widetilde{\bo Z} \triangleq \bo \Omega_\textsf{P}+\alpha \left( J\bo \Omega_\textsf{JD}-\bo \Omega_\textsf{JN}\right)+\sum_{m=1}^M \beta_m \bo \Omega_{\textsf P,m} \succeq 0, \\
&                         & & \alpha \ge 0, \; \bo \beta \ge 0.
\end{aligned} \label{def:Z:tilde}
\end{equation}
Also, we have the following complementary conditions (let \emph{tilde} denote respective values at optimality),
\begin{subequations}
\begin{align}
&\widetilde{\alpha} \left( \tr \left[\left( J\bo \Omega_\textsf{JD}-\bo \Omega_\textsf{JN}\right) \widetilde{\bo X} \right]+J \xi^2 \right) =0, \\
&\widetilde{\beta}_m \left(\tr \left[\bo \Omega_{\textsf P,m} \widetilde{\bo X} \right] - P^\textsf{C}_m\right) =0,\; m=1,\ldots,M, \\
&\tr\left[\widetilde{\bo X} \widetilde{\bo Z} \right] = 0. \label{comp:cond:3}
\end{align}
\end{subequations}
Without loss of generality, let $\widetilde{\bo X}=\widetilde{\bo Y} \widetilde{\bo Y}^T$ (such a decomposition is possible since $\widetilde{\bo X}$ is symmetric positive semidefinite). Also denote the columns of $\widetilde{\bo Y}$ as $\widetilde{\bo w}_l\in \mathbb R^L$ for $l=1,2,\ldots, L$, so that $\widetilde{\bo Y}=\left[\widetilde{\bo w}_1, \widetilde{\bo w}_2,\ldots,\widetilde{\bo w}_L\right]$. From \eqref{comp:cond:3}, we have
\begin{align}
\tr\left[\widetilde{\bo X} \widetilde{\bo Z} \right]=\sum_{l=1}^L \widetilde{\bo w}_l^T \widetilde{\bo Z} \widetilde{\bo w}_l=0,
\end{align}
which coupled with the fact that $\widetilde{\bo Z}\succeq 0$ implies that 
\begin{align}
\widetilde{\bo Z} \widetilde{\bo w}_l=0,\; \mbox{for all}\; l=1,\ldots,L.
\end{align}
Thus, from definition of $\widetilde{\bo Z}$ in \eqref{def:Z:tilde}, we conclude that $\widetilde{\bo w}_l$-s are the generalized eigenvectors that satisfy
\begin{align} 
\begin{split}
&\left(\widetilde{\bo \Omega} + \widetilde{\alpha} \left(J \bo \Omega_\textsf{JD}-\bo t \bo t^T\right)\right)\widetilde{\bo w}_l =\bo 0_L, \; \forall \, l,\; \mbox{with}\\
&\quad\quad\quad\quad \widetilde{\bo \Omega} \triangleq \bo \Omega_\textsf{P}+\sum_{m=1}^M \widetilde{\beta}_m \bo \Omega_{\textsf P,m},\\
&\quad\quad\quad\quad\; \bo t\triangleq\bo G \bo h \; (\mbox{note } \bo \Omega_\textsf{JN} \mbox{ in } \eqref{def:OPOJ}). \label{eigen:equation}
\end{split}
\end{align}
Hence it follows that for all $l$,
\begin{align}
 &\left(\left(\widetilde{\bo \Omega}+\widetilde{\alpha}  J \bo \Omega_\textsf{JD}\right)-\widetilde{\alpha} \bo t \bo t^T \right) \widetilde{\bo w}_l =\bo 0_L, \nonumber \\
\Leftrightarrow\quad &\left(\bo I_L- \left(\widetilde{\bo \Omega}+\widetilde{\alpha}  J \bo \Omega_\textsf{JD}\right)^{-1} \widetilde{\alpha} \bo t \bo t^T \right) \widetilde{\bo w}_l=\bo 0_L,  \label{app:relax:cond1} \\
\Rightarrow\quad &\widetilde{\bo w}_l \propto \left(\widetilde{\bo \Omega}+\widetilde{\alpha}  J \bo \Omega_\textsf{JD}\right)^{-1} \bo t, \label{wl}
\end{align}
where \eqref{app:relax:cond1} is because $\bo \Omega_\textsf{P}$ is positive definite and hence the matrix $\left(\widetilde{\bo \Omega}+\widetilde{\alpha}  J \bo \Omega_\textsf{JD}\right)$ is also positive definite and invertible, and  \eqref{wl} follows from the fact that both $\widetilde\alpha$ and $\bo t^T \widetilde{\bo w}_l$ has to be non-zero to satisfy \eqref{eigen:equation}. We thus conclude that
\begin{align}
\begin{split}
&\widetilde{\bo w}_l \; \mbox{is unique upto its norm}, \; \forall\, l, \\
&\Rightarrow\; \widetilde{\bo X}\; \mbox{is a rank-1 matrix},
\end{split}
\end{align}
thereby establishing Proposition \ref{result:SDR:lbl}.

\section{Proof of Proposition \ref{result:indiv:conn:lbl}} \label{app:indiv:conn}
Problem \eqref{prob:ideal:indiv:full} is equivalent to
\begin{equation}
\begin{aligned}
& \underset{\bo W}{\text{maximize}} & & \mathcal J_{\bo W}=\frac{\left( \bo g^T \bo W \bo h \right)^2}{\tr\left[\bo E_\textsf{g} \bo W \bo E_\textsf{x} \bo W^T \right] +\xi^2}, \\
&                            \text{subject to} & & \left[ \bo W \bo E_\textsf{x} \bo W^T \right]_{m,m} \le P^\textsf{C}_m, \, m=1,\ldots,M,
\end{aligned} \label{prob:ideal:indiv:full:W}
\end{equation}
in the sense that $J_{\bo W}$ and $\mathcal J_{\bo W}$ are monotonically related through
$J_{\bo W}=\frac{\mathcal J_{\bo W}}{1-\eta^2 \mathcal J_{\bo W}}$ and $\bo W_\textsf{opt}$ is the same for both problems. This is further equivalent to 
\begin{equation}
\begin{aligned}
& \underset{\bo V}{\text{maximize}} & & \mathcal J_{\bo V}=\frac{\left( \bo g^T \bo V \bo h_\textsf{x} \right)^2}{\tr\left[\bo E_\textsf{g} \bo V \bo V^T \right] +\xi^2}, \\
&                            \text{subject to} & & \left\| \bo v_m \right\|^2 \le P^\textsf{C}_m, \, m=1,\ldots,M,
\end{aligned} \label{prob:ideal:indiv:full:V}
\end{equation}
by defining $\bo V$, $\bo v_m$ and $\bo h_\textsf{x}$ such that
\begin{align}
\bo V\triangleq \bo W \bo E_\textsf{x}^{\frac{1}{2}}=\begin{bmatrix} \bo v_1^T \\ \vdots \\ \bo v_M^T \end{bmatrix}, \mbox{ and } \bo h_\textsf{x} &\triangleq\bo E_\textsf{x}^{-\frac{1}{2}} \bo h.
\end{align}
With a goal to reduce the number of optimization variables from $M N$ to $M$, we define the matrix transformation $\bo V \rightarrow \bo V_\textsf{x}$ as one that retains the norm of its individual row vectors but otherwise aligns the rows to $\bo h_\textsf{x}^T$, i.e.,
\begin{align}
\begin{split}
\bo V_\textsf{x}&\triangleq \bo t \frac{\bo h_\textsf{x}^T}{\left\| \bo h_\textsf{x} \right\|}, \; t_m\triangleq \left\| \bo v_m \right\|, \mbox{ so that} \\
\mathcal J_{\bo V_\textsf{x}}&=\left\| \bo h_\textsf{x} \right\|^2 \frac{ \left( \bo g^T \bo t \right)^2}{ \bo t^T \bo E_\textsf{g} \bo t +\xi^2}.
\end{split} \label{def:VxJVx}
\end{align}

We would need the following result to proceed further.
\begin{app:indiv:conn} \label{app:indiv:conn:lbl} 
When $\bo \Sigma_\textsf{g}$ is diagonal,
\begin{align}
\mathcal J_{\bo V}  \le \mathcal J_{\bo V_\textsf{x}}, \mbox{ for any } \bo V.
\end{align}
\end{app:indiv:conn}
\begin{IEEEproof}    
To prove Lemma \eqref{app:indiv:conn:lbl}, we will show that 
\begin{align}
\frac{\mathcal J_{\bo V}}{\left\| \bo h_\textsf{x} \right\|^2} 
\stackrel{\text{(a)}}{\le} \frac{\left\| \bo g^T \bo V \right\|^2}{\tr\left[\bo E_\textsf{g} \bo V \bo V^T \right] +\xi^2} 
\stackrel{\text{(b)}}{\le} \frac{ \left( \bo g^T \bo t \right)^2}{ \bo t^T \bo E_\textsf{g} \bo t +\xi^2}  
=\frac{\mathcal J_{\bo V_\textsf{x}}}{\left\| \bo h_\textsf{x}\right\|^2},
\end{align}
where (a) follows from definition of $\mathcal J_{\bo V}$ in \eqref{prob:ideal:indiv:full} and Cauchy-Schwartz inequality implying $\left( \bo g^T \bo V \bo h_\textsf{x} \right)^2\le || \bo g^T \bo V ||^2 || \bo h_\textsf{x}||^2$, and the last equality is due to definition of $\mathcal J_{\bo V_\textsf{x}}$ in \eqref{def:VxJVx}. Hence it remains to prove (b), which can be established by showing
\begin{subequations}
\begin{align}
\tr\left[ \left( \left( \bo t^T \bo \Sigma_\textsf{g} \bo t \right) \bo g \bo g^T - \left( \bo g^T \bo t \right)^2 \bo \Sigma_\textsf{g}  \right) \bo V \bo V^T \right] &\le 0, \label{app:indiv:conn:cond1} \\
\mbox{and }\; \xi^2\left(\left\| \bo g^T \bo V \right\|^2- \left( \bo g^T \bo t \right)^2 \right) &\le 0. \label{app:indiv:conn:cond2}
\end{align}
\end{subequations}
Define $\bar{\bo g}$, $\bar{\bo V}$ and its aligned equivalent $\bar{\bo V}_\textsf{x}$ as
\begin{align}
\begin{split}
\bar{\bo g} &\triangleq \bo \Sigma_\textsf{g}^{-\frac{1}{2}} \bo g, \;
\bar{\bo V} \triangleq
\bo \Sigma_\textsf{g}^{\frac{1}{2}} \bo V, \mbox{ so that} \\
\bar{\bo V}_\textsf{x}&=\overline{\bo t} \frac{\bo h_\textsf{x}^T}{\left\| \bo h_\textsf{x} \right\|}, \mbox{ where } \bar t_m=\left\| \bar{\bo v}_m \right\|, \mbox{ or}\\
\bar{\bo t}&=\bo \Sigma_\textsf{g}^{\frac{1}{2}} \bo t \mbox{ (since } \bo \Sigma_\textsf{g} \mbox{ is diagonal)}.
\end{split} 
\end{align}
Note that $\tr\left[\bar{\bo V} \bar{\bo V}^T \right]=\left\| \bar{\bo t} \right\|^2$. Condition \eqref{app:indiv:conn:cond1} is therefore equivalent to showing 
\begin{align}
 \tr\left[ \left( \left\| \bar{\bo t} \right\|^2 \bar{\bo g} \bar{\bo g}^T - \left( \bar{\bo g}^T \bar{\bo t} \right)^2  \right) \bar{\bo V} \bar{\bo V}^T \right] &\le 0, \nonumber \\
 \mbox{or equivalently } \quad \left\| \bar{\bo t} \right\|^2 \left(\left\| \bar{\bo g}^T \bar{\bo V} \right\|^2 - \left( \bar{\bo g}^T \bar{\bo t} \right)^2 \right) &\le 0,
\end{align}
which is similar to condition \eqref{app:indiv:conn:cond2}. Thus it remains to establish \eqref{app:indiv:conn:cond2}, which is true because
\begin{align}
\left\| \bo g^T \bo V \right\|^2 \stackrel{\text{(a)}}{=} \left\| \sum_{m=1}^M g_m \bo v_m \right\|^2  \stackrel{\text{(b)}}{\le} \left\| \sum_{m=1}^M g_m \left\| \bo v_m \right\| \right\|^2 \stackrel{\text{(c)}}{=} \left( \bo g^T \bo t \right)^2,
\end{align}
where (a) and (c) are due to definitions of $\bo v_m$  and $\bo t$ respectively, and (b) is due to Cauchy-Schwartz inequality, $\left\| \bo v_m^T \bo v_n \right\| \le \left\| \bo v_m \right\| \left\| \bo v_n \right\|$ for all $1 \le m,n \le M$. This completes the proof.
\end{IEEEproof}

Note that $\mathcal J_{\bo V_\textsf{x}}$ is a function of the vector $\bo t$, whose elements are non-negative (since $t_m$ is a norm). Therefore problem \eqref{prob:ideal:indiv:full:V}, in conjunction with Lemma \ref{app:indiv:conn:lbl}, is equivalent to 
\begin{equation}
\begin{aligned}
& \underset{\bo t}{\text{maximize}} & & \mathcal J_{\bo V_\textsf{x}}\left( \bo t \right), \\
&                            \text{subject to} & & t_m^2 \le P^\textsf{C}_m, \, m=1,\ldots,M,
\end{aligned} \label{prob:ideal:indiv:full:JVxt}
\end{equation}
through the relations $\bo V_\textsf{opt}=\bo t_\textsf{opt} \frac{\bo h_\textsf{x}^T}{\| \bo h_\textsf{x} \|}$ and $\mathcal J_{\bo V_\textsf{opt}}=\mathcal J_{\bo V_\textsf{x}}\left( \bo t_\textsf{opt} \right)$. Problem \eqref{prob:ideal:indiv:full:JVxt} is further equivalent to
\begin{equation}
\begin{aligned}
& \underset{\bo t}{\text{maximize}} & & F_{\bo t}=\frac{\left( \bo g^T \bo t \right)^2}{\bo t^T \bo \Sigma_\textsf{g} \bo t+\xi^2}, \\
&                            \text{subject to} & & t_m^2 \le P^\textsf{C}_m, \, m=1,\ldots,M,
\end{aligned} \label{prob:ideal:indiv:full:Ft}
\end{equation}
through the following relations between variables
$\mathcal J_{\bo V_\textsf{x}}\left( \bo t \right)=\left\| \bo h_\textsf{x} \right\|^2 \frac{F_{\bo t}}{1+F_{\bo t}}$, 
which proves Proposition \ref{result:indiv:conn:lbl}

\section{Proof of Lemma \ref{result:skewness:lbl}} \label{app:skewness}
We start by noting that for $\varrho \in (0,1)$
\begin{align}
\kappa(\varrho )&= \left( \frac{1-\varrho^{\frac{M}{2}} }{1-\varrho^{\frac{1}{2}} } \right)^2 \frac{1-\varrho}{1-\varrho^M} =\frac{1-\varrho^{\frac{M}{2}} }{1+\varrho^{\frac{M}{2}} } \frac{ 1+\varrho^{\frac{1}{2}} }{1-\varrho^{\frac{1}{2}} }. \label{app:kappa:rho}
\end{align}
We would show that $\frac{\ud \kappa(\varrho)}{\ud \varrho}>0$ for $\varrho \in (0,1)$. From \eqref{app:kappa:rho},
\begin{align}
\frac{\ud \kappa(\varrho)}{\ud \varrho}=\frac{ \varrho^{-\frac{1}{2}} - M\varrho^{\frac{M}{2}-1} - \varrho^{M-\frac{1}{2}} + M\varrho^{\frac{M}{2}} }{\left( 1+\varrho^{\frac{M}{2}} \right)^2 \left( 1-\varrho^{\frac{1}{2}} \right)^2},
\end{align}
the numerator of which can be rearranged as  
\begin{align}
&\varrho^{-\frac{1}{2}} \left( 1-\varrho^M \right)-M \varrho^{\frac{M}{2}-1} \left( 1-\varrho \right) \nonumber \\
&=\varrho^{-\frac{1}{2}} \left( 1-\varrho \right) \left(1+\varrho+\cdots+\varrho^{M-1}-M\varrho^{\frac{M-1}{2}} \right) \nonumber \\
&=\varrho^{-\frac{1}{2}} \left( 1-\varrho \right) \sum_{m=1}^{\lfloor \frac{M}{2} \rfloor} \left(\varrho^{\frac{m-1}{2}} -\varrho^{\frac{M-m}{2}} \right)^2,
\end{align}
which is evidently a positive quantity. This completes the proof.

\section{Proof of Proposition \ref{result:CG:indiv:lbl}} \label{app:CG:indiv}
We start with the cumulative-constraint case, for which we will use the results from Example 3. From Equation \eqref{Jopt:cyc:app} and corresponding to the distributed ($K=1$) and connected cases ($K=M$), we denote the constants $\{\mu, \nu\}$ with subscripts $1,2$, as $\{\mu_1, \nu_1\}$ and $\{\mu_2, \nu_2\}$ respectively, i.e.,
\begin{align}
\begin{split}
J_\textsf{opt}^\textsf{dist}(P)&=\frac{\phi}{\nu_1+\frac{\mu_1}{P_\xi}-\eta^2 \phi}, \mbox{ and } \\
J_\textsf{opt}^\textsf{conn}(P)&=\frac{\phi M}{\nu_2+\frac{\mu_2}{P_\xi}-\eta^2 \phi M},
\end{split} \label{Jopt:homo:dist:conn}
\end{align}
where $\phi=g_0^2 h_0^2 \alpha_\textsf{g} \alpha_\textsf{h} M$, $\nu_1=\sigma_\textsf{x}^2 g_0^2 \left(1+(M-1)\alpha_\textsf{g} \alpha_\textsf{x}\right)$,  $\mu_1=\sigma_\textsf{x}^2$, $\nu_2= \sigma_\textsf{x}^2 g_0^2 \left(1+(M-1)\alpha_\textsf{g}\right) \left(1+(M-1)\alpha_\textsf{x}\right)$ and $\mu_2=\sigma_\textsf{x}^2 \left(1+(M-1)\alpha_\textsf{x}\right)$. Applying the corresponding distortion terms (note $D=\left(\frac{1}{\eta^2}+J \right)^{-1}$) in \eqref{def:CG}, (note that $J_0=\frac{\phi M}{\nu_2-\eta^2 \phi M}$ therefore the denominator term of \eqref{def:CG} is $\eta^2-D_0= \frac{\eta^2 \phi}{\nu_2}$) the collaboration gain can be simplified as
\begin{align}
\textsf{CG}=\frac{\frac{M \mu_1-\mu_2}{M \mu_1}+P_\xi \frac{M \nu_1-\nu_2}{M \mu_1} }{\left( 1+ \frac{1}{P_\xi} \frac{ \mu_2 }{ \nu_2 } \right) \left( 1+ P_\xi \frac{ \nu_1 }{ \mu_1 } \right)}.  \label{CG:mu:nu}
\end{align}
Each of the fragments can be simplified further, $\frac{M \mu_1-\mu_2}{M \mu_1}=\frac{1}{M}(M-1)(1-\alpha_\textsf{x})$, $\frac{M \nu_1-\nu_2}{M \mu_1}=\frac{g_0^2}{M}(M-1)(1-\alpha_\textsf{g})(1-\alpha_\textsf{x})$, $\frac{ \mu_2 }{ \nu_2 }=\frac{1}{g_0^2 \left(1+(M-1)\alpha_\textsf{g}\right)}$ and $\frac{ \nu_1 }{ \mu_1 }=g_0^2 \left(1+(M-1)\alpha_\textsf{g} \alpha_\textsf{x}\right)$. Replacing these fragments in \eqref{CG:mu:nu}, defining $P_\textsf{g}=P_\xi g_0^2$ and dividing both numerator and denominator by $(1+P_\textsf{g})$  leads to Equation \eqref{CG:indiv} (with $\kappa$ replaced by $M$).

For the individual-constraint case, we would use Examples 5 (distributed) and 6 (connected) to compute the collaboration gain. First we show that all the constraints are active for both the distributed and connected cases if condition \eqref{CG:indiv:cond} is satisfied. If we apply Proposition \ref{soln:special:indiv:lbl} to a homogeneous problem with $a_1=\cdots=a_M$ and $b_1=\cdots=b_M$, the active constraint condition $\Phi_{M-1}d_M\ge 1$ simplifies to 
\begin{align}
\frac{\sum_{m=1}^{M-1} c_m^2+\frac{\xi^2}{b_m} }{ \sum_{m=1}^{M-1} c_m }\ge c_M. \label{cond:active:homo}
\end{align}
For the distributed case (Example 5), we refer to problem \eqref{Jopt:homo:indiv} to find that $b_m=\sigma_\textsf{x}^2 g_0^2 (1-\alpha_\textsf{g} \alpha_\textsf{x})$ and $c_m=\frac{\sqrt{P_m^\textsf{C}}}{\sigma_\textsf{x}}$, so that inequality \eqref{cond:active:homo} explicitly evaluates to condition \eqref{CG:indiv:cond}.  For the connected case (specialized version of Example 6 for homogeneous parameters), we refer to Equation \eqref{Jopt:indiv:conn} to note that $b_m=g_0^2 (1-\alpha_\textsf{g})$ and $c_m=\sqrt{P_m^\textsf{C}}$, so that inequality \eqref{cond:active:homo} evaluates to
\begin{align}
\frac{\sum_{m=1}^{M-1} P_m^\textsf{C}+\frac{\xi^2}{g_0^2 (1-\alpha_\textsf{g})} }{ \sum_{m=1}^{M-1} \sqrt{P_m^\textsf{C}} }\ge \sqrt{P_M^\textsf{C}},
\end{align}
which is clearly true if condition \eqref{CG:indiv:cond} holds (since $\alpha_\textsf{x} \in [0,1]$). 

As regards collaboration gain, we proceed as we did in the cumulative case. For the distributed (Equation \eqref{Jopt:homo:indiv:active}) and connected (Equation \eqref{Jopt:indiv:conn} with homogeneous parameters) cases, we can rearrange the terms of $J_\textsf{opt}(P)$ to express them in the form of \eqref{Jopt:homo:dist:conn}, where the various constants are now $\phi=g_0^2 h_0^2 \alpha_\textsf{g} \alpha_\textsf{h} \kappa$, $\nu_1=\sigma_\textsf{x}^2 g_0^2 \left(1+(\kappa-1)\alpha_\textsf{g} \alpha_\textsf{x}\right)$,  $\mu_1=\sigma_\textsf{x}^2$, $\nu_2= \sigma_\textsf{x}^2 g_0^2 \left(1+(\kappa-1)\alpha_\textsf{g}\right) \left(1+(M-1)\alpha_\textsf{x}\right)$ and $\mu_2=\sigma_\textsf{x}^2 \left(1+(M-1)\alpha_\textsf{x}\right)$. However, in this case $J_0$ is obtained not just by letting $P_\xi \rightarrow \infty$ (thereby letting $\mu$ vanish), but also by setting $\kappa=M$ in both $\phi$ and $\nu$, i.e., $J_0=\frac{\phi(\kappa=M) M}{\nu_2(\kappa=M)-\eta^2 \phi(\kappa=M) M}$. Therefore the denominator term of \eqref{def:CG} is $\eta^2-D_0= \frac{\eta^2 \phi(\kappa=M)}{\nu_2(\kappa=M)}=\frac{\eta^2 \phi}{\nu_2} \frac{M}{\kappa} \frac{\left(1+(\kappa-1)\alpha_\textsf{g}\right)}{\left(1+(M-1)\alpha_\textsf{g}\right)}$, which is just a scaled version of $\frac{\eta^2 \phi}{\nu_2}$. Adjusting \eqref{CG:mu:nu} for this scaling and rest of the derivation remaining similar, we obtain
\begin{align}
\textsf{CG}=\frac{\frac{\kappa}{M} \frac{\left(1+(M-1)\alpha_\textsf{g}\right)}{\left(1+(\kappa-1)\alpha_\textsf{g}\right)} \left( \frac{M \mu_1-\mu_2}{M \mu_1}+P_\xi \frac{M \nu_1-\nu_2}{M \mu_1} \right) }{\left( 1+ \frac{1}{P_\xi} \frac{ \mu_2 }{ \nu_2 } \right) \left( 1+ P_\xi \frac{ \nu_1 }{ \mu_1 } \right)},  \label{CG:mu:nu:2}
\end{align}
which is precisely Equation \eqref{CG:indiv}, thereby completing the proof.

\section{Summary of main results (Tables \ref{tbl:cum} and \ref{tbl:indiv})}

\newsavebox\JGeneral \begin{lrbox}{\JGeneral} \begin{minipage}{0.1\textwidth} \begin{align*}
J&=\bo h^T\bo G^T\left(\bo \Omega_\textsf{JD}+\frac{\bo \Omega_\textsf{P}}{P_\xi} \right)^{-1} \bo G \bo h
\end{align*}  \end{minipage} \end{lrbox}
\newsavebox\wGeneral \begin{lrbox}{\wGeneral} \begin{minipage}{0.1\textwidth} \begin{align*}
\bo w&\propto\left(\bo \Omega_\textsf{JD}+\frac{\bo \Omega_\textsf{P}}{P_\xi} \right)^{-1} \bo G \bo h
\end{align*}  \end{minipage} \end{lrbox}
\newsavebox\JCSI \begin{lrbox}{\JCSI} \begin{minipage}{0.1\textwidth} \begin{align*}
J&=\bo h^T \left(\widetilde{\bo \Sigma}+\frac{\bo \Gamma_\textsf{P}}{P_\xi} \right)^{-1} \bo h
\end{align*}  \end{minipage} \end{lrbox}
\newsavebox\wCSI \begin{lrbox}{\wCSI} \begin{minipage}{0.1\textwidth} \begin{align*}
\bo w&\propto\bo \Omega_\textsf{P}^{-1}\bo G \bo \Gamma_\textsf{P}\left(\widetilde{\bo \Sigma}+\frac{\bo \Gamma_\textsf{P}}{P_\xi} \right)^{-1} \bo h
\end{align*}  \end{minipage} \end{lrbox}
\newsavebox\JPer \begin{lrbox}{\JPer} \begin{minipage}{0.1\textwidth} \begin{align*}
J&=\bo h^T \left(\bo \Sigma+\frac{\bo \Gamma_\textsf{P}}{P_\xi} \right)^{-1} \bo h
\end{align*}  \end{minipage} \end{lrbox}
\newsavebox\wPer \begin{lrbox}{\wPer} \begin{minipage}{0.1\textwidth} \begin{align*}
\bo w&\propto\bo \Omega_\textsf{P}^{-1}\bo G \bo \Gamma_\textsf{P}\left(\bo \Sigma+\frac{\bo \Gamma_\textsf{P}}{P_\xi} \right)^{-1} \bo h
\end{align*}  \end{minipage} \end{lrbox}
\newsavebox\condCHE \begin{lrbox}{\condCHE} \begin{minipage}{0.1\textwidth} \begin{align*}
\bo A&=\mathcal C(K),       \\    \bo h&=h_0\sqrt{\alpha_\textsf{h}} \bo 1, \\ \bo \Sigma_\textsf{h}&=h_0^2 (1-\alpha_\textsf{h}) \bo I,\\
\bo \Sigma&=\sigma^2\bo R(\rho), \\ \bo g&=g_0\sqrt{\alpha_\textsf{g}} \bo 1, \\ \bo \Sigma_\textsf{g}&=g_0^2 (1-\alpha_\textsf{g}) \bo I
\end{align*}  \end{minipage} \end{lrbox}
\newsavebox\JCHE \begin{lrbox}{\JCHE} \begin{minipage}{0.1\textwidth} \begin{align*}
&J=\frac{h_0^2}{\sigma^2} \left[\rho_N+\widetilde{\alpha_\textsf{h}}\left(\rho_N+\frac{\gamma}{N}\right) \right. \\
&\qquad \qquad\left.+\frac{1}{N}\left(\widetilde{\alpha_\textsf{g}}+\frac{1}{P_\xi g_0^2 \alpha_\textsf{g}}\right)\left\{\gamma+\rho_K +\widetilde{\alpha_\textsf{h}}\left(\rho_K+\frac{\gamma}{K}\right)\right\}\right]^{-1}, \\
&\;\rho_t\triangleq \rho+\frac{1-\rho}{t},\; \gamma\triangleq\frac{\eta^2h_0^2}{\sigma^2},\; \widetilde{\alpha}\triangleq\frac{1}{\alpha}-1. \quad\quad\quad\;\;\;\; \bo w\propto\bo 1_L. \\
&\;\mbox{Note:}\; K=1\Rightarrow \mbox{Distributed},\; K=N\Rightarrow \mbox{Connected}\\
&\quad\quad\;\;\alpha_\textsf{g}=1\Rightarrow \mbox{Perfect CSI},\; \alpha_\textsf{h}=1\Rightarrow \mbox{Perfect OGI}
\end{align*}  \end{minipage} \end{lrbox}
\newsavebox\conddist \begin{lrbox}{\conddist} \begin{minipage}{0.1\textwidth} \begin{align*}
\bo A&=\bo I,  \\ \bo \Sigma &\mbox{ is diagonal} \\ \bo \Sigma_\textsf{h}&=0, \\ \bo \Sigma_\textsf{g}&=0
\end{align*}  \end{minipage} \end{lrbox}
\newsavebox\Jdist \begin{lrbox}{\Jdist} \begin{minipage}{0.1\textwidth} \begin{align*}
&J=\sum_{n=1}^N \frac{h_n^2}{\sigma_n^2}\left[1+\frac{1+\gamma_n}{P_\xi g_n^2}\right]^{-1},\\
&\;\sigma_n^2\triangleq \Sigma_{n,n},\; \gamma_n\triangleq \frac{\eta^2 h_n^2}{\sigma_n^2}
\end{align*}  \end{minipage} \end{lrbox}
\newsavebox\wdist \begin{lrbox}{\wdist} \begin{minipage}{0.1\textwidth} \begin{align*}
\bo W&\propto \textsf{diag}\left([v_1,v_2,\ldots,v_N]\right),\\
v_n&=\frac{h_n}{g_n\sigma_n^2}\left[1+\frac{1+\gamma_n}{P_\xi g_n^2}\right]^{-1}
\end{align*}  \end{minipage} \end{lrbox}
\newsavebox\Jconn \begin{lrbox}{\Jconn} \begin{minipage}{0.1\textwidth} \begin{align*}
&J= \widetilde{J} \left[1+\frac{1+\eta^2  \widetilde{J}}{\bo g^T \left(\bo \Sigma_\textsf{g}+\frac{\bo I}{P_\xi}\right)^{-1} \bo g}\right]^{-1}, \\
&\; \widetilde{J} \triangleq \bo h^T \widetilde{\bo \Sigma}^{-1} \bo h
\end{align*}  \end{minipage} \end{lrbox}
\newsavebox\wconn \begin{lrbox}{\wconn} \begin{minipage}{0.1\textwidth} \begin{align*}
\bo W\propto \bo u \bo v^T, \; \bo u&=\left(\bo \Sigma_\textsf{g}+\frac{\bo I}{P_\xi}\right)^{-1} \bo g,\\
\bo v&=\widetilde{\bo \Sigma}^{-1} \bo h
\end{align*}  \end{minipage} \end{lrbox}
\newsavebox\defn \begin{lrbox}{\defn} \begin{minipage}{0.1\textwidth} \begin{align*}
\mbox{Other definitions: }\quad \bo \Gamma_\textsf{P}\triangleq \left(\bo G^T \bo \Omega_\textsf{P}^{-1} \bo G\right)^{-1},\; \widetilde{\bo \Sigma}\triangleq \bo \Sigma+\eta^2\bo \Sigma_\textsf{h},\; \bo R(\rho)\triangleq \left((1-\rho)\bo I+\rho\bo 1\bo 1^T\right). 
\end{align*}  \end{minipage} \end{lrbox}

\begin{table*}
  \caption{Main results: cumulative power constraint (examples 1, 2 and 3) }
  \begin{center}
    \begin{tabular}{|l|l|l|l|} \hline
      Cases         & Conditions                      & \specialcell{Optimal (equivalent)\\Fisher Information} & \specialcell{Optimal weights are\\proportional to} \\ \hline \hline
      \specialcell{(A) \\ General    }  &                                      & \usebox{\JGeneral}                                                 & \usebox{\wGeneral} \\ \hline
      \specialcell{(B) \\ Perfect CSI  } & $\bo \Sigma_\textsf{g}=0$  & \usebox{\JCSI}                                                       & \usebox{\wCSI} \\ \hline
      \specialcell{(C) \\  Perfect OGI,\\Perfect CSI,\\ \cite{KarISIT12}} & \specialcell{$\bo \Sigma_\textsf{h}=0$,\\$\bo \Sigma_\textsf{g}=0$}  & \usebox{\JPer}  & \usebox{\wPer} \\ \hline
      \specialcell{(D) \\  Distributed,\\ Uncorrelated, \\Perfect OGI,\\Perfect CSI,\\ \cite{Xiao08}} & \usebox{\conddist}&  \usebox{\Jdist} & \usebox{\wdist}    \\  \hline
      \specialcell{(E) \\ Connected} & $\bo A=\bo 1 \bo 1^T$ &  \usebox{\Jconn} & \usebox{\wconn}    \\  \hline
      \specialcell{(F) \\ Cycle topology,\\ Homogeneous, \\Equicorrelated-\\ ($\bo \Sigma, \bo E_\textsf{h}, \bo E_\textsf{g}$)} & \usebox{\condCHE}& \multicolumn{2}{|l|}{ \usebox{\JCHE}}              \\  \hline
      \multicolumn{4}{|c|}{\usebox{\defn}} \\ \hline
\end{tabular} 
\end{center}   
\label{tbl:cum}
\end{table*}

\newsavebox\JdistI \begin{lrbox}{\JdistI} \begin{minipage}{0.1\textwidth} \begin{align*}
J=&F_\textsf{opt}\left(\bo a,\bo b,\bo c \right), \\
a_m&= g_m h_m,\\ 
b_m&= g_m^2\sigma_m^2, \; \sigma_m^2=\left[\bo \Sigma\right]_{m,m}, \\ 
c_m&= \sqrt{\frac{P_m^\textsf{C}}{\sigma_m^2+\eta^2 h_m^2}}
\end{align*}  \end{minipage} \end{lrbox}
\newsavebox\wdistI \begin{lrbox}{\wdistI} \begin{minipage}{0.1\textwidth} \begin{align*}
\bo W&= \textsf{diag}\left( \bo t_\textsf{opt}\right)
\end{align*}  \end{minipage} \end{lrbox}
\newsavebox\distHEI \begin{lrbox}{\distHEI} \begin{minipage}{0.1\textwidth} \begin{align*}
\bo A&=\bo I,   \\ \bo h&=h_0\sqrt{\alpha_\textsf{h}} \bo 1, \\ \bo \Sigma_\textsf{h}&=h_0^2 (1-\alpha_\textsf{h}) \bo I,\\
\bo \Sigma&=\sigma^2\bo R(\rho), \\ \bo g&=g_0\sqrt{\alpha_\textsf{g}} \bo 1, \\ \bo \Sigma_\textsf{g}&=g_0^2 (1-\alpha_\textsf{g}) \bo I
\end{align*}  \end{minipage} \end{lrbox}
\newsavebox\JdistHEI \begin{lrbox}{\JdistHEI} \begin{minipage}{0.1\textwidth} \begin{align*}
J=&\left[\frac{1}{F_\textsf{opt}\left(\bo a,\bo b,\bo c \right)}+\frac{\rho \sigma^2}{\alpha_\textsf{h} h_0^2}\right]^{-1}, \\
\bo a&= \sqrt{\alpha_\textsf{g} \alpha_\textsf{h}} g_0 h_0 \bo 1, \\
\bo b&= \sigma_\textsf{x}^2 g_0^2 (1-\alpha_\textsf{g} \alpha_\textsf{x}) \bo 1, \\
c_m&=\sqrt{\frac{P_m^\textsf{C}}{\sigma_\textsf{x}^2}}
\end{align*}  \end{minipage} \end{lrbox}
\newsavebox\JconnI \begin{lrbox}{\JconnI} \begin{minipage}{0.1\textwidth} \begin{align*}
J=&\widetilde{J} \left[1+\frac{1+\eta^2  \widetilde{J}}{ F_\textsf{opt}\left(\bo a,\bo b,\bo c \right) }\right]^{-1}, \\
\widetilde{J}&\triangleq \bo h^T \widetilde{\bo \Sigma}^{-1} \bo h, \\
\bo a&= \bo g,\; \bo b=\textsf{diag}\left(\bo \Sigma_\textsf{g} \right), c_m=\sqrt{P_m^\textsf{C}}
\end{align*}  \end{minipage} \end{lrbox}
\newsavebox\wconnI \begin{lrbox}{\wconnI} \begin{minipage}{0.1\textwidth} \begin{align*}
\bo W=&\kappa \bo t_\textsf{opt} \bo v^T, \\
 \bo v&=\widetilde{\bo \Sigma}^{-1} \bo h,\\
\kappa&=\frac{1}{\sqrt{\widetilde{J}(1+\eta^2 \widetilde{J})}}
\end{align*}  \end{minipage} \end{lrbox}
\newsavebox\defnI \begin{lrbox}{\defnI} \begin{minipage}{0.1\textwidth} \begin{align*}
\mbox{Other definitions: }\quad \widetilde{\bo \Sigma}\triangleq \bo \Sigma+\eta^2\bo \Sigma_\textsf{h},\; \bo R(\rho)\triangleq \left((1-\rho)\bo I+\rho\bo 1\bo 1^T\right),\; \sigma_\textsf{x}^2=\sigma^2+\eta^2 h_0^2, \; \alpha_\textsf{x}=\frac{\rho \sigma^2+\alpha_\textsf{h} \eta^2 h_0^2 }{\sigma_\textsf{x}^2}
\end{align*}  \end{minipage} \end{lrbox}

\begin{table*}
  \caption{Main results: individual power constraint (Examples 4, 5 and 6) }
  \begin{center}
    \begin{tabular}{|l|l|l|l|} \hline
      Cases         & Conditions                      & \specialcell{Optimal (equivalent)\\Fisher Information} & \specialcell{Optimal weights} \\ \hline \hline
      \specialcell{(A) \\ Distributed,\\ Uncorrelated, \\Perfect OGI,\\Perfect CSI,\\ \cite{Jing09}} & \usebox{\conddist}&  \usebox{\JdistI} & \usebox{\wdistI}    \\  \hline
      \specialcell{(B) \\ Distributed,\\ Homogeneous, \\Equicorrelated-\\ ($\bo \Sigma, \bo E_\textsf{h}, \bo E_\textsf{g})$} & \usebox{\distHEI}&  \usebox{\JdistHEI}  & \usebox{\wdistI}             \\  \hline
      \specialcell{(C) \\ Connected,\\Uncorrelated-\\channel-gain} & \specialcell{$\bo A=\bo 1 \bo 1^T$,\\ $\bo \Sigma_\textsf{g}$ is diagonal} &  \usebox{\JconnI} & \usebox{\wconnI}    \\  \hline
      \multicolumn{4}{|c|}{\usebox{\defnI}} \\ \hline
\end{tabular} 
\end{center}   \label{tbl:indiv}
\end{table*}

\bibliographystyle{IEEEtran}
\bibliography{thesis}

\end{document}